\documentstyle[epsfig,a4p,12pt,rotating]{article}

\begin{document}         

\newcommand{\Rparity}{$R$-parity}
\newcommand{\Rp}  {$R_{p}$}
\newcommand{\lb}  {$\lambda$}
\newcommand{\lbp} {$\lambda^{'}$}
\newcommand{\lbpp} {$\lambda^{''}$}

\newcommand{\ee}{{\mathrm e}^+ {\mathrm e}^-}
\newcommand{\sq}{\tilde{\mathrm q}}
\newcommand{\seff}{\tilde{\mathrm f}}
\newcommand{\sele}{\tilde{\mathrm e}}
\newcommand{\supq}{\tilde{\mathrm u}}
\newcommand{\sdown}{\tilde{\mathrm d}}
\newcommand{\sfer}{\tilde{\mathrm f}}
\newcommand{\sstrange}{\tilde{\mathrm s}}
\newcommand{\scharm}{\tilde{\mathrm c}}
\newcommand{\sbottom}{\tilde{\mathrm b}}
\newcommand{\ellp}{\ell^+}
\newcommand{\ellm}{\ell^-}
\newcommand{\sell}{\tilde{\ell}}
\newcommand{\snu}{\tilde{\nu}}
\newcommand{\smu}{\tilde{\mu}}
\newcommand{\stau}{\tilde{\tau}}
\newcommand{\chp}{\tilde{\chi}^+_1}
\newcommand{\chip}{\tilde{\chi}^+_1}
\newcommand{\chim}{\tilde{\chi}^-_1}
\newcommand{\chpm}{\tilde{\chi}^\pm_1}
\newcommand{\chipm}{\tilde{\chi}^\pm_1}
\newcommand{\nt}{\tilde{\chi}^0}
\newcommand{\qq}{{\mathrm q}\bar{\mathrm q}}
\newcommand{\sleppair}{\sell^+ \sell^-}
\newcommand{\nunu}{\nu \bar{\nu}}
\newcommand{\mumu}{\mu^+ \mu^-}
\newcommand{\tautau}{\tau^+ \tau^-}
\newcommand{\ellell}{\ell^+ \ell^-}
\newcommand{\nulqq}{\nu \ell {\mathrm q} \bar{\mathrm q}'}
\newcommand{\MZ}{M_{\mathrm Z}}

\newcommand {\stopm}         {\tilde{\mathrm{t}}_{1}}
\newcommand {\stopn}         {\tilde{\mathrm{t}}}
\newcommand {\stops}         {\tilde{\mathrm{t}}_{2}}
\newcommand {\stopbar}       {\bar{\tilde{\mathrm{t}}}_{1}}
\newcommand {\stopx}         {\tilde{\mathrm{t}}}
\newcommand {\sneutrino}     {\tilde{\nu}}
\newcommand {\slepton}       {\tilde{\ell}}
\newcommand {\stopl}         {\tilde{\mathrm{t}}_{\mathrm L}}
\newcommand {\stopr}         {\tilde{\mathrm{t}}_{\mathrm R}}
\newcommand {\stoppair}      {\tilde{\mathrm{t}}_{1}
\bar{\tilde{\mathrm{t}}}_{1}}
\newcommand {\gluino}        {\tilde{\mathrm g}}

\newcommand {\chin}          {\tilde{\chi }^{0}_{1}}
\newcommand {\neutralino}    {\tilde{\chi }^{0}_{1}}
\newcommand {\neutrala}      {\tilde{\chi }^{0}_{2}}
\newcommand {\neutralb}      {\tilde{\chi }^{0}_{3}}
\newcommand {\neutralc}      {\tilde{\chi }^{0}_{4}}
\newcommand {\bino}          {\tilde{\mathrm B}^{0}}
\newcommand {\wino}          {\tilde{\mathrm W}^{0}}
\newcommand {\higginoa}      {\tilde{\rm H_{1}}^{0}}
\newcommand {\higginob}      {\tilde{\mathrm H_{1}}^{0}}
\newcommand {\chargino}      {\tilde{\chi }^{\pm}_{1}}
\newcommand {\charginop}     {\tilde{\chi }^{+}_{1}}
\newcommand {\KK}            {{\mathrm K}^{0}-\bar{\mathrm K}^{0}}
\newcommand {\ff}            {{\mathrm f} \bar{\mathrm f}}
\newcommand {\bstopm} {\mbox{$\boldmath {\tilde{\mathrm{t}}_{1}} $}}
\newcommand {\Mt}            {M_{\mathrm t}}
\newcommand {\mscalar}       {m_{0}}
\newcommand {\Mgaugino}      {M_{1/2}}
\newcommand {\rs}            {\sqrt{s}}
\newcommand {\WW}            {{\mathrm W}^+{\mathrm W}^-}
\newcommand {\MGUT}          {M_{\mathrm {GUT}}}
\newcommand {\Zboson}        {{\mathrm Z}^{0}}
\newcommand {\Wpm}           {{\mathrm W}^{\pm}}
\newcommand {\allqq}         {\sum_{q \neq t} q \bar{q}}
\newcommand {\mixang}        {\theta _{\mathrm {mix}}}
\newcommand {\thacop}        {\theta _{\mathrm {Acop}}}
\newcommand {\cosjet}        {\cos\thejet}
\newcommand {\costhr}        {\cos\thethr}
\newcommand {\djoin}         {d_{\mathrm{join}}}
\newcommand {\mstop}         {m_{\stopm}}
\newcommand {\msell}         {m_{\sell}}
\newcommand {\mchi}          {m_{\neutralino}}
\newcommand {\pp}{p \bar{p}}

\newcommand{\epair}{\mbox{${\mathrm e}^+{\mathrm e}^-$}}
\newcommand{\mupair}{\mbox{$\mu^+\mu^-$}}
\newcommand{\taupair}{\mbox{$\tau^+\tau^-$}}
\newcommand{\qpair}{\mbox{${\mathrm q}\overline{\mathrm q}$}}
\newcommand{\eeee}{\mbox{\epair\epair}}
\newcommand{\eemumu}{\mbox{\epair\mupair}}
\newcommand{\eetautau}{\mbox{\epair\taupair}}
\newcommand{\eeqq}{\mbox{\epair\qpair}}
\newcommand{\fs}{ final states}
\newcommand{\epairf}{\mbox{\epair\fs}}
\newcommand{\mupairf}{\mbox{\mupair\fs}}
\newcommand{\taupairf}{\mbox{\taupair\fs}}
\newcommand{\qpairf}{\mbox{\qpair\fs}}
\newcommand{\eeeef}{\mbox{\eeee\fs}}
\newcommand{\eemumuf}{\mbox{\eemumu\fs}}
\newcommand{\eetautauf}{\mbox{\eetautau\fs}}
\newcommand{\eeqqf}{\mbox{\eeqq\fs}}
\newcommand{\ffff}{four fermion final states}
\newcommand{\llnunu}{\mbox{\lpair\nul\nubar}}
\newcommand{\lnuqq}{\mbox{\lept\nubar\qpair}}
\newcommand{\zee}{\mbox{Zee}}
\newcommand{\zzg}{\mbox{ZZ/Z$\gamma$}}
\newcommand{\wenu}{\mbox{We$\nu$}}

\newcommand{\el}{\mbox{${\mathrm e}^-$}}
\newcommand{\selem}{\mbox{$\tilde{\mathrm e}^-$}}
\newcommand{\smum}{\mbox{$\tilde\mu^-$}}
\newcommand{\staum}{\mbox{$\tilde\tau^-$}}
\newcommand{\slept}{\mbox{$\tilde{\ell}^\pm$}}
\newcommand{\sleptm}{\mbox{$\tilde{\ell}^-$}}
\newcommand{\lept}{\mbox{$\ell^-$}}
\newcommand{\Hl}{\mbox{$\mathrm{L}^\pm$}}
\newcommand{\Hm}{\mbox{$\mathrm{L}^-$}}
\newcommand{\Hnu}{\mbox{$\nu_{\mathrm{L}}$}}
\newcommand{\nul}{\mbox{$\nu_\ell$}}
\newcommand{\nubar}{\mbox{$\overline{\nu}_\ell$}}
\newcommand{\spair}{\mbox{$\tilde{\ell}^+\tilde{\ell}^-$}}
\newcommand{\lpair}{\mbox{$\ell^+\ell^-$}}
\newcommand{\staupair}{\mbox{$\tilde{\tau}^+\tilde{\tau}^-$}}
\newcommand{\smupair}{\mbox{$\tilde{\mu}^+\tilde{\mu}^-$}}
\newcommand{\selepair}{\mbox{$\tilde{\mathrm e}^+\tilde{\mathrm e}^-$}}
\newcommand{\ch}{\mbox{$\tilde{\chi}^\pm_1$}}
\newcommand{\chpair}{\mbox{$\tilde{\chi}^+_1\tilde{\chi}^-_1$}}
\newcommand{\chm}{\mbox{$\tilde{\chi}^-_1$}}
\newcommand{\chmp}{\mbox{$\tilde{\chi}^\pm_1$}}
\newcommand{\chz}{\mbox{$\tilde{\chi}^0_1$}}
\newcommand{\dch}{\mbox{\chm$\rightarrow$\chz\lept\nubar}}
\newcommand{\dslept}{\mbox{\sleptm$\rightarrow$\chz\lept}}
\newcommand{\dH}{\mbox{\Hm$\rightarrow$\lept\nubar\Hnu}}
\newcommand{\mch}{\mbox{$m_{\tilde{\chi}^\pm_1}$}}
\newcommand{\mslept}{\mbox{$m_{\tilde{\ell}}$}}
\newcommand{\mstau}{\mbox{$m_{\staum}$}}
\newcommand{\msmu}{\mbox{$m_{\smum}$}}
\newcommand{\msele}{\mbox{$m_{\selem}$}}
\newcommand{\mchz}{\mbox{$m_{\tilde{\chi}^0_1}$}}
\newcommand{\dm}{\mbox{$\Delta m$}}
\newcommand{\dmch}{\mbox{$\Delta m_{\ch-\chz}$}}
\newcommand{\dmslept}{\mbox{$\Delta m_{\slept-\chz}$}}
\newcommand{\dmhl}{\mbox{$\Delta m_{\Hl-\Hnu}$}}
\newcommand{\w}{\mbox{W$^\pm$}}

\newcommand{\acopc}{\mbox{$\phi^{\mathrm{acop}}$}}
\newcommand{\acolc}{\mbox{$\theta^{\mathrm{acol}}$}}
\newcommand{\acop}{\mbox{$\phi_{\mathrm{acop}}$}}
\newcommand{\acol}{\mbox{$\theta_{\mathrm{acol}}$}}
\newcommand{\pt}{\mbox{$p_{t}$}}
\newcommand{\pz}{\mbox{$p_{\mathrm{z}}^{\mathrm{miss}}$}}
\newcommand{\ptevt}{\mbox{$p_{t}^{\mathrm{miss}}$}}
\newcommand{\ptaxic}{\mbox{$a_{t}^{\mathrm{miss}}$}}
\newcommand{\stevt}{\mbox{$p_{t}^{\mathrm{miss}}$/\Ebeam}}
\newcommand{\staxic}{\mbox{$a_{t}^{\mathrm{miss}}$/\Ebeam}}
\newcommand{\dptaxic}{\mbox{missing $p_{t}$ wrt. event axis \ptaxic}}
\newcommand{\cosevt}{\mbox{$\mid\cos\theta_{\mathrm{p}}^{\mathrm{miss}}\mid$}}
\newcommand{\axicos}{\mbox{$\mid\cos\theta_{\mathrm{a}}^{\mathrm{miss}}\mid$}}
\newcommand{\pthet}{\mbox{$\theta_{\mathrm{p}}^{\mathrm{miss}}$}}
\newcommand{\athet}{\mbox{$\theta_{\mathrm{a}}^{\mathrm{miss}}$}}
\newcommand{\dcosevt}{\mbox{$\mid\cos\theta\mid$ of missing p$_{t}$}}
\newcommand{\daxicos}{\mbox{$\mid\cos\theta\mid$ of missing p$_{t}$ wrt. event
axis}}
\newcommand{\efdsw}{\mbox{$x_{\mathrm{FDSW}}$}}
\newcommand{\acopf}{\mbox{$\Delta\phi_{\mathrm{FDSW}}$}}
\newcommand{\acopm}{\mbox{$\Delta\phi_{\mathrm{MUON}}$}}
\newcommand{\acopt}{\mbox{$\Delta\phi_{\mathrm{trk}}$}}
\newcommand{\po}{\mbox{$E_{\mathrm{isol}}^\gamma$}}
\newcommand{\qprod}{\mbox{$q1$$*$$q2$}}
\newcommand{\lcode}{lepton identification code}
\newcommand{\nctro}{\mbox{$N_{\mathrm{trk}}^{\mathrm{out}}$}}
\newcommand{\necao}{\mbox{$N_{\mathrm{ecal}}^{\mathrm{out}}$}}
\newcommand{\mout}{\mbox{$m^{\mathrm{out}}$}}
\newcommand{\nctec}{\mbox{\nctro$+$\necao}}
\newcommand{\gfract}{\mbox{$F_{\mathrm{good}}$}}
\newcommand{\zz}       {\mbox{$|z_0|$}}
\newcommand{\dz}       {\mbox{$|d_0|$}}
\newcommand{\sint}      {\mbox{$\sin\theta$}}
\newcommand{\cost}      {\mbox{$\cos\theta$}}
\newcommand{\mcost}     {\mbox{$|\cos\theta|$}}
\newcommand{\dedx}     {\mbox{$dE/dx$}}
\newcommand{\wdedx}     {\mbox{$W_{dE/dx}$}}
\newcommand{\xe}     {\mbox{$x_E$}}

\newcommand{\ssix}     {\mbox{$\sqrt{s}$~=~161~GeV}}
\newcommand{\sthree}     {\mbox{$\sqrt{s}$~=~130--136~GeV}}
\newcommand{\mrecoil}     {\mbox{$m_{\mathrm{recoil}}$}}
\newcommand{\llmass}     {\mbox{$m_{ll}$}}
\newcommand{\sml}{\mbox{Standard Model \lpair$\nu\nu$ events}}
\newcommand{\sme}{\mbox{Standard Model events}}
\newcommand{\sig}{events containing a lepton pair plus missing transverse momentum}
\newcommand{\wpair}{\mbox{$W^+W^-$}}
\newcommand{\dW}{\mbox{W$^-\rightarrow$\lept\nubar}}
\newcommand{\dsele}{\mbox{\selem$\rightarrow$\chz e$^-$}}
\newcommand{\eeeell}{\mbox{\epair$\rightarrow$\epair\lpair}}
\newcommand{\eell}{\mbox{\epair\lpair}}
\newcommand{\llgam}{\mbox{$\ell\ell(\gamma)$}}
\newcommand{\nunugam}{\mbox{$\nu\bar{\nu}\gamma\gamma$}}
\newcommand{\acope}{\mbox{$\Delta\phi_{\mathrm{EE}}$}}
\newcommand{\nee}{\mbox{N$_{\mathrm{EE}}$}}
\newcommand{\eesum}{\mbox{$\Sigma_{\mathrm{EE}}$}}
\newcommand{\at}{\mbox{$a_{t}$}}
\newcommand{\spp}{\mbox{$p$/\Ebeam}}
\newcommand{\acoph}{\mbox{$\Delta\phi_{\mathrm{HCAL}}$}}

\newcommand{\roots}     {\sqrt{s}}
%
%
\newcommand{\thrust}    {T}
\newcommand{\nthrust}   {\hat{n}_{\mathrm{thrust}}}
\newcommand{\thethr}    {\theta_{\,\mathrm{thrust}}}
\newcommand{\phithr}    {\phi_{\mathrm{thrust}}}
\newcommand{\acosthr}   {|\cos\thethr|}
\newcommand{\thejet}    {\theta_{\,\mathrm{jet}}}
\newcommand{\acosjet}   {|\cos\thejet|}
\newcommand{\thmiss}    { \theta_{\mathrm{miss}} }
\newcommand{\cosmiss}   {| \cos \thmiss |}

\newcommand{\Evis}      {E_{\mathrm{vis}}}
\newcommand{\Rvis}      {E_{\mathrm{vis}}\,/\roots}
\newcommand{\Mvis}      {m_{\mathrm{vis}}}
\newcommand{\Rbal}      {R_{\mathrm{bal}}}

\newcommand{\Ecm}{\mbox{$E_{\mathrm{cm}}$}}
\newcommand{\Ebeam}{\mbox{$E_{\mathrm{beam}}$}}
\newcommand{\ipb}{\mbox{pb$^{-1}$}}
\newcommand{\wrt}{with respect to}
\newcommand{\sm}{Standard Model}
\newcommand{\smb}{Standard Model background}
\newcommand{\smp}{Standard Model processes}
\newcommand{\smc}{Standard Model Monte Carlo}
\newcommand{\mc}{Monte Carlo}
\newcommand{\btb}{back-to-back}
\newcommand{\tp}{two-photon}
\newcommand{\tpb}{two-photon background}
\newcommand{\tpp}{two-photon processes}
\newcommand{\lp}{lepton pairs}
\newcommand{\vto}{\mbox{$\tau$ veto}}
\newcommand{\gsim}{\;\raisebox{-0.9ex}
           {$\textstyle\stackrel{\textstyle >}{\sim}$}\;}
\newcommand{\lsim}{\;\raisebox{-0.9ex}{$\textstyle\stackrel{\textstyle<}
           {\sim}$}\;}
\newcommand{\degree}    {^\circ}

\newcommand{\phiacop}   {\phi_{\mathrm{acop}}}


%
%
\newcommand{\ZP}[3]    {Z. Phys. {\bf C#1} (#2) #3.}
\newcommand{\PL}[3]    {Phys. Lett. {\bf B#1} (#2) #3.}
\newcommand{\etal}     {{\it et al}.,\,\ }
\newcommand{\PhysLett}  {Phys.~Lett.}
\newcommand{\PRL} {Phys.~Rev.\ Lett.}
\newcommand{\PhysRep}   {Phys.~Rep.}
\newcommand{\EuroPhys}  {Eur.~Phys. \ J.}
\newcommand{\PhysRev}   {Phys.~Rev.}
\newcommand{\NPhys}  {Nucl.~Phys.}
\newcommand{\NIM} {Nucl.~Instr.\ Meth.}
\newcommand{\CPC} {Comp.~Phys.\ Comm.}
\newcommand{\ZPhys}  {Z.~Phys.}
\newcommand{\IEEENS} {IEEE Trans.\ Nucl.~Sci.}
%
%
\newcommand{\OPALColl}  {OPAL Collab.}
\newcommand{\ALEPHColl}  {ALEPH Collab.}
\newcommand{\JADEColl}  {JADE Collab.}
%
\newcommand{\onecol}[2] {\multicolumn{1}{#1}{#2}}
\newcommand{\ra}        {\rightarrow}   


 


%
%
\newcommand{\PPEnum}    {CERN-EP/99-043}
\newcommand{\PNnum}     {OPAL Physics Note PN-359}
\newcommand{\TNnum}     {OPAL Technical Note TN-xxx}
\newcommand{\Date}      {March 16th, 1999}

\newcommand{\Author}    {S.~Braibant, I.~Fleck, M.~Fierro, S.~Shotkin-Gascon,\\
P.~Giacomelli, G.~P\'asztor}
\newcommand{\MailAddr}  {Sylvie.Braibant@cern.ch or Paolo.Giacomelli@cern.ch}
\newcommand{\EdBoard}   {F.~Fiedler, M.Gruw\'{e}, T.~Junk, B.~List}
\newcommand{\DraftVer}  {Version 3.0}
\newcommand{\DraftDate} {\today}
\newcommand{\TimeLimit} {Wednesday, March 3rd, 19.00pm.}


\begin{titlepage}
%
%
\begin{center}
    \large
    EUROPEAN LABORATORY FOR PARTICLE PHYSICS 
\end{center}
\begin{flushright}
    \large
    \PPEnum\\
    \Date
\end{flushright}

%
%
\bigskip
\bigskip
\bigskip
\bigskip
\begin{center}
    \huge\bf\boldmath
 Search for R-Parity Violating Decays \\
      of Scalar Fermions at LEP

\normalsize

\vspace{0.5cm}

\LARGE

The OPAL Collaboration \\

\end{center}

\vspace{1.0cm}

%
%
\begin{abstract}

A search for pair produced scalar fermions with couplings 
that violate \Rparity\  has been performed 
using a data sample corresponding to an integrated luminosity of
56~pb$^{-1}$ at a centre-of-mass energy of $\sqrt{s}=$ 183 GeV 
collected with the OPAL detector at LEP. 
An important consequence of \Rparity\ breaking interactions is that 
the lightest supersymmetric particle is expected to be unstable. 
Searches for \Rparity\ violating decays of charged sleptons, 
sneutrinos and stop quarks have been 
performed under the assumptions that the 
lightest supersymmetric particle decays promptly and that only one of 
the \Rparity\ violating couplings is dominant for each of the decay 
modes considered. Such processes would yield multi-leptons, 
jets plus leptons or multi-jets, with 
or without 
missing energy, in the final state.
No significant excess of such events has been observed.
Limits on the production cross-sections of 
scalar fermions in \Rparity\ violating scenarios are obtained. 
Mass exclusion regions are also presented in the framework
of the Constrained Minimal Supersymmetric Standard Model.
\end{abstract}

\vspace{1.5cm}

\begin{center}
{\large  (Submitted to E. Phys. C.)}
\end{center}

\bigskip
\smallskip
 
\bigskip
\bigskip
\bigskip
\smallskip



\end{titlepage}

\begin{center}{\Large        The OPAL Collaboration
}\end{center}\bigskip
\begin{center}{
G.\thinspace Abbiendi$^{  2}$,
K.\thinspace Ackerstaff$^{  8}$,
G.\thinspace Alexander$^{ 23}$,
J.\thinspace Allison$^{ 16}$,
N.\thinspace Altekamp$^{  5}$,
K.J.\thinspace Anderson$^{  9}$,
S.\thinspace Anderson$^{ 12}$,
S.\thinspace Arcelli$^{ 17}$,
S.\thinspace Asai$^{ 24}$,
S.F.\thinspace Ashby$^{  1}$
% ...
}\end{center}\bigskip
\bigskip
$^{  1}$School of Physics and Astronomy, University of Birmingham,
Birmingham B15 2TT, UK
\newline
$^{  2}$Dipartimento di Fisica dell' Universit\`a di Bologna and INFN,
I-40126 Bologna, Italy
\newline
% ...
$^{  5}$Cavendish Laboratory, Cambridge CB3 0HE, UK
\newline
% ...
$^{  8}$CERN, European Organisation for Particle Physics,
CH-1211 Geneva 23, Switzerland
\newline
$^{  9}$Enrico Fermi Institute and Department of Physics,
University of Chicago, Chicago IL 60637, USA
\newline
% ...
$^{ 12}$Indiana University, Department of Physics,
Swain Hall West 117, Bloomington IN 47405, USA
\newline
% ...
$^{ 16}$Department of Physics, Schuster Laboratory, The University,
Manchester M13 9PL, UK
\newline
$^{ 17}$Department of Physics, University of Maryland,
College Park, MD 20742, USA
\newline
% ...
$^{ 23}$Department of Physics and Astronomy, Tel Aviv University,
Tel Aviv 69978, Israel
\newline
$^{ 24}$International Centre for Elementary Particle Physics and
Department of Physics, University of Tokyo, Tokyo 113-0033, and
Kobe University, Kobe 657-8501, Japan
\newline
% ...
\newline
\newpage

\section{Introduction}
\label{sec:intro}

In Supersymmetric (SUSY)~\cite{ref:SUSY} models each elementary
particle is accompanied by a
supersymmetric partner whose spin differs by half a unit.
Most of the searches for these supersymmetric particles (``sparticles'') 
are performed within the Minimal Supersymmetric extension of the 
\sm\ (MSSM)~\cite{ref:MSSM}, assuming \Rparity\ conservation. 
\Rparity~\cite{ref:rparity} is a new multiplicative quantum number defined as 
$R_{p}=(-1)^{2S+3B+L}$ where $S$, $B$ and $L$ are the spin, baryon and
lepton number of the particle, respectively. 
\Rparity\ discriminates between ordinary
and supersymmetric particles: \Rp\ = +1 for the \sm\ particles 
and \Rp\ = --1 for their supersymmetric partners.
\Rparity\ conservation implies that supersymmetric particles are 
always pair produced and always decay through
cascade decays to ordinary particles and the lightest supersymmetric 
particle (LSP). In this context, the LSP is often assumed to be
the lightest neutralino, $\neutralino$, which is then expected to be 
stable and to escape detection due to its weakly interacting nature. 
The  characteristic signature of the supersymmetric \Rparity\ 
conserving decays is therefore missing energy. 

In this paper, the possible direct manifestations of \Rparity\
breaking couplings via processes with 
distinct signatures are studied. 
If \Rparity\ is violated, sparticles can decay directly to \sm\ 
particles. Therefore, the signatures sought in the analyses of 
this paper differ from the 
missing energy signatures of \Rparity\ conserving processes. 

With the MSSM particle content,  \Rparity\ violating interactions are 
parametrised with a gauge-invariant superpotential that includes the 
following Yukawa coupling terms~\cite{ref:dreiner1}: 

\begin{eqnarray}
{\sl W}_{RPV}  = 
    \lambda_{ijk}      L_i L_j {\overline E}_k
 +  \lambda^{'}_{ijk}  L_i Q_j {\overline D}_k
 +  \lambda^{''}_{ijk} {\overline U}_i {\overline D}_j {\overline D}_k, 
\label{lagrangian}
\end{eqnarray}
where $i,j,k$ are the generation indices of the superfields 
$L, Q,E,D$ and $U$. $L$ and $Q$ are lepton and quark left-handed doublets,  
respectively. 
$\overline E$, $\overline D$ and $\overline U$ are right-handed 
singlet charge-conjugate superfields for the charged 
leptons and down- and up-type quarks, respectively. 
The interactions corresponding to these superpotential terms are
assumed to respect the gauge symmetry SU(3)$_{\rm C}$ $\times$ 
SU(2)$_{\rm L}$ $\times$ U(1)$_{\rm Y}$ of the \sm.
The \lb$_{ijk}$ are 
non-vanishing only if $i < j$, so that at least two different 
generations are coupled in the purely leptonic vertices. 
The \lbpp$_{ijk}$ are 
are non-vanishing only for for $ j < k $.
The \lb\ and \lbp\ couplings both violate lepton number ($L$) conservation
and the \lbpp\ couplings violate baryon number ($B$) conservation. 
There are nine \lb\ couplings for the triple lepton vertices, 27 \lbp\ 
couplings for the lepton-quark-quark vertices and nine \lbpp\ couplings for the 
triple quark vertices. There are therefore a total of 45 new 
\Rparity\ violating couplings.  
In the constrained MSSM framework~\footnote{
        The constrained MSSM implies a common gaugino mass
        and a common sfermion mass at the GUT scale.},
there are five initial parameters completely determining all sparticle masses 
and couplings. 

Recently, supersymmetric models with \Rparity\ violation (RPV) have 
attracted considerable theoretical and phenomenological interest
(see for instance~\cite{ref:dreiner1}). 
Indeed, there exist no theoretical or experimental arguments excluding 
\Rparity\ violation~\cite{ref:opal_rpv_gauginos,ref:aleph_rpv_lle, 
ref:aleph_rpv_lqd}. 
Therefore, it is 
important to consider the phenomenology of possible \Rparity\ 
violating scenarios. The branching ratios of some of the \Rparity\ 
violating decay modes can be comparable or even larger than 
\Rparity\ conserving modes. For example, this could be the case 
for the scalar top 
quark (``stop") decay modes to third-generation fermions.  

From the experimental point of view, there are several 
upper bounds\footnote{All quoted limits are given for a sparticle mass
of 100~GeV.}
on the \Rparity\ violating Yukawa couplings, \lb, \lbp\ and \lbpp. 
A list of upper limits on individual couplings
can be found 
in~\cite{ref:bhatta,ref:barger1,ref:agashe,ref:godbole,ref:ellis}.  
Most of the upper limits on the couplings are of 
${\cal{O}} (10^{-2})$, but there also exist 
some more stringent limits. 
For instance, \lbp$_{111} < 10^{-4}$ from
neutrinoless double beta decay~\cite{ref:mohapatra}, 
\lbpp$_{112} < 10^{-6}$~\cite{ref:goity} from  double 
nucleon decay and \lbpp$_{113} < 10^{-4}$~\cite{ref:goity} from
limits on
$n-\overline{n}$ oscillation.  
Most of the couplings are constrained by
experimental results but
most of these upper bounds are still high compared to the sensitivity 
attainable with direct searches at LEP (of ${\cal{O}} (10^{-5})$). 
Furthermore the simultaneous presence 
of the couplings \lbpp\ ($B$-violating) and \lbp\ ($L$-violating) is
forbidden since it would allow fast squark-mediated proton decay at 
tree level. The experimental non-observation of proton decay 
places strong bounds on the product of these two couplings, i.e., 
\lbp $\times$ \lbpp $<$ 10$^{-10}$~\cite{ref:smir}. 

Although pair production is not required with \Rparity\ violation, 
only searches for \Rparity\ violating decays of pair-produced scalar 
fermions  (``sfermions"), such as the charged and neutral scalar leptons  
and scalar top quark,
are presented in this paper.
Their production is fully determined by gauge couplings and their masses. 
Supersymmetric particles can also be singly produced and, for example,
indirect limits from the OPAL two-fermion pair-production 
cross-section measurements are given in~\cite{opal_2fermion}.

Two different scenarios are probed. In the first scenario, the decays 
of sfermions via the lightest neutralino, $\nt_1$, are considered, 
where $\nt_1$ is
treated as the LSP and assumed to decay via an \Rparity\ violating
interaction.
These are denoted ``indirect decays''. 
SUSY cascade decays via particles other than the LSP are not 
considered.
In the second scenario,
``direct'' decays
of sparticles to \sm\ particles are investigated.  
In this case, the sparticle considered is assumed
to be the LSP, such that \Rparity\ conserving decay modes do not
contribute. 
In both scenarios, it is assumed that only one of the 
45 Yukawa-like couplings is non zero at a time. 

Only values of the Yukawa-like \lb-couplings larger than 
$ {\cal{O}} (10^{-5})$ are relevant to this analysis. 
For smaller couplings,  the lifetime of sparticles would be
sufficiently long 
to produce a secondary decay vertex, clearly detached 
from the primary vertex, or even outside the detector. These topologies
have not been considered in this paper, but decays outside the 
detector have been treated elsewhere \cite{ref:stable-part}.

In this paper, the data produced in $\ee$ collisions at LEP and 
collected with the OPAL detector 
during 1997 at a
centre-of-mass energy of 
$\sqrt{s} \simeq$183~GeV are analysed. These data correspond 
to an integrated luminosity of
about 56~pb$^{-1}$. 
The production and \Rparity\ violating decays
of $\sell$, $\snu$ via \lb\ and \lbp\ and $\stopx$ via \lbp\ and 
\lbpp\ are described 
in Section~\ref{sec:production}, 
together with the possible signal topologies 
resulting from these processes. The 
signal and background Monte Carlo simulations used in the 
different analyses are described in Section~\ref{sec:MC},
and a short description of the 
OPAL detector follows in Section~\ref{sec:opaldet}. 
Sections~\ref{sec:multileptons}, \ref{sec:2jets2leptons},
\ref{sec:jetsleptons}  
,\ref{sec:multijetsemiss} and \ref{sec:multijetsnoemiss}
describe the specific analyses optimised to search for \Rparity\ violating
processes. The physics interpretation is given in 
Section~\ref{sec:results} which presents 
cross-section limits and interpretations in the MSSM.

\section{Sparticle Production and Decays}
\label{sec:production}

In this section, the production and decay modes of different sfermion
species are discussed. 
The decay modes that result from \lb, \lbp\ and \lbpp\
couplings are presented. 
Table~\ref{tab:relation} summarises 
the production and decay mechanisms as well as the coupling 
involved in the decay, 
the final state topologies searched for, 
and the analysis names as used in 
Sections~\ref{sec:multileptons}, \ref{sec:2jets2leptons}, 
\ref{sec:jetsleptons}, 
\ref{sec:multijetsemiss} and \ref{sec:multijetsnoemiss}. 
In the indirect decays, the particles
resulting from the $\nt_1$ decay are put in parentheses. 

\begin{table}[htbp]
\begin{center}
\begin{tabular}{|ll|l|l|c|}
\hline
Production and Decay  & & Coupling & Topology &  Analysis \\
\hline

\hline 
$ \sell^+ \sell^- \rightarrow $ &  
$ \nu  \ell   $  $ \nu  \ell   $ & \lb\ direct &
2 $\ell$ + $E_{miss}$ 
& (A) \ref{sec:2leptons}\\

\hline
$ \snu \snu \rightarrow  $ $\nu \chin $ $ \nu \chin \ra $ & 
$\nu (\nu \ellp \ellm)$  $\nu (\nu \ellp \ellm)$  &  
\lb\ indirect & 
4 $\ell$ + $E_{miss}$ & 
(B) \ref{sec:4leptons} \\

$ \snu\snu \rightarrow $ & 
$ \ellp \ellm  $  $  \ellp \ellm  $  &  \lb\ direct  & 
4 $\ell$  & 
(C) \ref{sec:4leptons} \\ 

\hline
$ \sell^+ \sell^- \rightarrow $   
$ \ell^+ \chin  \ell^- \chin$  $\ra$  &
$\ellp (\nu \ellp \ellm)$  $\ellm (\nu \ellp \ellm)$ &  \lb\ indirect & 
6 $\ell$ + $E_{miss}$ & 
(D) \ref{sec:6leptons} \\
\hline

\hline
$ \stoppair \rightarrow $   &
$ e^+ q  $  $ e^- q  $ &  \lbp\ direct & 
2 e + 2 jets  &
(E) \ref{sec:stopemu} \\

$ \stoppair \rightarrow $   &
$ \mu^+ q  $  $ \mu^- q  $  & \lbp\ direct & 
2 $\mu$ + 2 jets  &
(E) \ref{sec:stopemu} \\

$ \stoppair \rightarrow $   &
$ \tau^+ q  $  $ \tau^- q  $ & \lbp\ direct  & 
2 $\tau$ + 2 jets  &
(F) \ref{sec:stoptau} \\

\hline

\hline
$ \stau^+ \stau^- \rightarrow $   
$ \tau^+  \chin \tau^- \chin$   $\ra$  &
$\tau^+ (\ell q q)$ $\tau^- (\ell q q)$ &  \lbp\ indirect  & 
$\tau$ + jets &
(F) \ref{sec:jetstau} \\
&
$ \tau^+ (\nu q q)$ $ \tau^- (\ell q q)$  & \lbp\ indirect & 
$\tau$ + jets &
(F) \ref{sec:jetstau} \\
&
$ \tau^+ (\nu q q)$ $  \tau^- (\nu q q)$  & \lbp\ indirect & 
$\tau$ + jets &
(F) \ref{sec:jetstau} \\


$ \sell^+ \sell^- \rightarrow $   
$ \ell^+  \chin    \ell^- \chin$   $\ra$  &
$ \ell^+  (\ell q q)$ $ \ell^- (\ell q q)$ &  \lbp\ indirect & 
$\ell$ + jets &
(G) \ref{sec:jetsemu} \\
&
$ \ell^+ (\nu q q)$ $\ell^- (\ell q q)$  & \lbp\ indirect & 
$\ell$ + jets &
(G) \ref{sec:jetsemu} \\
&
$ \ell^+ (\nu q q) $ $\ell^- (\nu q q)$ &  \lbp\ indirect & 
$\ell$ + jets &
(G) \ref{sec:jetsemu} \\

\hline

\hline 
$ \snu \snu \rightarrow  $ $\nu \chin $ $ \nu \chin \ra $ & 
$ \nu (\nu q q)$  $\nu (\nu q q )$ & \lbp\  indirect & 
4 jets + $E_{miss}$  &
(H) \ref{sec:multijetsemiss} \\

\hline 
$ \snu \snu \rightarrow$   & 
 $ q q \; q q$  & \lbp\  direct & 
4 jets   &
(I) \ref{sec:multijetsnoemiss-sl} \\
$ \sell^+ \sell^- \rightarrow $ &  
  $q q \; q q$ &  \lbp\ direct & 
4 jets   &
(I) \ref{sec:multijetsnoemiss-sl} \\
$ \tilde{q}\bar{\tilde{q}} \rightarrow $ &  
  $q q \; q q$ &  \lbpp\ direct & 
4 jets   &
(I) \ref{sec:multijetsnoemiss-sq} \\

\hline
\end{tabular}
\end{center}
\caption{\it
List of production and decay mechanisms 
of the channels that are covered by the various analyses
described in this paper. The couplings and decay type searched for in
each analysis and the corresponding topologies are described in the
second and third columns, respectively.
The corresponding section number is indicated
in the last column.} 
\label{tab:relation}
\end{table}

The charged lepton, $\ell^\pm$, is either an electron or a muon.  
Different  analyses are applied when the charged lepton is an electron or a 
muon (denoted ``electron channel" and ``muon channel") or when 
it is a tau (denoted ``tau channel"). Each analysis is optimised
regarding the number of jets or charged leptons expected in the final states.

 


If the mass of the scalar charged lepton (``slepton") 
is less than the beam energy, sleptons may be 
pair produced in electron-positron collisions  
through $s$-channel processes involving a $\Zboson$ 
or a $\gamma$. Scalar electrons (``selectrons," $\sele$) 
may also be produced through $t$-channel neutralino exchange. This may
enhance their production cross-section compared to those for the 
scalar muons (``smuons," $\smu$) and scalar taus (``staus," $\stau$).
Similarly, neutral scalar leptons (``sneutrinos'') may be
pair-produced via the $s$-channel or through $t$-channel chargino exchange.

Sleptons and sneutrinos may decay directly to \sm\ particles through
the \lb$_{ijk} L_i L_j {\overline E}_k$ operator. The possible
decays are:

\begin{center}
$\sell^-_{iL} \rightarrow \overline{\nu}_j \ell^-_k$, \quad 
$\sell^-_{jL} \rightarrow \overline{\nu}_i \ell^-_k$, \quad
$\sell^-_{kR} \rightarrow \nu_i \ell^-_j, \nu_j \ell^-_i$ 
\end{center}

\begin{center}
$\snu_{i} \rightarrow \ell^+_j \ell^-_k$, \quad
$\snu_{j} \rightarrow \ell^+_i \ell^-_k$ 
\end{center}
where $\sell^-_{iL}$ denotes a left-handed slepton of the $i^{th}$
generation and $\sell^-_{kR}$ denotes a right-handed slepton of the $k^{th}$
generation. 


If the slepton or sneutrino decays directly via 
the \lbp$_{ijk} L_i Q_j {\overline D}_k$ 
operator\footnote{
Right-handed sleptons cannot decay via the operator
\lbp$_{ijk} L_i Q_j {\overline D}_k$.}, the decay modes are:

\begin{center}
$\sell^-_{iL} \rightarrow  \overline{u}_j d_k$, \quad  
$\snu_{iL} \rightarrow \overline{d}_j d_k$
\end{center}
where $d_{k}$ denotes a down-type quark of the $k^{th}$ generation, 
$u_{j}$ denotes an up-type quark of the $j^{th}$ generation and
$d_{j}$ denotes a down-type quark of the $j^{th}$ generation.


Sleptons and sneutrinos may also decay indirectly 
to $\nt_1$ plus the corresponding charged or neutral 
lepton\footnote{Decays like $\sell \rightarrow \nt_2 \ell$ or 
$\sell \rightarrow \chargino \nu$ are not considered here but the
appropriate branching ratios are taken into account for 
interpretation of the results.}:

\begin{center}
$\sell \rightarrow \nt_1 \ell$, \quad 
$\snu \rightarrow \nt_1 \nu$ 
\end{center}

The $\nt_1$ may
subsequently decay violating \Rparity\ with a \lb, \lbp\ or \lbpp\ coupling
through an intermediate slepton or sneutrino. 
In the case of a non-vanishing \lb\ coupling,  
the $\neutralino$ decays proceeding via the
\lb$_{ijk} L_i L_j {\overline E}_k$ operator are:

\begin{center}
$\nt_1 \rightarrow  \ell^-_i \nu_j \ell^+_k$, \quad
$\nt_1 \rightarrow  \ell^+_i \overline{\nu}_j \ell^-_k$, \quad
$\nt_1 \rightarrow   \nu_i \ell^-_j \ell^+_k$, \quad 
$\nt_1 \rightarrow   \overline{\nu}_i \ell^+_j \ell^-_k$
\end{center}


In the case of a non-vanishing \lbp\ coupling, the
$\neutralino$ decays proceeding via the
\lbp$_{ijk} L_i Q_j {\overline D}_k$ operator are:

\begin{center}
$\nt_1 \rightarrow  \ell^-_i u_j \overline{d}_k$, \quad
$\nt_1 \rightarrow  \ell^+_i \overline{u}_j d_k$, \quad
$\nt_1 \rightarrow  \nu_i d_j \overline{d}_k$, \quad
$\nt_1 \rightarrow  \overline{\nu}_i \overline{d}_j d_k$,
\end{center}


In the case of a non-vanishing \lbpp\ coupling, the
$\neutralino$ decays proceeding via the
\lbpp$_{ijk} {\overline U}_i {\overline D}_j {\overline D}_k$ operator
are:

\begin{center}
$\nt_1 \rightarrow  u_i d_j d_k$, \quad
$\nt_1 \rightarrow  {\overline u}_i {\overline d}_j {\overline d_k}$
\end{center}

If the mass of the scalar top quark (``stop") is smaller than the beam energy,
stop quarks may be produced in pairs in e$^+$e$^-$ collisions
via $s$-channel $\Zboson$ or  $\gamma$ exchange.
Due to the mixing of the left- and right-handed stop, 
$ \stopl $ and $ \stopr $, the observable
$ \stopm = \stopl \cos \theta_{ \stopx\ } +
\stopr \sin \theta_{ \stopx\ } $
could become very light, even the lightest
supersymmetric particle. 
The coupling of the $ \stopm $ 
to the $ \Zboson $
boson is determined by the mixing angle $ \theta_{ \stopx\ } $,
whose value
is determined by the top quark mass and the soft SUSY breaking parameters.
The $ \stopm $  decouples from the $\Zboson $ if
$\cos^2 \theta_{ \stopx }  = \frac{4}{3} \sin^2 \bar{\theta}_{\rm{W}}$ 
($\theta_{ \stopx } \simeq 0.98 $ radian),
where
$\bar{\theta}_{\rm{W}}$ is the effective weak mixing angle.
For this value of $\theta_{\stopx}$,
$ \stoppair $
may only be produced via a virtual 
$\gamma$ and the expected cross-section is therefore reduced.
 
For the purpose of \Rparity\ violating searches, the stop quark is 
assumed to be the lightest supersymmetric particle and only direct
decays are considered. 
Only 9 of the 27 \lbp\ parameters are relevant: 
$\lambda_{i3k}^{'}$, $i,k = 1,2,3$, 
as the stop is contained  in the 
SU(2) doublet field but not in the down type singlet field.
 
If the stop decays via 
the \lbp$_{ijk} L_i Q_j {\overline D}_k$ operator, 
the decay modes are:

\begin{center}
$\stopx_{jL} \rightarrow \ell^+_i d_k$
\end{center}



If the stop decays via 
the \lbpp$ {\overline U}_i {\overline D}_j {\overline D}_k $ 
operator, the decay modes are:

\begin{center}
$\stopx_{iR} \rightarrow \overline{d}_j \; \overline{d}_k$
\end{center}

Under the assumption of \Rparity\ violation, the strength of the coupling
and the decay width of a sfermion are determined only by its mass and 
the \lb, \lbp\ and \lbpp\ parameters
if the  sparticle is the LSP.
If the sparticle is not the LSP, both the \Rparity\ conserving and 
the \Rparity\ violating decay modes are accessible. 

In the analyses described in this paper, tracks are required to come from the 
interaction vertex. Analyses would become inefficient for decay lengths 
larger than some centimeters. For very long lifetimes, the LSP decays
outside the detector, and in the case it is neutral, 
the event topology would be exactly the same
as the \Rp\ conserving case.

For sleptons and sneutrinos, the decay widths 
are given by~\cite{ref:barger2,ref:carena}:

\[ 
\Gamma (\sell^-_{i} \rightarrow \overline{\nu}_j \ell^-_k, 
\tilde{\nu}_i \rightarrow \ell^+_j \ell^-_k) = 
\frac{1}{16 \pi} \lambda^2_{ijk} 
m_{\sell, \tilde{\nu}} \; , \qquad 
\Gamma (\sell^-_{i} \rightarrow  \overline{u}_j d_k, 
\tilde{\nu}_i \rightarrow \bar{d}_j d_k) = 
\frac{3}{16 \pi} \lambda^{'}{}^2_{ijk} 
m_{\sell,\tilde{\nu} } \; ,
\] 
neglecting quark and lepton masses. 

Similarly, the \Rparity\ violating decay of the stop has a decay 
width~\cite{ref:dreiner2} of:

\[
\Gamma (\stopl \rightarrow \ell^+_i d_k) = 
\frac{1}{16 \pi} \lambda{'}{}^2_{i3k} 
m_{\stopx\ }
\]




Under the conservative assumptions of a sparticle mass of 
45 GeV\footnote{This number takes into account the indirect limits obtained
from the study of the Z$^0$ width at LEP1.}  
and a decay length of 0.1 mm the analyses presented in this paper 
would be sensitive to 
\lb\ couplings larger than ${\cal O} (10^{-5})$. 

\section{Monte Carlo Simulation}
\label{sec:MC}

Monte Carlo samples corresponding to the 
charged slepton, sneutrino and stop 
pair-production processes as well as  
Monte Carlo samples used to estimate the background levels
due to \smp\ were simulated.
All generated events were processed
through the full simulation of the OPAL detector~\cite{ref:GOPAL},
and the same analysis chain was applied to simulated events
as to the data.


The simulation of the signal events has been done at $\sqrt{s}$~=183~GeV 
with the Monte Carlo program SUSYGEN~\cite{ref:SUSYGEN}.
Charged and neutral sleptons decaying directly or indirectly 
via \lb\ or \lbp\ have been produced
for the mass values of  45, 70 and 90 GeV. Five
masses (45, 60, 75, 80 and 90 GeV) were used for the sneutrino direct
decays via \lbp. Stop events were simulated at 6 different stop
masses (45, 55, 65, 75, 85 and 90 GeV). 
Samples of 1000 or 2000 events were generated for each relevant coupling.
 
For the indirect decays, events were produced with 
$\Delta m = m_{\sfer} - m_{\nt_1} = m_{\sfer}/2$.
Additional samples were simulated for $m_{\sfer}$ = 90~GeV and  
$\Delta m = m_{\sfer} - m_{\nt_1} = 5$~GeV to account for changes in
the event topologies from the model parameters. The values of  
$\Delta m$ were chosen to cover a large range for a limited number of 
Monte Carlo events. 
To estimate the systematic errors related to different 
gaugino mixings, extra samples of pair-produced selectrons and 
electron-sneutrinos were simulated with five 
different sets of SUSY parameters.

Events were produced for each of the nine possible \lb\
couplings. Events were also simulated for each lepton flavour 
corresponding to the first index of
\lbp. The quark flavour corresponding to the second and third index
of \lbp\ were fixed to the first and second generation, with a few
samples containing bottom qaurks for systematic checks. 

\label{sec:higgs}

For the stop decaying via the \lbp\ coupling into a quark and a lepton
all nine combinations of quark and lepton flavours
in the final state were generated.
The production and decay of the stop is simulated as described 
in~\cite{ref:stoppaper}.
The stops are hadronised to form colourless hadrons
and associated fragmentation particles,
according to the Lund string fragmentation scheme 
(JETSET 7.4)~\cite{ref:JETSET1, ref:JETSET2}.
For the decay, a colour string was stretched between the spectator 
quark and the quark from the stop decay. Further hadronisation 
was also done using the Lund scheme.
The fragmentation function of Peterson~\cite{ref:peterson} has been used.
Events were simulated 
with the mixing angle $\theta_{\stopx\ }$
set to zero.



The main sources of background arise from 
\sm\ four-fermion,
two-photon and two-fermion (lepton-pair and multi-hadronic) 
processes. 
For two-photon processes, the PHOJET~\cite{ref:PHOJET} and 
HERWIG~\cite{ref:herwig} generators have been used to simulate
hadronic final states.
The Vermaseren~\cite{ref:VERMASEREN} generator was
used to estimate the background contribution from all 
two-photon $\ee \ell^+ \ell^-$
final states.
All other four-fermion final states, other than two-photon 
$\ee \ell^+ \ell^-$,  were simulated 
with grc4f~\cite{ref:grace4f}, which takes into 
account all interfering four-fermion diagrams. 
For the two-fermion final states, BHWIDE~\cite{ref:BHWIDE} 
was used for the ee$(\gamma)$ 
final state and KORALZ~\cite{ref:KORALZ} for the 
$\mu \mu$ and the $\tau \tau$ states. The multi-hadronic events, 
${\rm qq}(\gamma)$, 
were simulated using PYTHIA~\cite{ref:JETSET1}.


For small contributions to background final states with six or more
primary fermions, no Monte Carlo generator exists. These final states
are therefore not included in the background Monte Carlo
samples. Consequently, the background could be slightly
underestimated, which would lead to a conservative approach when
calculating upper bounds applying background subtraction.

\section{The OPAL Detector}
\label{sec:opaldet}

A complete description of the  OPAL detector can be found 
in Ref.~\cite{ref:OPAL-detector} and only a brief overview is given here.

The central detector consists of
a system of tracking chambers
providing charged particle tracking
over 96\% of the full solid 
angle\footnote
   {The OPAL coordinate system is defined so that the $z$ axis is in the
    direction of the electron beam, the $x$ axis is horizontal 
    and points towards the centre of the LEP ring, and  
    $\theta$ and $\phi$
    are the polar and azimuthal angles, defined relative to the
    $+z$- and $+x$-axes, respectively. The radial coordinate is denoted
    as $r$.}
inside a 0.435~T uniform magnetic field parallel to the beam axis. 
It is composed of a two-layer
silicon microstrip vertex detector, a high precision drift chamber,
a large volume jet chamber and a set of $z$ chambers measuring 
the track coordinates along the beam direction. 
A lead-glass electromagnetic (EM)
calorimeter located outside the magnet coil
covers the full azimuthal range with excellent hermeticity
in the polar angle range of $|\cos \theta |<0.82$ for the barrel
region  and $0.81<|\cos \theta |<0.984$ for the endcap region.
The magnet return yoke is instrumented for hadron calorimetry (HCAL)
and consists of barrel and endcap sections along with pole tip detectors that
together cover the region $|\cos \theta |<0.99$.
Four layers of muon chambers 
cover the outside of the hadron calorimeter. 
Electromagnetic calorimeters close to the beam axis 
complete the geometrical acceptance down to 24 mrad, except
for the regions where a tungsten shield is present to protect
the detectors from synchrotron radiation.
These include 
the forward detectors (FD) which are
lead-scintillator sandwich calorimeters and, at smaller angles,
silicon tungsten calorimeters (SW)~\cite{ref:SW}
located on both sides of the interaction point.
The gap between the endcap EM calorimeter and the FD
is instrumented with an additional lead-scintillator 
electromagnetic calorimeter, called the gamma-catcher.

To be considered in the analyses, tracks in the central detector and clusters
in the electromagnetic calorimeter were required to satisfy the normal quality
criteria employed in OPAL's analysis of Standard Model (SM) lepton
pairs~\cite{ref:leptpairs}.
\section{Multi-lepton Final States}
\label{sec:multileptons}

This section describes the searches for purely leptonic final states
that may result from pair production of neutral or charged sleptons,
involving subsequent direct or indirect \lb\ decays 
(see Table~\ref{tab:relation}). 

\subsection{Event and Track Selection}
\label{sec:preselections}

The event preselection and lepton identification are 
described in~\cite{ref:slept161}.
Multi-hadronic, cosmic and Bhabha 
scattering events were vetoed~\cite{ref:slept161}.

At the preselection level, it was also required that
the ratio of the number of tracks satisfying the
quality criteria described in~\cite{ref:leptpairs}
to the total number of reconstructed tracks be greater than 0.2
to reduce backgrounds from beam-gas and beam-wall events.
The visible energy, the visible mass and the total transverse momentum
of the event were calculated using the method 
described in~\cite{ref:OPAL-Higgs}.
Finally, the number of good charged tracks
was required to be at least two.

Only tracks with $|\cos \theta| <$ 0.95 were considered  
for lepton identification. A track was considered ``isolated'' 
if the total energy of other charged tracks within 
$10\degree$ of
the lepton candidate was less than 2~GeV.
A track was selected as an electron candidate if one of the following three
algorithms was satisfied: {\it (i)} the output probability of the 
neural net algorithm described in
\cite{ref:NN} was larger than 0.8;
{\it (ii)} the electron selection 
algorithm as described in \cite{ref:elecbarrel}
for the barrel region or in \cite{ref:elecendcap} for
the endcap region was satisfied;
{\it (iii)} $0.5 < E/p < 2.0$,  where $p$ is 
the momentum of the electron candidate
and $E$ is the energy of the electromagnetic calorimeter cluster
associated with the track. 
A track was selected as a muon candidate according to the 
criteria employed in OPAL's
analysis of \sm\ muon pairs~\cite{ref:leptpairs}.
That is, the track had associated activity in the muon chambers or hadron
calorimeter strips or it had a high momentum but was associated with only a
small energy deposit in the electromagnetic calorimeter.
Tau candidates were selected by requiring that there were at most three 
tracks within a 35$^\circ$ cone.
The invariant mass computed using all good tracks and EM clusters
within the above cone had to be less than 3~GeV.
For muon and electron candidates, the momentum was estimated 
from the charged track momentum measured in the central detector, while
for tau candidates the momentum was estimated from the vector sum of
the measured momenta of the charged tracks within the tau cone.

Tracks resulting from photon conversion
were rejected using the algorithm described in
\cite{ref:conversion}. For the two- and six-lepton final states, 
the large background from two-photon
processes was reduced by requiring that  
the total energy deposited in each silicon tungsten calorimeter be
less than 5~GeV, be less than 5~GeV in each forward calorimeter, and
be less than 5~GeV in each side of the gamma-catcher.
In addition to the requirement that there be no 
unassociated electromagnetic cluster with an energy larger 
than 
25~GeV in the event, it was also required that there be no 
unassociated 
hadronic clusters with an energy larger than 10~GeV. 

\subsection{Final States with Two Leptons plus Missing Energy}
\label{sec:2leptons} 

Final states with two charged leptons and missing energy may 
result from direct slepton decays via a \lb\ coupling.
The analysis 
was optimised to retain good signal efficiency
while reducing the background, mainly due to 
$\ell \ell \nu \nu$ final states from $\WW$ production and to 
two-photon processes.
The following criteria were applied in addition to those described in 
Section~\ref{sec:preselections}.

\begin{description}

\item[(A1)]
Events had to contain 
exactly two identified 
and oppositely-charged leptons, each with a transverse 
momentum with respect to the beam axis greater than 2~GeV.

\item[(A2)]
The background from two-photon processes and ``radiative return" events
($\ee \ra {\mathrm Z} \gamma$, where the $\gamma$ escapes
down the beam pipe) was reduced  by requiring
that the polar angle of the missing momentum, $\thmiss$,
satisfy \mbox{$\cosmiss < 0.9$}. 

\item[(A3)]
To reduce further the residual background from 
Standard Model lepton pair events,
it was required that $\Mvis /\sqrt{s} < 0.80$, 
where $\Mvis$ is the event visible mass.

\item[(A4)]
The acoplanarity angle\footnote
   {The acoplanarity angle, $\acop$,
    is defined as 180$\degree$ minus the angle
    between the two lepton momentum vectors  
    projected 
    into the $x-y$ plane.}
($\acop$) between the two leptons was required to be
greater than 10$\degree$ in order to reject Standard Model leptonic events,
and smaller than 175$\degree$ in order to reduce the
background due to photon conversions.
The acoplanarity angle distribution is shown
in Figure~\ref{fig:multilepton}~(a)
after cuts (A1) to (A3).
The acollinearity angle\footnote
   {The acollinearity angle, $\acol$,
    is defined as 180$\degree$ minus the space-angle
    between the two lepton momentum vectors.}
($\acol$) was also required to be greater than 10$\degree$
and smaller than 175$\degree$.

\item[(A5)]
Cuts on $a_t^{\mathrm {miss}}$ 
and $p_t^{\mathrm {miss}}$ were applied; 
$a_t^{\mathrm {miss}}$ is the component
of the missing momentum vector perpendicular to the 
event thrust axis in the
plane transverse to the beam axis and $p_t^{\mathrm {miss}}$ 
is the missing transverse momentum. 
The cuts on $a_t^{\mathrm {miss}}$ and $p_t^{\mathrm {miss}}$ are complementary 
and reject some two-photon events with high transverse momentum.
The full description of these cuts
can be found in~\cite{ref:slept161}.


\end{description}

In order to maximise the detection efficiencies,  
events were accepted if they passed the above selection criteria or if
they passed the selection of $\WW$ pair 
events~\cite{ref:wwpaper} where both W's decay leptonically. 
The preselection and detector status criteria described in
Section~\ref{sec:preselections} were imposed in both cases.
There are 75 events selected with 
79.7 events expected from all Standard Model processes considered
(75.2 from $\WW$ events).

\begin{description}

\item[(A6)]
At this stage the background from two-photon processes and 
$\WW$ production was reduced by categorizing the events
in different classes 
according to the flavour of the leptons expected in the final state,
as can be seen in Table~\ref{tab:2leptons}.
Events were further selected by applying cuts on the momentum
of the two leptons as described in~\cite{ref:slept161} in both the right-
and left-handed slepton searches.

\end{description}

\begin{table}[htbp]
\begin{center}
\begin{tabular}{|r||c||c|c|c|}
\hline
Final State & Eff. (\%) & Selected Events & Tot. bkg MC  & 4-f \\
\hline
\hline
$ee     + E_{T {\mathrm miss}}$   & 58--76  & 11 & 13.8 & 13.5 \\
$\mu\mu + E_{T {\mathrm miss}}$   & 57--81  & 10 & 11.3 & 11.0 \\
$\tau\tau + E_{T {\mathrm miss}}$ & 30--50  & 10 & 15.5 & 12.5 \\
$ee$     or $e\mu$  or $\mu\mu   + E_{T {\mathrm miss}}$   & 65--80  & 39 & 52.2 & 51.0 \\
$ee$     or $e\tau$ or $\tau\tau + E_{T {\mathrm miss}}$   & 58--71  & 39 & 51.9 & 48.6 \\
$\mu\mu$ or $\mu\tau$ or $\tau\tau + E_{T {\mathrm miss}}$  & 58--72  & 40 & 52.0 & 48.6 \\
\hline
\end{tabular}
\end{center}
\caption{\it 
Detection efficiencies (in \%), events selected and background
predicted for the lepton-pair plus missing energy channel and 
for slepton masses between 45 and 90~GeV. 
The deficit of events selected in the data compared to the background 
expectations is interpreted as a downward statistical fluctuation.
The number of events in the last three rows are largely correlated, 
as many final states are shared.
}
\label{tab:2leptons}
\end{table}

The detection efficiencies are summarised in Table~\ref{tab:2leptons}.  
The efficiencies are quoted for slepton 
masses between 45 and 90~GeV. 
Detection efficiencies were
estimated separately for right- and left-handed $\sele$, $\smu$ and
$\stau$.  The first three lines of Table~\ref{tab:2leptons} refer to
left-handed sleptons while the other lines refer to right-handed 
sleptons.
Indeed, due
to the structure of the corresponding \lb\ term in the  
Lagrangian of equation (\ref{lagrangian}), these
particles are expected to yield different final states. 
The 
expected background from all Standard Model
processes considered is normalised to the data luminosity of 56.5 pb$^{-1}$.
As can be seen in Table~\ref{tab:2leptons}, 
most of the background remaining comes from 4-fermion processes,
expected to be dominated by $\WW$ doubly-leptonic decays.

Due to beam-related backgrounds and to 
incomplete modelling of two-photon
processes, there is poor agreement between the data
and Monte Carlo expectation in the early stages of some of the analyses. 
When the two-photon processes have been effectively reduced
after specific cuts (for instance, a cut on the missing 
transverse momentum), the agreement between data and Monte Carlo is good. 


\subsection{Final States with Four Leptons with or without Missing Energy }
\label{sec:4leptons}

Final states with four charged leptons and no missing energy 
may result from direct sneutrino decays while the final states with
missing energy may result from indirect sneutrino  decays via a \lb\ coupling. 
Two analyses have been developed and optimised separately for these two final
states. 
No specific cut on the
lepton flavour present in the final state was applied.
To be independent of the type of decay and \lb\ coupling the two
analyses were at the end combined.  

\begin{figure}[htbp]
\centering
\epsfig{file=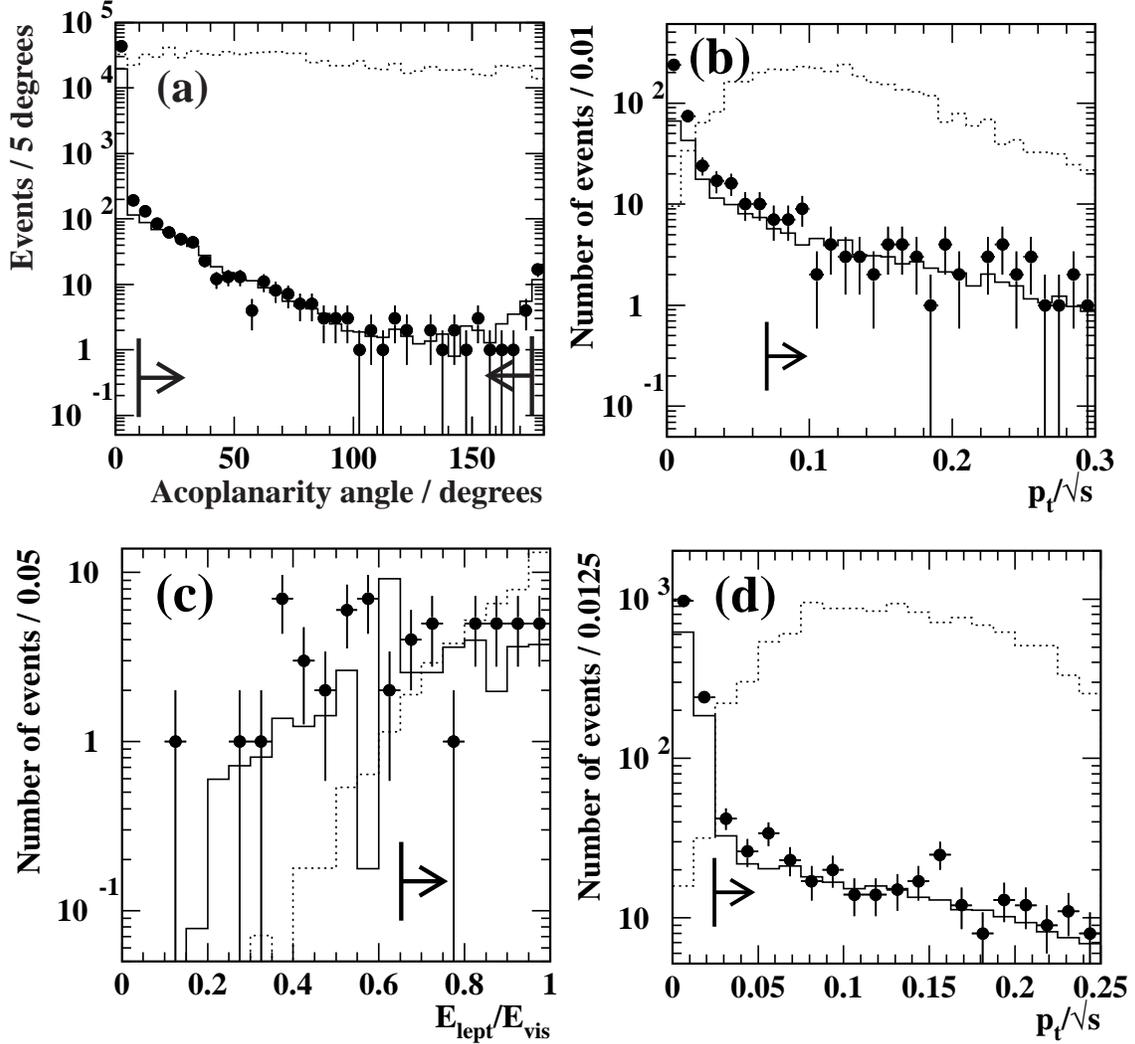       ,width=15.0cm}
\caption[]{\sl
  (a) Two lepton and missing energy search (Analysis A): Distribution of the
  acoplanarity angle.
  The dotted histogram 
  shows signal Monte Carlo events for direct decays of $\sele$ via \lb$_{121}$
  with $m_{\sele} = 70$~GeV.
  (b) Four lepton and missing energy search (Analysis B): Distribution
  of the event 
  transverse momentum calculated without the hadron calorimeter.
  The dotted histogram 
  shows signal Monte Carlo events for indirect decays of $\snu$ with 
  $m_{\snu} = 70$~GeV and for \lb$_{233}$.
  (c) Four lepton and no missing energy search (Analysis C): 
  Distribution of the sum of the energies of the 
  identified leptons divided by the total visible energy.
  The dotted histogram 
  shows signal Monte Carlo events for direct decays of $\snu$ with 
  $m_{\snu} = 70$~GeV and for \lb$_{121}$. 
  (d) Six lepton with missing energy search (Analysis D): 
  Distribution of the event 
  transverse momentum calculated without the hadron calorimeter.
  The dotted histogram 
  shows signal Monte Carlo events for indirect decays of $\smu$ with 
  $m_{\smu} = 70$~GeV and for \lb$_{233}$. 
  Data are shown as points and the sum of all Monte Carlo background 
  processes is shown as the solid line. 
  The simulated signal events have arbitrary normalisation. 
  The arrows point into the regions accepted by the cuts.
 } 
\label{fig:multilepton}
\end{figure}

The following criteria were applied to select a possible signal in the
four leptons plus missing energy topology:

\begin{description}

\item[(B1)]
Events were required to have at least three charged tracks with a transverse 
momentum with respect to the beam axis greater than 1.0~GeV.

\item[(B2)]
The event transverse momentum calculated without the hadron calorimeter
was required to be larger than 0.07 $\times \rs$.
This distribution 
is shown in Figure~\ref{fig:multilepton}~(b) 
after cut (B1) has been applied.

\item[(B3)]
Events had to contain at least three identified 
isolated leptons each with a transverse 
momentum with respect to the beam axis greater than 1.5~GeV.

\item[(B4)]
 It was also required that $\Evis /\sqrt{s} < 1.1$,
where $\Evis$ is the event visible energy.

\item[(B5)]
The total leptonic energy, defined as the sum of the 
energies of all identified leptons,
was required to be greater than $0.5 \times \Evis$.

\item[(B6)]
The background from two-photon processes and ``radiative return" events
($\ee \ra {\mathrm Z} \gamma$, where the $\gamma$ escapes
down the beam pipe) was reduced  by requiring
that the polar angle of the missing momentum direction, $\thmiss$,
satisfies \mbox{$\cosmiss < 0.9$}. 

\item[(B7)]
To reduce further the total background from Standard Model lepton 
pair events, it was required that the energy sum of the two most 
energetic leptons be smaller than 0.75~$\times \Evis$.

\end{description}

To select final states without missing energy, the following 
requirements were imposed:

\begin{description}

    

\item[(C1)]
Events had to contain at least three identified 
isolated leptons each with a transverse 
momentum with respect to the beam axis greater than 1.5~GeV.

\item[(C2)]
It was also required that $0.65 < \Evis /\sqrt{s} < 2.0$.

\item[(C3)]
The total leptonic energy, defined as the sum of the 
energy of all identified leptons,
was required to be greater than $0.65 \times \Evis$.
This distribution
is shown
in Figure~\ref{fig:multilepton}(c), after cuts (C1) have been 
applied.

\item[(C4)]
To reduce the residual four-fermion background,
pairs were formed with the four most energetic tracks, and the invariant
mass was computed for each pair. Events were selected if 
one of the three possible pairings satisfies 
$ |m_{i,j} - m_{k,l}|/(m_{i,j} + m_{k,l}) < 0.4 $, were 
$m_{i,j}$ is the invariant mass of the pair $(i,j)$. Only pairs 
with invariant mass $m_{i,j}$ greater than 20 GeV were used in the computation.

\item[(C5)]
To reduce further the total background from Standard Model lepton 
pair events, it was required that the energy sum of the two most 
energetic leptons be smaller than 0.75~$\times \Evis$.

\end{description}

The two analyses were then combined. Events passing either set of criteria 
were accepted as candidates. 
Detection efficiencies 
range from 34\% to 
80\% for direct sneutrino decays, 
and from 13\% to 
58\% for indirect sneutrino decays, 
for sneutrino masses between 45 and 90~GeV. 
The expected background is 
estimated to be 2.5 events. 
There is one candidate event selected in the data.

\subsection{Final States with Six Leptons plus Missing Energy}
\label{sec:6leptons}

An analysis has been designed to select events with
six charged leptons and missing energy in the final state. These 
topologies may  for example result  
from indirect slepton decays with a \lb\ coupling.

The following criteria were applied:

\begin{description}

\item[(D1)]
To reduce the background from 
two-photon and di-lepton processes,
it was required that $0.1 < \Evis /\sqrt{s} < 0.7$.
\item[(D2)]
The event longitudinal momentum
was required to be smaller than 0.9 $\times {p_{\mathrm{vis}}}$, 
where ${p_{\mathrm{vis}}}$ is the event total momentum.

\item[(D3)]
The event transverse momentum calculated without the hadron calorimeter
was required to be larger than 0.025 $\times \rs$.
This distribution 
is shown in Figure~\ref{fig:multilepton}(d) 
after cuts (D1) and (D2) have been applied.

\item[(D4)]
Events with less than five charged tracks 
with a transverse momentum with respect 
to the beam axis larger than 0.3~GeV were rejected.  

\item[(D5)]
Events had to contain at least three well-identified 
isolated leptons; at least two of them  with a transverse 
momentum with respect to the beam axis greater than 1.5~GeV, and the
third one with a transverse 
momentum with respect to the beam axis greater than 0.3~GeV.

\item[(D6)]
The total leptonic energy, 
was required to be greater than $0.2 \times \Evis$. 

\end{description}

Detection efficiencies 
range from 40\% to 88\% for indirect selectron decays,
from 59\% to 93\% for indirect smuon decays and
from 33\% to 70\% for indirect stau decays, for 
slepton masses between 45 and 90~GeV.
The total background expectation is 1.7 events. 
There is one candidate event selected in the data.

\subsection{Inefficiencies and Systematic Errors}
\label{sec:syserr}

Variations in the efficiencies were estimated with events 
generated with $\Delta m$ = 5~GeV, as described in 
Section~\ref{sec:MC}.

The inefficiency due to forward detector false vetoes caused by
beam-related backgrounds or detector noise was estimated from a study
of randomly triggered beam crossings to be 
3.2\%. The quoted efficiencies 
take this effect into account. 

The systematic errors on the number of signal events expected that
have been considered are:
the statistical error on the determination of the efficiency from the 
Monte Carlo simulation (typically less than 2\%);
the systematic error on the integrated luminosity of 0.4\%; 
the uncertainty due to the interpolation of the efficiencies,
estimated to be 4.0\% and the lepton identification uncertainty, 
estimated to be 2.4\% for the muons, 3.9\% for the electrons and
4.7\% for the taus. The systematic error arising from the modelling of
the variables used in the multi-lepton final state selections is 
smaller than the lepton identification uncertainties.
The systematic error due to the trigger efficiency is
negligible because of the high lepton transverse momentum requirement.
The total systematic error was calculated by summing in quadrature 
the individual errors and is incorporated into the limit calculation
using the method described in Ref.~\cite{ref:cousins}.

The systematic error on the number of expected background events from 
SM processes has a negligible effect when computing limits.

\section{Final States with Two Jets and Two Leptons }
\label{sec:2jets2leptons}

\subsection{Electron and Muon Channels}
\label{sec:stopemu}

In this section, the analysis for the selection of the final state of two electrons
or two muons plus two jets and no missing energy is described. 
These final states may result
from the direct decay of pair-produced stops via a \lbp\ coupling.
In contrast to the purely leptonic final states described in the previous
section, the topologies searched for in this analysis involve hadronic jets; 
more stringent cuts are needed to obtain a purer lepton sample.  
Particles are considered as electrons or muons if they are either identified by
the  selection algorithms described in~\cite{ref:elecbarrel} 
and~\cite{ref:elecendcap}, or by an algorithm used for selecting 
semileptonic W decays, as
described in~\cite{ref:wwpaper}.

Events were preselected by requiring the following criteria to be 
satisfied (the same criteria were also used for the analysis presented 
in Section~\ref{sec:jetsleptons}):

The fraction of good tracks had to be greater than 0.2, 
to reduce beam-gas
and beam-wall background events.
Events with fewer than 
seven good charged tracks were not considered in order to reduce the
background from Bhabha scattering.
Events had to contain at least one identified 
electron or muon with a momentum 
greater than 3~GeV, to
reduce the background  from final states with 
low energy leptons (e or $\mu$).
To reduce background from two-photon processes, it was required that 
the visible energy
normalised to the centre-of-mass energy,
$R_{\rm vis} =  E_{\rm vis}/ \sqrt{s} >$~0.3.

\begin{figure}
\begin{center}
\epsfig{file=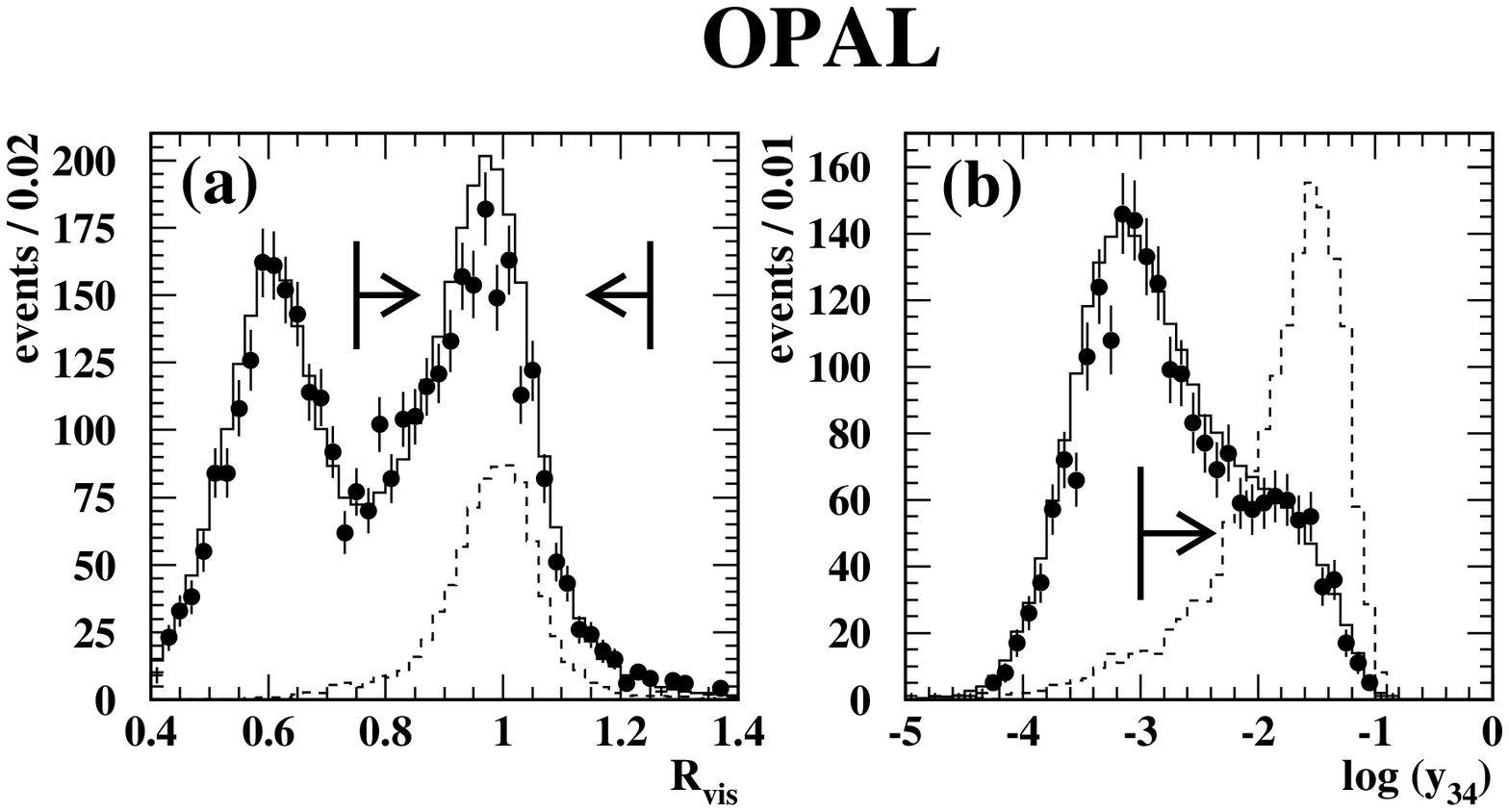,width=15.0cm}
\caption[]{\sl
Stop search (Analysis E): (a) Visible energy R$_{vis}$ 
after the preselection and (b) jet resolution 
$y_{34}$ after cut (E1).
Data are shown as points and the sum of all Monte Carlo background  
processes is shown as the solid line. 
The dashed histogram 
shows signal Monte Carlo events for direct decays of $\stopm$ with 
$m_{\stopm} = 85$~GeV and for \lbp$_{ij3}, (i=1,2)$.
The scale of the signal MC is arbitrary.
The arrows point into the regions accepted by the cuts.
}
\label{fig:ivor_stop}
\end{center}
\end{figure}

The following cuts are then applied:

\begin{description}

\item [(E1)]
The visible energy had to be close to the centre-of-mass energy,
$0.75 < R_{\rm vis} < 1.25$.
Figure~\ref{fig:ivor_stop}(a) shows the visible energy distribution. 
\item [(E2)]
It was required that four jets be reconstructed using the Durham~\cite{ref:durham} algorithm,
with $y_{34} > 0.001$, where $y_{34}$ is the cut parameter between 3
and 4 jets. 
Both hadronic and leptonic objects are used in the jet reconstruction.
Figure~\ref{fig:ivor_stop}(b) shows the $y_{34}$ distribution. 
\item [(E3)]
Events had to contain at least one pair of identified oppositely-charged 
lepton candidates of the same flavour.
\item [(E4)]
To make use of the signal topology of two leptons and 
two jets,
where a lepton and a jet stem from the same object, 
a five-constraint (5C) kinematic fit was
performed for the two possible 
combinations of each lepton with each jet.
The kinematic constraints are: the vector sum of all 
momenta has to be
equal to zero, the total energy of all objects has to 
be equal to the 
centre-of-mass energy and the masses of the two 
reconstructed particles
have to be equal.
From the three most energetic leptons of the same 
flavour, 
the two most isolated\footnote{
The most isolated lepton is the one with the largest 
angle to the closest track.}
were selected and the rest of the event was 
reconstructed as two jets.  
The combination with the highest fit 
probability was
selected. The probability for the fit, based on the 
$\chi^2$, was
required to be larger than 0.01.

\item [(E5)]
The momentum of the most energetic lepton had to be 
greater than 15~GeV and the momentum of the second most 
energetic lepton had to be greater than 10~GeV.
\item [(E6)]
It was required that there be no charged track 
within $15^{\circ}$ of the most energetic 
lepton candidate.
\end{description}

These cuts yield an efficiency of more than 50\% for a stop mass 
of 65~GeV, which rises to approximately 65~\% for masses above 85~GeV.
No candidate event is selected in the data. 
The expected background is 0.9 events for final states with two electrons
and 0.6 events for final states with two muons.
The largest background results in both cases from WW events.

The following systematic errors have been considered:

\begin{enumerate}

\item 
The statistical error from the limited size of the Monte Carlo samples.

\item 
The error due to the interpolation of efficiencies for mass values 
between the generated stop masses, which was estimated to be less than 4\%. 

\item 
A 4\%  error due to the lepton identification 
for the electron 
and a 2\%  error for the muon channel.

\item 
The fragmentation of the stop has been simulated using the fragmentation 
function from Peterson {\it et al.} with the $\epsilon$ parameter 
extrapolated from measurements of charm and bottom~\cite{ref:opalstop}.
To check the model dependence of the fragmentation,
it has also been performed using the function from 
Bowler~\cite{ref:bowler}.
No significant change in the efficiency due to the difference in the 
fragmentation function has 
been found. The difference is at most 0.5\%, where a variation of the 
$\epsilon$ parameter of the $\stopx\ $ in the Peterson {\it et al.} scheme 
is included. This error on $\epsilon_{\stopx}$
is propagated from the error of $\epsilon_{\rm b}$ and the error on the b-quark
mass as described in detail in Ref.~\cite{ref:opalstop}.

\item 
The signal events have been produced for a zero mixing angle between the two
stop eigenstates. The mixing angle describes the coupling between the stop 
and the Z$^0$, and therefore the energy distribution of the initial state 
radiation depends on this mixing angle. 
To check the dependence of the detection efficiency on this angle, 
events have been generated with $\theta_{\stopx\ } = 0.98$, 
where the stop decouples from the Z$^0$.
The change in efficiency is less than 0.5\% for the two extreme cases.

\item 
The Fermi motion of the spectator quark in the stop-hadron influences 
its  measured mass. The Fermi motion has been increased from 220~MeV to
520~MeV and the efficiency changes by no more than 1\%, which is taken as a 
systematic error.

\item 
The systematic error on the measured luminosity is 0.4\%. 

\item
The systematic error due to the uncertainty in the trigger efficiency was 
estimated to be negligible, because of the requirement of at least seven good
tracks.

\end{enumerate}

The systematic error on the expected number of 
background events has been 
estimated to be less than 20\% for all cases by varying 
the cut values
by the experimental resolution.


\subsection{Tau Channel}
\label{sec:stoptau}

This section describes the analysis used to search for the final state 
consisting of two $\tau$-leptons and
two jets, which may result from the direct decay of a stop
via a coupling $\lambda^\prime$.
The backgrounds come predominantly from 
($Z/\gamma)^*\rightarrow q\bar{q}(\gamma)$ and SM four-fermion processes.

\begin{figure}[htbp]
\centering
\epsfig{file=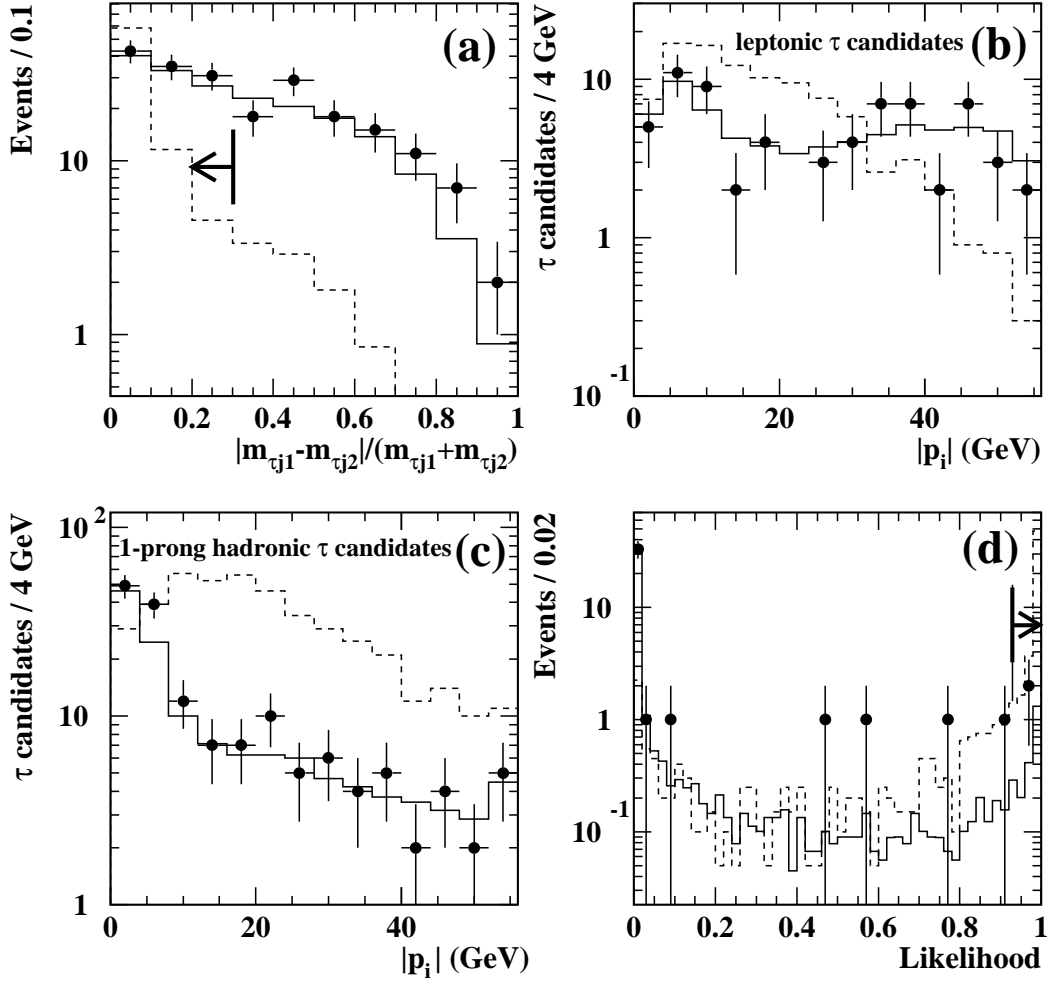,width=15.5cm}     
\caption[]{\label{fig:twotaufourjet}\sl
         Search for jets plus at least two $\tau$ leptons (Analysis F): 
Distributions of relevant
quantities for data (points), estimated Standard Model background (full histogram)
normalised to the integrated luminosity of the data, and a simulated signal
(dashed histogram, arbitrary normalisation) corresponding to 
$m_{\stopm}=75$~GeV (direct decay).
(a) Distribution of the difference in invariant mass of the tau-jet
systems scaled by their sum after cut (F2); events to the
left of the arrow indicating the cut position are accepted. Figures (b) and 
(c) show
the difference in the distributions of the same likelihood input variable
for two different categories of $\tau$ candidates, after cut (F4):
(b) The momentum of leptonic $\tau$ candidates;
(c) the momentum of 1-prong hadronic $\tau$ candidates. 
The likelihood distribution is shown in (d) after cut (F6).   
The arrows point into the regions accepted by the cuts.
 
}
\end{figure}

The selection begins with the identification of $\tau$ lepton candidates, 
identical
to that in~\cite{ref:smpaper}, using three algorithms 
designed to identify electronic, muonic and hadronic $\tau$-lepton decays.
An average of 2.3 $\tau$ candidates per signal event are identified.
The original $\tau$ lepton direction is approximated
by that of the visible decay products.
The following 
requirements,
similar to those described in Ref.~\cite{ref:mssmpaper} up to (F4),
are then imposed:
%
\begin{description}
\item[(F1)]  
Events are required to contain at least nine charged tracks, and must have at 
least two $\tau$ lepton
candidates, including at least one pair where 
each $\tau$ has electric charge $|q|=1$ and the charges sum to zero.
Pairs not fulfilling these requirements are not considered further.

\item[(F2)]
Events must have no more than a total of 20 GeV of energy
deposited in the forward detector, gamma catcher, and 
silicon-tungsten calorimeter; a missing momentum vector
satisfying $|\cos\theta_{\rm miss}| < 0.97$,
a total transverse momentum of at least 2\% of $\sqrt{s}$, and a scalar sum
of all track and cluster transverse momenta larger than 40~GeV.
  
\item[(F3)]
Events must contain at least
three jets reconstructed using
the cone algorithm as in~\cite{ref:smpaper}\footnote{Here, single 
electrons and muons from $\tau$ lepton decays are allowed to be recognised 
as low-multiplicity ``jets''.}, and no energetic isolated photons\footnote{
An energetic 
isolated
photon is defined as an electromagnetic cluster
with energy larger than 15~GeV and no track within a cone
of $30^\circ$ half-angle.}.

\item[(F4)]
Events must contain no track or cluster with 
energy exceeding $0.3\sqrt{s}$. 
\end{description}
For events surviving these requirements, the hadronic part of the event
corresponding to each surviving $\tau$ lepton candidate pair, composed 
of those tracks and clusters not having been
identified as belonging to the pair (henceforth referred to as the
``rest of the event'' or RoE),
is then split into two jets 
using the Durham~\cite{ref:durham}
algorithm.  Two 
pairings between the two $\tau$ candidates and the jets are 
possible.
The invariant masses $m_{\tau j}$ of the two resulting $\tau$-jet systems 
within each
pairing are then calculated 
using only
the $\tau$ lepton and jet momentum directions and requiring energy and
momentum conservation. The pairing scheme 
with the smaller difference between 
$m_{\tau j1}$ and 
$m_{\tau j2}$ 
is then chosen. 
In order for a $\tau$ candidate pair to be considered 
further, the following 
requirements on $m_{\tau j1}$ and 
$m_{\tau j2}$  are imposed, consistent
with the hypothesis of the decay of two heavy objects of identical mass: 
\begin{description}
\item[(F5)]
Both $m_{\tau j1}$ and $m_{\tau j2}$ must be
at least 30 GeV.
\item[(F6)]
The difference in invariant masses must be no more than
30\% of their sum, i.e.
        $|m_{\tau j1}-m_{\tau j2}|/|m_{\tau j1}+m_{\tau j2}|\leq 0.3$.
\end{description}
The distribution of 
$|m_{\tau j1}-m_{\tau j2}|/|m_{\tau j1}+m_{\tau j2}|$  is shown in
Fig.~\ref{fig:twotaufourjet}~(a) for the data,
the backgrounds, and
for a signal sample with $m_{\tilde{t}}=75$ GeV. 
The resolution 
on $m_{\tau j}$ is typically below 5 GeV, except very close to
the kinematic limit.

A likelihood method similar to that described in~\cite{ref:opal_rpv_gauginos}
is then
applied to those events satisfying the above requirements, 
in order to select a final $\tau$ candidate pair for each
event from those surviving, and to suppress further the remaining background.

Distributions of two of the input variables as well as that of 
${\cal L}$ are shown in 
Figures~\ref{fig:twotaufourjet}~(b) to (d).
In each event, the $\tau$-candidate pair with the highest value of ${\cal L}$
is chosen, and the following
requirement is then made:

\begin{description}
\item[(F7)] ${\cal L}>0.93$
\end{description}

Two events survive the selection while the background, almost all
from four-fermion processes, is 
estimated to be
2.07 events for an integrated luminosity of 55.8 pb$^{-1}$.
The reconstructed $\tau$-jet masses 
are 78.9 and 87.9 GeV for the first selected event and 71.7 
and 67.2 GeV for the second one.

The detection efficiencies for 
stop masses between 55 and 90 GeV range from 30 to 40\%,
while that for 45 GeV is approximately 22\%.


These efficiencies are affected by the following
relative uncertainties:
Monte Carlo statistics, typically 2.5 to 3.5\%;
uncertainty in the tau-lepton preselection efficiency, 1.2\%;
uncertainty in the modelling of the other preselection variables, 2.0\%;
uncertainties in the modelling of the likelihood input variables, 10.0\%;
uncertainties in the modelling of fragmentation and hadronisation, 6.0\%;
and uncertainty on the integrated luminosity, 0.5\%~\cite{ref:lumino}.
Taking these uncertainties as independent and adding them in quadrature
results in a total relative systematic uncertainty  of 12.3\% 
The systematic uncertainty in the number of expected background events was 
estimated to be 18\%.


\section{Final States with more than Two Jets and at Least two Charged Leptons}
\label{sec:jetsleptons}

\subsection{Indirect Selectron and Smuon Decays}
\label{sec:jetsemu}

This section describes the event selection for final states from the indirect 
decay of selectrons and smuons via the coupling \lbp.
The final state consists of two leptons of the same flavour from 
the sleptons plus the decay products of the two $\chin$'s.
These will be two jets plus a neutral or charged lepton for each $\chin$.
This results in seven different final states for each slepton flavour, as 
shown in Table~\ref{tab:sleptons_ivor}.
Electrons and muons are identified as described in Section~\ref{sec:stopemu}.
To identify taus in the final states an Artificial Neural Net based on
tracks~\cite{tauID} is used, rather than the selection presented in 
Section~\ref{sec:stoptau} designed specifically for events with two 
$\tau$'s. 

The preselection is the same as described in Section~\ref{sec:2jets2leptons}.
The selection cuts are as follows:

\begin{description}
\item [(G1)]
A cut on the visible energy scaled by the centre-of-mass energy
in  the range $0.5 < E_{\mathrm vis}/\sqrt{s} < 1.2$,
depending on the expected number of neutrinos,
is applied.
In addition a cut on the angle of the missing momentum 
with respect to the beam direction
at $|\cos \theta| <$~0.95
is performed,
if some missing momentum is expected.
\item [(G2)]
The jets in the event have been reconstructed using the Durham algorithm.
The jet resolution $y_{45}$  
at which the number of jets changes from 4 to 5 jets,
is required to be greater than  0.002.
This cut takes into account the high multiplicity of the signal events.
\item [(G3)]
To reduce the background from W pair production
for events with  missing momentum, 
a single-constraint kinematic fit has been performed.
The inputs to the fit are the momenta of the lepton and 
the neutrino, 
taking the missing momentum to be the momentum of the 
neutrino, and the  
rest of the event reconstructed into 2 jets. 
The lepton is taken to be the
most energetic muon or electron in the case of smuon or 
selectron production,
respectively.
 The invariant mass is 
calculated (a) for the lepton and the  neutrino system 
and (b) for the two 
jet system, letting the masses of both systems be 
independent.
The reconstructed mass of at least one  system has to be 
outside a mass
window of  70~GeV$< m < 90$~GeV, or the probability for 
the fit has
to be less than 0.01.

\item [(G4)]
For the topologies with no charged lepton from the 
$\chin$ decay,
the background from W pair production is reduced further 
by a kinematic fit 
on the invariant mass of two pairs of jets, when 
reconstructing the 
whole event into 4 jets.
This kinematic fit assumes energy and momentum 
conservation and
the same mass for both jet pairs. 
From the three possible jet pairings, the one with the 
highest fit
probability is chosen.
The reconstructed mass of the jet pairs has to be 
outside a mass
window of  70~GeV$< m < 90$~GeV, or the probability for 
the fit has
to be less than 0.01.

\item [(G5)]
At least two leptons of the flavour of the slepton  
have to be identified.
To have sensitivity also to small mass differences 
between the slepton and
the 
$\chin$, the required momentum has to be greater than 
4~GeV
for both muons  
in the smuon case and the required energy greater than 4~GeV and 3~GeV 
for the two
electrons
in the selectron case, respectively.

\item [(G6)]
In addition to the leptons required in (G5), also the leptons from 
the $\chin$ decay have to be identified. If two additional charged leptons are 
expected, both have to be identified, if they have a different flavour than 
the slepton.
If two taus are expected, only
one, being different from the leptons in cut (G5), has to be identified.
If a total of four leptons of the same flavour is expected, 
including those in cut (G5), only three of them have to be identified.

If only one additional lepton is expected, it has to be identified.

The energy or momentum of the most energetic lepton has to be above
a cut value varying between 8 and 15~GeV, depending on the topology.
If a total of four leptons is required, for the second most energetic 
an energy or momentum larger than a cut value varying between 3~GeV and 4~GeV, 
depending on the topology, is required.
\item [(G7)]
To make use of the isolation of the leptons in the signal, one or two
of the identified leptons, depending on the expected topology,
are required to be isolated.
The isolation criterion is that  
there be no charged track within a cone of 
half opening angle $\phi$, such that $|\cos \phi| =$~0.99,
around the track of the lepton. 
\end{description}

These selections give efficiencies between 45 and 85\% for final states
without taus, and around 30\% for final states with taus, all for 
slepton masses greater than 70~GeV. 
The expected backgrounds and the numbers of events observed for each 
final state 
are shown in Table~\ref{tab:sleptons_ivor}.

\begin{table}[tbp]
\begin{center}
\begin{tabular}{|l||c|c|}
\hline
Final State & Selected Events & Tot. bkg MC  \\
\hline
\hline
$\smu^+ \smu^- \ra $ & & \\
\hline
$ \mu^+ \mu^- eqq eqq $ & 2 & 0.69 \\
$ \mu^+ \mu^- \mu qq \mu qq $ & 1 & 0.67 \\
$ \mu^+ \mu^- \tau qq \tau qq $ & 1 & 1.10 \\
$ \mu^+ \mu^- eqq \nu qq $ & 3  & 1.05 \\
$ \mu^+ \mu^- \mu qq \nu qq $ & 1 & 0.95 \\
$ \mu^+ \mu^- \tau qq \nu qq $ & 0 & 0.58 \\
$ \mu^+ \mu^- \nu qq \nu qq $ & 0 & 0.91 \\
\hline
\hline
$\tilde{e}^+ \tilde{e}^- \ra $ & & \\
\hline
$ e^+ e^- eqq eqq $ & 1 & 0.29 \\
$ e^+ e^- \mu qq \mu qq $ & 1 & 0.37 \\
$ e^+ e^- \tau qq \tau qq $ & 3 & 1.09 \\
$ e^+ e^- eqq \nu qq $ & 1 & 0.52 \\
$ e^+ e^- \mu qq \nu qq $ & 2 & 1.10 \\
$ e^+ e^- \tau qq \nu qq $ & 3 & 0.81 \\
$ e^+ e^- \nu qq \nu qq $ & 0 & 1.13 \\
\hline
\end{tabular}
\end{center}
\caption{\it 
Number of events remaining after the selection cuts and the 
expected backgrounds from all Standard Model processes. 
The main contribution to the total
background comes 
from $\WW$ leptonic decays (4-fermion processes); multi-hadronic events
contribute up to 30\% and other processes are negligible.}
\label{tab:sleptons_ivor}
\end{table}

\subsubsection*{Systematic Errors}

For the lepton identification, a systematic error of 4\% was estimated for the 
electrons, 3\% for the muons and 3\% for the taus.
For the interpolation of the efficiency between the generated mass points,
a systematic error of 4\% has been assigned.
From the studies on the fragmentation in Section~\ref{sec:2jets2leptons}
the systematic error for this analysis is estimated to be less than 1\%. 
The systematic error on the measured luminosity is 0.4\%.
The systematic error due to the uncertainty in the trigger efficiency 
is negligible, because of the requirement of at least seven good
tracks.
The statistical error on the determination of the efficiency from the MC
samples has also been treated as a systematic error.
The systematic error on the expected number of background events has been 
estimated to be less than 20\% for all cases.

\subsection{Stau Indirect Decays}
\label{sec:jetstau}

If requirements (F5) and (F6) described in 
Section~\ref{sec:stoptau} are suppressed, then the
same analysis as that for the stop search in the tau channel 
can be used to search for the indirect decay of staus
via the coupling \lbp,
where now the final state consists
of two $\tau$ leptons plus four jets and two additional leptons.
In this case, the reference distributions are regenerated in light of
the different topology of this signal, and the minimum required value
of the resulting likelihood discriminant ${\cal L}$ 
(cf. (F7)) is relaxed to 0.9. 
No events survive the selection
while the background expectation rises slightly to 2.27 events. 
The detection efficiencies range from 12\% for final states with two
taus, four quarks plus missing energy and $m_{\stau} =$~45~GeV, to 54\%
for final states with two
taus, four quarks plus two electrons and $m_{\stau} =$~70~GeV.   
The systematic uncertainties are evaluated in the same way as for
the stop search as described in Section~\ref{sec:stoptau}, and are
similar in magnitude.

\section{Final States with Four Jets plus Missing Energy}
\label{sec:multijetsemiss}

Indirect decays of sneutrinos via $\lambda^\prime$ coupling
can lead to final states with four jets and large missing energy
due to the four undetected neutrinos. The dominant backgrounds come from
four-fermion processes and radiative or mis-measured two-fermion events.
The selection procedure is described below:

\begin{description}
\item[(H0)] The event has to be classified as a multi-hadron final-state as 
described in \cite{LEP2MH}.
\item[(H1)] The visible energy of the event is required to be less than 
0.75$\sqrt{s}$.   
\item[(H2)] To reject two-photon and radiative two-fermion events
the transverse momentum should be larger than 10 GeV, 
the total energy measured in the forward calorimeter, gamma-catcher 
and silicon tungsten calorimeter should be less than 20 GeV, 
and the missing momentum should not point along the beam 
direction ($| \cos \theta_{\mathrm{miss}}| <$ 0.96).   
\item[(H3)] The events are forced into four jets using the Durham 
jet-finding algorithm, and rejected if the jet resolution parameter 
$y_{34}$ is less than 0.0008. 
\item[(H4)] An additional cut is applied 
against semi-leptonic four-fermion events, vetoing on  
isolated leptons being present in the event.
The lepton identification is based on 
an Artificial Neural Network routine (ANN)~\cite{tauID},
which was originally designed to identify tau leptons 
but is efficient for electrons and muons, as well.
If at least one lepton candidate is found, 
with ANN output larger than 0.97,
the event is rejected. 
\item[(H5)] Finally, a likelihood selection
is employed to classify the remaining events
as two-fermion, four-fermion or signal processes.
The method and the likelihood variables are described 
in~\cite{ref:opal_rpv_gauginos}, with the restiction that the 
minimum number of charged tracks and the minimum number of 
electromagnetic clusters in a jet are replaced by the aplanarity of 
the event~\cite{dpar}.
The event is rejected if its likelihood output is less than 0.9.
\end{description}

\begin{figure}
\begin{center}
\epsfig{file=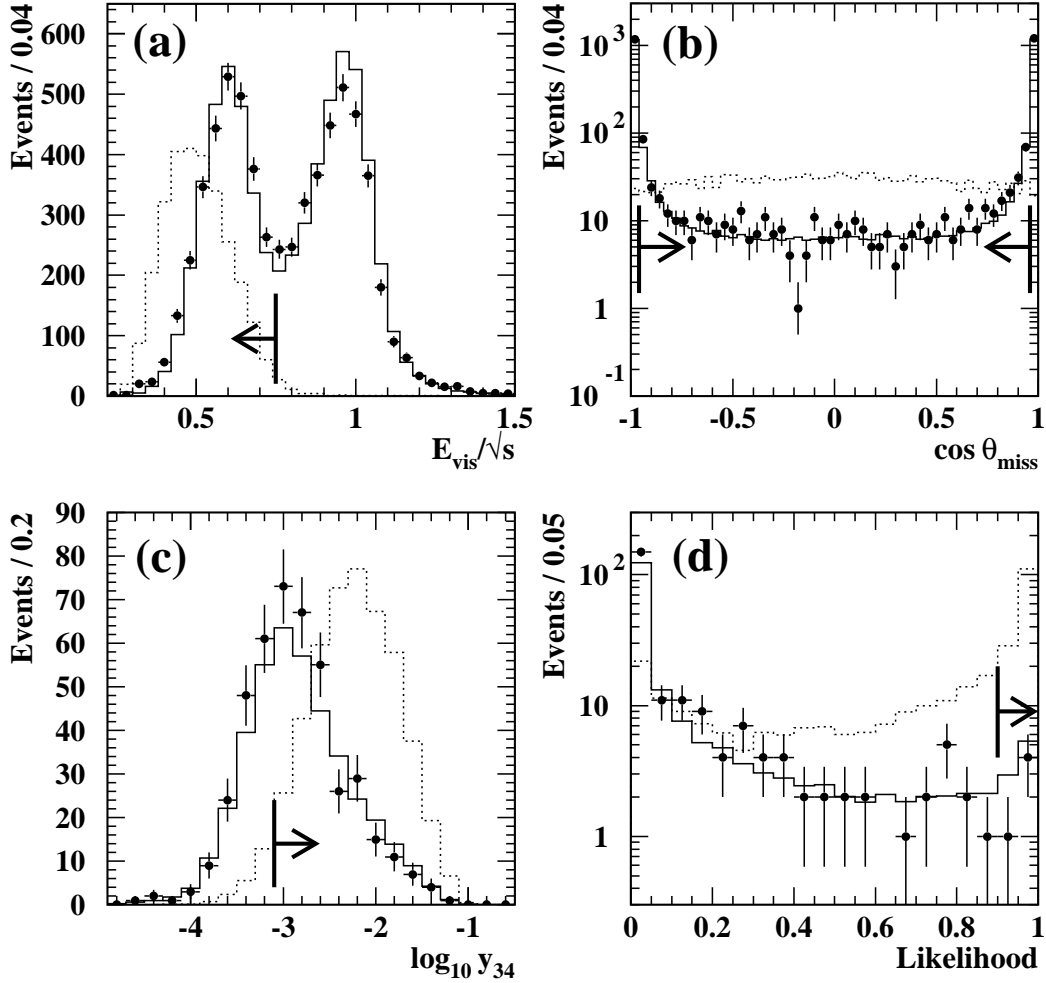,width=15.5cm} 
\end{center}
\caption{ \sl Four jets plus missing energy search (Analysis H): 
Distributions for data (points), for the estimated Standard Model background  
(full histogram) and  for a sum of simulated 
signals (dotted histogram).
Figure (a) shows the visible energy, 
E$_{vis}$, divided by the centre of mass energy, $\sqrt{s}$, for multi-hadron
events after cut (H0). 
In Figure (b) the distribution of 
the cosine of the polar angle of the missing momentum
vector is plotted after cut (H1).  
In Figure (c) the logarithm of the jet resolution, $y_{34}$, 
at which the number of reconstructed jets changes between 4 and 3, is shown
after cut (H2) has been applied.
Figure (d) shows the final selection using the likelihood output.
The arrows indicate the accepted regions in each plot.
The Standard Model background is normalised to the integrated luminosity of the
data, while the normalisation of the  signal distribution is arbitrary. 
}
\label{fig:gabi_fig}
\end{figure}

Figure~\ref{fig:gabi_fig} shows experimental plots 
for the data, the estimated background and simulated signal
events. 

After all cuts, 5 events are selected in the data sample, while 
8.17$\pm$0.31$\pm$1.32 events 
are expected from Standard Model processes, of which 75\% originate 
from four-fermion processes. The signal detection  efficiency varies 
between 5\% and 34\% for sneutrino masses between 45 -- 90 GeV
for $\lambda^\prime_{121}$ and $\lambda^\prime_{123}$ couplings
if the mass difference is one half of the sneutrino mass.
For a small mass difference ($\approx$5 GeV),  
the efficiency is more than doubled.

The small efficiency for light sneutrino masses is the result of 
initial-state radiation and the larger boost  of the particles, which make the
event similar to the QCD two-fermion background.

 
The expected signal rates are affected by the following
uncertainties: Monte Carlo statistics, 3.3 -- 13.9\%;
statistical and systematic uncertainties on the luminosity measurement,
0.3 and 0.4\%;
uncertainties on  modelling of the kinematic variables, 6.7\%; and
on the lepton veto, 1.0\%.

The background estimate has the following errors:  Monte Carlo
statistics, 3.7\%;
modelling of the hadronisation process estimated by comparing different event
generators, 5.3\%; uncertainty on the lepton veto, 
1\%; and modelling of the kinematic variables, 14.9\%.

The inefficiency due to the forward energy veto is found to be 1.8\%.


\section{Final States with Four Jets without Missing Energy}
\label{sec:multijetsnoemiss}

Direct decays of sleptons (squarks) via $\lambda^\prime$
($\lambda^{\prime\prime}$) coupling
can result in final states with four well-separated, 
high multiplicity hadronic jets and large visible energy. 
The background comes from q$\bar{\mathrm q}(\gamma)$ events with hard gluon
emission and four-fermion processes, predominantly 
W$^+$W$^-$ $\rightarrow$ qqqq.

The analysis closely follows our published selection
for H$^+$H$^-$ $\rightarrow$ qqqq \cite{tauID}. 
First, well-defined four-jet events are
selected; then a set of variables are combined using a likelihood technique.

The preselection consists of the following steps:
\begin{description}
\item[(I0)] The event has to be classified as a multi-hadron final-state as 
described in \cite{LEP2MH}.
\item[(I1)] To reduce the radiative two-fermion background, 
the effective centre-of-mass energy of the 
event,$\sqrt{s'}$~\cite{sprime}, 
is required to be greater than 150 GeV.
\item[(I2)] To ensure that the events are well-contained, the   
visible energy should be greater than 0.7$\sqrt{s}$.   
\item[(I3)] The events are forced into four jets using the Durham 
jet-finding algorithm, and rejected if the jet resolution parameter 
$y_{34}$ is less than 0.0025. Moreover, all jets must contain at least one
charged particle.
\item[(I4)] A four-constraint kinematic fit, applied to the jet four-momenta
requiring energy and momentum conservation (4C-fit), 
should yield a $\chi^2$-probability larger than $10^{-5}$.
\item[(I5)] To test the compatibility 
with pair-produced equal mass objects and
to obtain the best possible di-jet mass
resolution, the jet four-momenta are refitted requiring energy and momentum
conservation and equal di-jet masses (5C-fit). 
The event is kept if at least one of the three di-jet combinations
has a $\chi^2$-probability larger than $10^{-5}$.
\end{description} 

To separate the signal from the background events surviving the above selection
a likelihood technique is applied. Three event classes are defined: signal,
two-fermion and four-fermion. 

\subsection{Sleptons}
\label{sec:multijetsnoemiss-sl}

We have used the H$^+$H$^-$ $\rightarrow$ 
c$\bar{\mathrm s}\bar{\mathrm c}$s 
MC samples to produce the signal reference histograms.
This is possible because of the similarities between charged Higgs
and smuon, stau, muon- and tau-sneutrino decays.
Since selectrons and electron-sneutrinos can also be produced in 
$t$-channel-exchange processes, 
their event properties (especially the angular distributions) are different,
and we have used dedicated MC samples 
with $\lambda^{\prime}_{121}$ and $\lambda^{\prime}_{123}$ couplings
to produce these reference histograms.

The following variables were used as input to the likelihood calculation:
\begin{list}{$\bullet$}{\itemsep=0pt \parsep=0pt \topsep=-5pt \leftmargin=30pt}
\item the cosine of the polar angle of the thrust axis;
\item the cosine of the smallest jet-jet angle;
\item the difference between the largest and smallest jet energy
after the 4C-fit;
\item the smallest di-jet mass difference after the 4C-fit;
\item the cosine of the 
di-jet production angle multiplied by the sum of the jet charges
for the combination with the highest $\chi^2$-probability given by
the 5C-fit.
\end{list}
Events were accepted if their likelihood output was larger than 0.5, 0.55 and
0.6 for selectrons, electron-sneutrinos and other sleptons, 
respectively.

The numbers of selected data and expected background events are listed in 
Table~\ref{table:4jet} for the different selections. 
Since the background is dominated by W$^+$W$^-$ production (82--87\%),
the mass distributions are peaked around the W$^\pm$ boson mass. 
No excess (unexpected accumulation) was observed in the data.
Figure~\ref{fig:gabi_fig2}a shows, as an example, the mass distribution
of the selected events for the data, 
the estimated background and simulated selectron events.

\begin{table}[ht]
\begin{center}
\begin{tabular}{|l|c|c|}
\hline
                   & Data & Background    \\ \hline \hline
Preselection       & 454  & 445.4$\pm$2.3 \\ \hline \hline
Selectron          & 55   &  55.4$\pm$0.8 \\ \hline
Electron-sneutrino & 41   &  49.1$\pm$0.7 \\ \hline
Other sleptons    & 50   &  48.8$\pm$0.7 \\ \hline
Squarks            & 7    &   8.8$\pm$0.3 \\ \hline  
\end{tabular}
\end{center}
\caption{\it 
The numbers of selected data and expected background events
in the four-jet channel after the preselection and at the end of the
different selections. Only the statistical error is indicated.}
\label{table:4jet}
\end{table}

The di-jet mass resolution using the 5C-fit is 0.6--1.6 
GeV, depending on the sparticle mass and decay. 
Events in a 2$\sigma$ mass window around the test
mass were selected. The efficiencies vary between 11.3\% and 34.3\%
within such a mass window for sparticle masses between 50 and 75 GeV,
depending on the sparticle mass and decay.

The signal detection efficiency is subject to the following inefficiencies and
systematic errors:
the statistical error due to the limited number of Monte Carlo events,
4.4--17.7\%;  
the uncertainty on modelling 
the kinematic variables used in the analysis, 3\%;
and additionally  for the smuon, muon-sneutrino, stau and tau-sneutrino 
selection, the inefficiency due to the differences between the slepton
and the charged Higgs boson simulation, 0--12\%.

The background estimate has the following uncertainties:
the statistical error due to the limited number of Monte Carlo events,
1.5\%;  
the statistical and systematic error on the luminosity measurement, 0.3 and
0.4\%; 
the uncertainty on modelling the SM background processes, estimated by comparing
different event generators, 2\%; and 
the kinematic variables used in the analysis, 4.9\%.

\subsection{Squarks}
\label{sec:multijetsnoemiss-sq}

Squarks are expected to hadronize resulting in a final state with six jets,
from which the two spectator jets have small energy, at least for heavy squarks,
and therefore it is still possible to reconstruct the squark pair events into
four jets.

To produce the signal reference histograms, we have used 
dedicated squark samples generated by SUSYGEN
with $\lambda^{\prime\prime}_{121}$ and $\lambda^{\prime\prime}_{123}$
couplings. Since jets originating from squark decays are 
narrower than the ones coming from Standard Model sources, 
in addition to the five input variables used in the slepton searches,
two new variables are introduced:
\begin{list}{$\bullet$}{\itemsep=0pt \parsep=0pt \topsep=-5pt \leftmargin=30pt}
\item the smallest boosted jet thrust; 
\item the highest jet mass.
\end{list}
The events are rejected if their likelihood output is less than 0.95.

\begin{figure}
\begin{center}
\epsfig{file=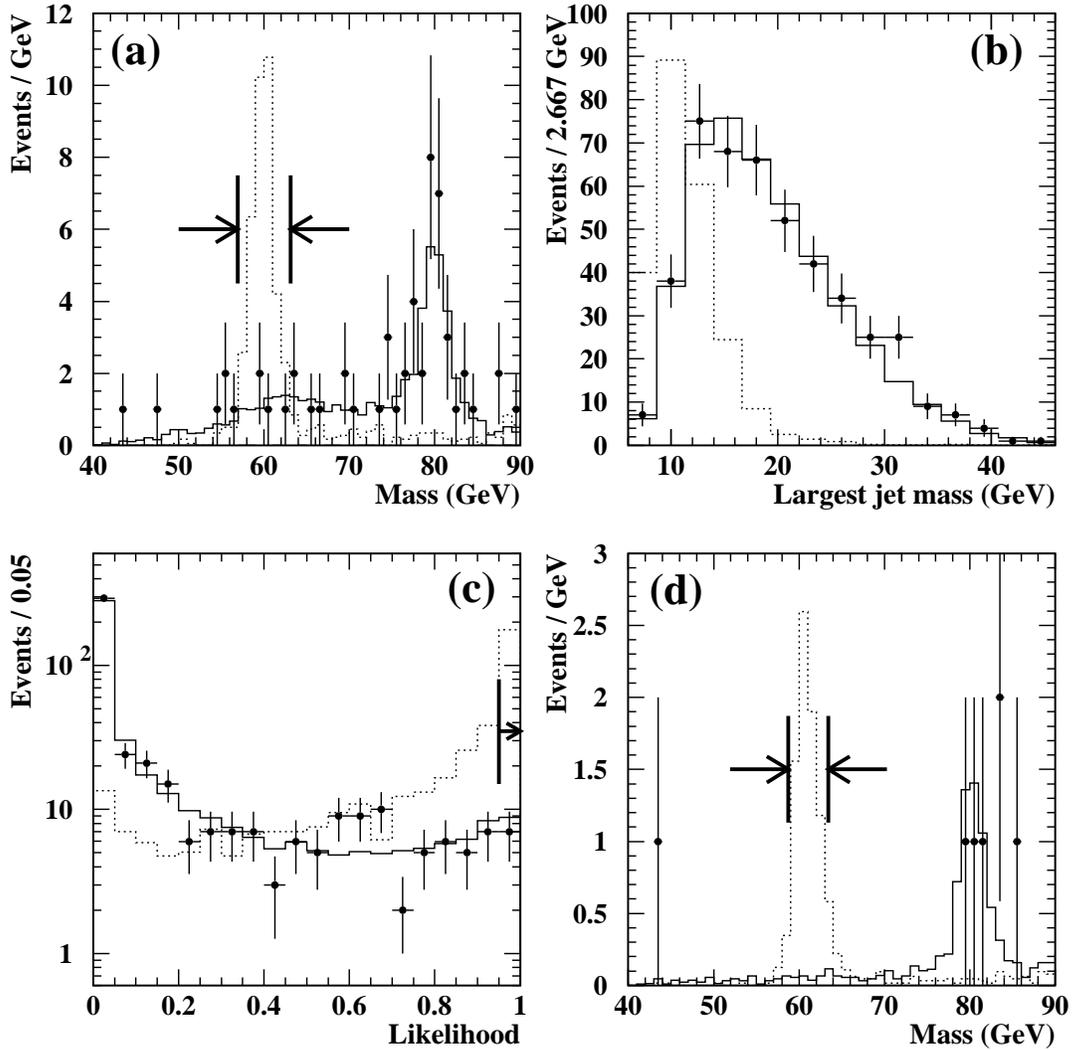,width=14.4cm} 
\end{center}
\caption{ \sl Four jets search (Analysis I): 
Distributions for data (points), for the estimated SM background  
(full histogram) and  for simulated 
signal events (dotted histogram).
{\it Selectron search:}
Figure (a) shows the mass distribution of selected events.
The mass window for a 60 GeV selectron is indicated by arrows.
{\it Squark search:} 
In Figure (b) one of the likelihood reference distributions, 
the largest jet mass, is plotted.
In Figure (c) the selection on the likelihood output can be seen.
In Figure (d) the mass distribution of selected events is plotted.  
The arrows indicate the mass window for a 60 GeV squark.
The SM background is normalised to the integrated luminosity of the
data, while the normalisation of the  signal distribution is arbitrary. 
}
\label{fig:gabi_fig2}
\end{figure}

Figures~\ref{fig:gabi_fig2}b-d show experimental plots
for the data, the estimated background and simulated signal
events. 
The numbers of selected data and expected background events are listed in 
Table~\ref{table:4jet}. 
Since the background is dominated by W$^+$W$^-$ production (93.3\%),
the mass distribution is peaked around the W$^\pm$ boson mass. 
No unexpected accumulation of events is observed in the data.

The di-jet mass resolution using the 5C kinematic fit is 0.45--1.2
GeV, depending on the squark mass and decay. A systematic shift of the 
reconstructed mass (up to +2.2 GeV for squark masses of 45 GeV)
is observed, which is taken into
account when applying the 2$\sigma$ mass window.
The signal detection  efficiencies within the mass windows vary 
between 14.1\% and 29.8\% for squark masses of 45--90 GeV.

The signal detection efficiency is subject to the following inefficiencies and
systematic errors:
the statistical error due to the limited number of Monte Carlo events,
4.9--7.8\%;  and
the uncertainty on modelling 
the kinematic variables used in the analysis, 13.2\%.

The effect of
different fragmentation and hadronization models has been tested 
comparing SUSYGEN and a special stop generator~\cite{ref:stopgen} used in 
OPAL stop searches~\cite{ref:stoppaper}. 
It was found that SUSYGEN produces wider
(more SM-like) jets,
and our efficiency would be more than a factor of two higher
for events generated by the stop generator. 
Thus our efficiency estimates 
using SUSYGEN are considered to be conservative.

The background estimate has the following uncertainties:
the statistical error due to the limited number of Monte Carlo events,
3.6\%;  
the statistical and systematic error on the luminosity measurement, 0.3 and
0.4\%; 
the uncertainty on modelling the SM background processes, 
estimated by comparing different event generators, 20.4\%; and 
the kinematic variables used in the analysis, 23.8\%.

The result of the slepton and squark
analyses is combined with previous searches performed at
$\sqrt{s}$=130--172 GeV for pair-produced, equal mass scalar particles (charged
Higgs bosons)~\cite{ch172} in order to increase the sensitivity for low mass
sleptons and squarks.
These previous searches are assumed to be equally efficient
for slepton, squark and charged Higgs search.
This hypothesis has been tested using slepton (squark) Monte Carlo
samples generated at $\sqrt{s}=172$ GeV for several $\lambda^\prime$
($\lambda^{\prime\prime}$) couplings with sparticle masses of 45, 55 and 70 GeV.
The efficiencies are found to be consistent within the statistical errors
except for the squark samples, where a relative 20\% increase
in the efficiency is observed. 
Conservatively, this gain is not taken into account.

\section{Interpretation}
\label{sec:results}

No significant excess of events in the 
data with respect to the expected background
has been observed for all analyses listed in Table~\ref{tab:relation}.
Production cross-section and mass limits have therefore been computed. 
These limits also take into account 
indirect limits obtained from the study of the Z$^0$ width at LEP1 and 
therefore concern only sparticle masses above 45 GeV.

Two approaches are used to present sfermion production limits.
In the first one, upper limits
on production cross-sections as functions of the sfermion 
masses are calculated with minimal model assumptions.
These upper limits in general do not depend
on the details of SUSY models, except for the assumptions 
that the sparticles are pair-produced and that only one \lb-like coupling
at a time is nonzero, as stated in Section~\ref{sec:intro}.
In the second approach, limits on the sfermion masses were calculated 
in the framework of the Constrained MSSM where mass limits are derived 
using the following parameters: 
$m_0$, the common sfermion mass at the GUT scale;
$M_2$, the SU(2) gaugino mass parameter 
at electroweak scales\footnote{We assume that 
      $M_1$, the U(1) gaugino mass at electroweak scales,
      is related to $M_2$ by the usual gauge unification
      condition:  $M_1 =  \frac{5}{3} \tan^2 \theta_{\mathrm W} M_2$.}; 
$\mu$, the mixing parameter of the two Higgs doublets and
$\tan \beta= v_2/v_1$, the ratio of the vacuum expectation values for 
the two Higgs doublets. 
For the indirect sfermion decays, we have used the branching ratios
for the decay $\tilde{f} \rightarrow f \nt_1$ predicted 
by the MSSM, and we have conservatively assumed no 
experimental sensitivity to any other decay mode.
The branching ratio for direct decay is always treated as equal 
to 1, as we allow 
only one \lb\ coupling to be different from zero at a time.
The MSSM mass exclusion 
plots presented in the following sections are computed for 
$\tan \beta =$~1.5 and $\mu =$~--200~GeV. This choice of parameters 
is rather conservative as sfermion production cross-sections generally 
increase for larger values of $\tan \beta $ or $|\mu|$.

In the indirect decay of a sfermion, $\tilde{f} \rightarrow f \nt_1$, 
via a \lbp\
coupling, the $\neutralino$ decays either as:

\begin{equation}
\nt_1 \rightarrow  \ell^-_i u_j \overline{d}_k \; , \;\;\;\;
\nt_1 \rightarrow  \ell^+_i \overline{u}_j d_k \; ,
\label{ref:decays1}
\end{equation}

or as:

\begin{equation}
\nt_1 \rightarrow  \nu_i d_j \overline{d}_k \; , \;\;\;\;
\nt_1 \rightarrow  \overline{\nu}_i \overline{d}_j d_k
\label{ref:decays2}
\end{equation}

This leads to final states
with two fermions from the sfermion decay plus the $\nt_1$ decay products:
\begin{enumerate}
\item
Four jets and two charged leptons if both $\nt_1$ decay via
(\ref{ref:decays1})
\item
Four jets and missing energy if both $\nt_1$ decay via
(\ref{ref:decays2})
\item
Four jets, one charged lepton and one neutrino if one $\nt_1$ decays via
(\ref{ref:decays1}) and the other via (\ref{ref:decays2}).
\end{enumerate}  

The relative branching ratios
of the neutralino into a final state with a charged or a neutral
lepton 
depends on the mass of the sneutrinos, the mass of the 
sleptons and on the components of the gaugino (Wino or Higgsino). 
To avoid a dependence of the results on the MSSM parameters, the 
branching ratio of $\nt_1$ to charged leptons and jets (\ref{ref:decays1}) 
was varied between
0 and 1. The branching ratio of $\nt_1$ to neutrinos 
and jets (\ref{ref:decays2})
was varied accordingly between 1 and 0. The combination of these two
branching ratios fixes the branching ratio for one $\nt_1$ decaying
via  (\ref{ref:decays1}) and the other via (\ref{ref:decays2}).
A likelihood ratio method~\cite{likelihood} was used 
to determine an upper limit for the 
cross-section. This method combines the individual 
analyses looking for the different final states possible for one given
\lbp\ coupling and assigns greater weight 
to those with a higher expected sensitivity, taking into account the expected 
number of background events.
This results in a cross-section limit as a function of the 
branching ratio and the sfermion mass. By taking the 
worst limit at each sfermion mass, a limit independent of the branching ratio
is determined.
For the direct decays, the final states are fully
determined by the indices of the coupling considered. 

In the following sections, cross-sections limits are shown for the
various direct and indirect decays studied in this paper, see 
Table~\ref{tab:relation}. 
In each cross-section plot,
only the curve corresponding to the worst cross-section limit is
shown amongst all possible cross-section limits resulting from the
couplings considered. The coupling yielding the worst
cross-section limit is indicated in each plot. 
Generally, the best excluded cross-section comes from final states
with a maximum number of muons and no taus, while the worst results come from
final states with many taus, due to their lower detection efficiency.

In the MSSM framework, the exclusion regions for the indirect
decays are valid for $\Delta m = m_{\sell} - m_{\nt_1} \ge 5$~GeV except
for the indirect decays of staus via \lbp\ which are
valid for $\Delta m = m_{\sell} - m_{\nt_1} \ge 22.5$~GeV. In this
particular case there is not enough sensitivity to place limits in
the small $\Delta m$ region. 
The exclusion region for the direct decays is independent of  
$\Delta m$. 

All limits presented here are quoted at the 95\%~C.L.
The inefficiencies due to different angular distributions 
(possible for selectron or electron sneutrino pair production via the 
$t$-channel) of produced sfermions and decay products 
were estimated for five different MSSM parameter sets, representing different 
neutralino field contents (gaugino/higgsino) and couplings,
and calculated separately for each analysis.
The selection efficiencies may vary by up to 10\%.
In interpreting the results, 
a conservative approach was adopted by choosing the lowest 
efficiencies in the limit calculation.
The systematic and the statistical 
errors were added in quadrature and then subtracted when using the 
number of background events.

\subsection{Selectron Limits}

\begin{figure}[htbp]
\centering
\begin{tabular}{c}
\epsfig{file=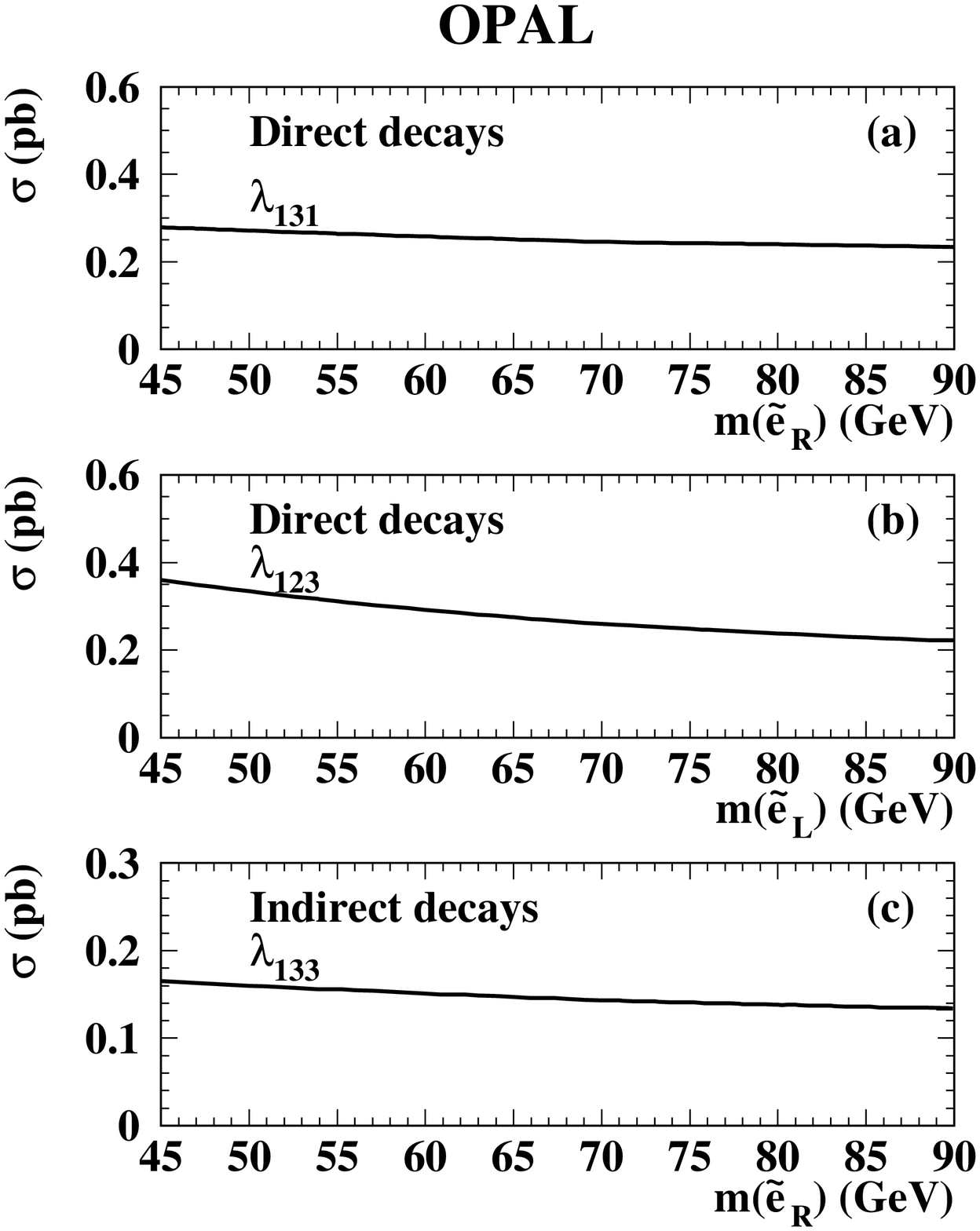     ,width=15.0cm} \\
\end{tabular}
\caption[]{\sl
Selectron decays via a \lb\ coupling: Upper limits at the 95\% C.L.
on the pair-production 
cross-sections 
for (a) the direct decay of a right-handed $\sele_R$, 
(b) the direct decay of a left-handed $\sele_L$ and
(c) the indirect decay of a $\sele_R$.
Only the worst limit curve is shown and the \lb\ corresponding to it is 
indicated.} 
\label{fig:cross_selectron_lb}
\end{figure}

Figure~\ref{fig:cross_selectron_lb} shows 
upper limits on the cross-sections of pair-produced $\sele$ followed
by a decay via a \lb\ coupling:
for (a) the direct decay of a right-handed $\sele_R$,  
(b) the direct decay of a left-handed $\sele_L$ and
(c) the indirect decay of a $\sele_R$. 
The production cross-section for left-handed sfermion is always 
larger that for right-handed sfermions, therefore we have 
conservatively quoted results for right-handed sfermions only.
For all cases, 
the worst upper limit on the cross-section is 0.36~pb.

\begin{figure}[htbp]
\centering
\begin{tabular}{c}
\epsfig{file=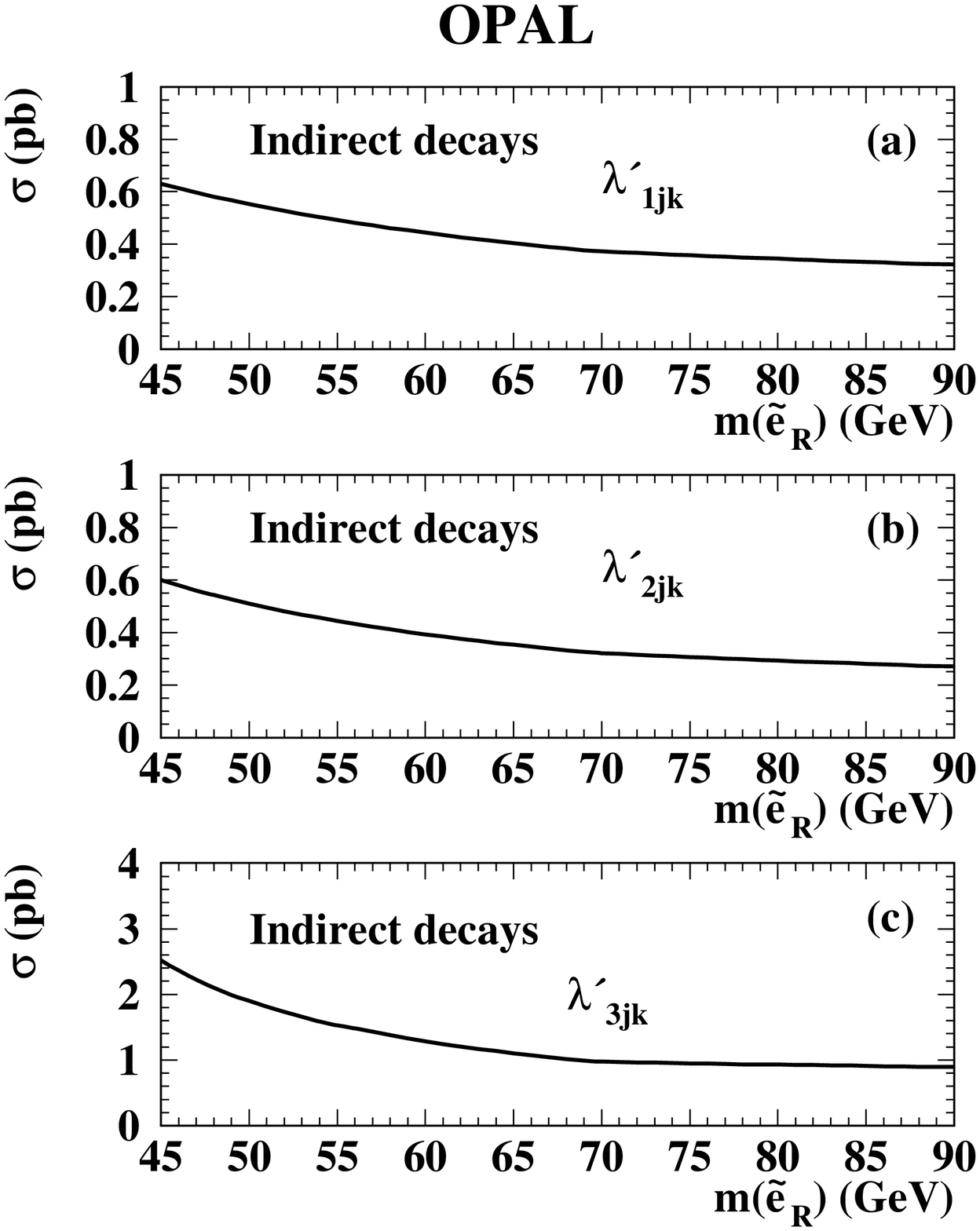     ,width=15.0cm} \\
\end{tabular}
\caption[]{\sl
Indirect selectron decays via a \lbp\ coupling:
Upper limits at the 95\% C.L. on the pair-production cross-sections
for (a) the indirect decay of a $\sele_R$ in the electron
channel, 
(b) the indirect decay of a $\sele_R$ in the muon channel and
(c) the indirect decay of a $\sele_R$ in the tau channel. 
} 
\label{fig:cross_selectron_lbp}
\end{figure}

Figure~\ref{fig:cross_selectron_lbp} shows 
upper limits on the cross-sections of pair-produced $\sele$ followed
by a decay via a \lbp\ coupling:
for (a) the indirect decay of a $\sele_R$ in the electron
channel, 
(b) the indirect decay of a $\sele_R$ in the muon channel and
(c) the indirect decay of a $\sele_R$ in the tau channel. For all cases, 
the weakest upper limit on the cross-section is 2.5~pb.

\begin{figure}[htbp]
\centering
\begin{tabular}{c}
\epsfig{file=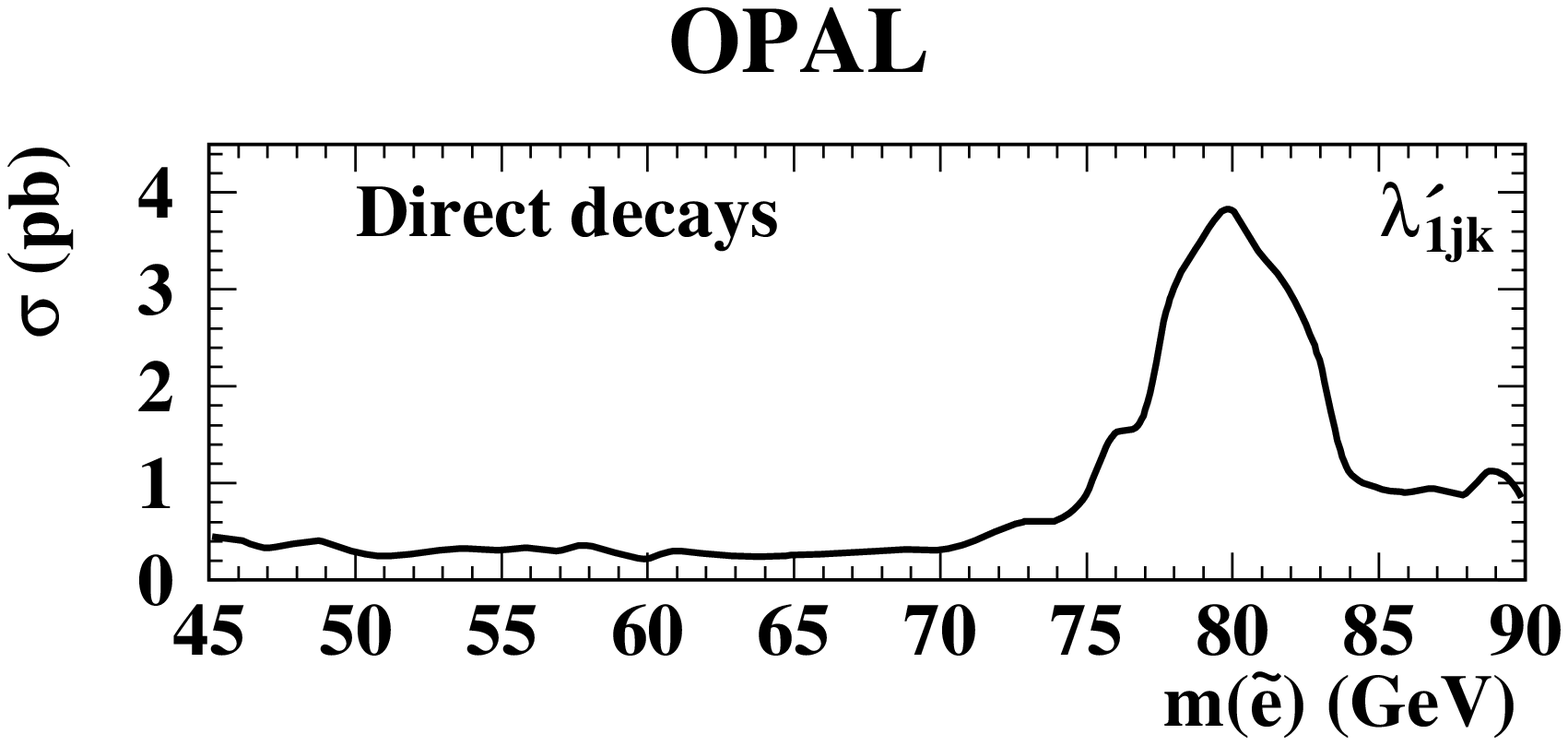,width=15.0cm} \\
\end{tabular}
\caption[]{\sl
Direct selectron decays via a \lbp\ coupling:
Upper limits at the 95\% C.L. 
on the pair-production cross-sections of $\sele$. 
} 
\label{fig:cross_sele_4jets}
\end{figure}

Figure~\ref{fig:cross_sele_4jets} shows 
upper limits on the cross-sections of pair-produced $\sele$ directly
decaying via a \lbp\ coupling to a four-jet final state. The peak 
structure visible in the figure at approximately the mass of the 
W-boson comes from irreducible background due to WW pair-production.

\begin{figure}[htbp]
\centering
\epsfig{file=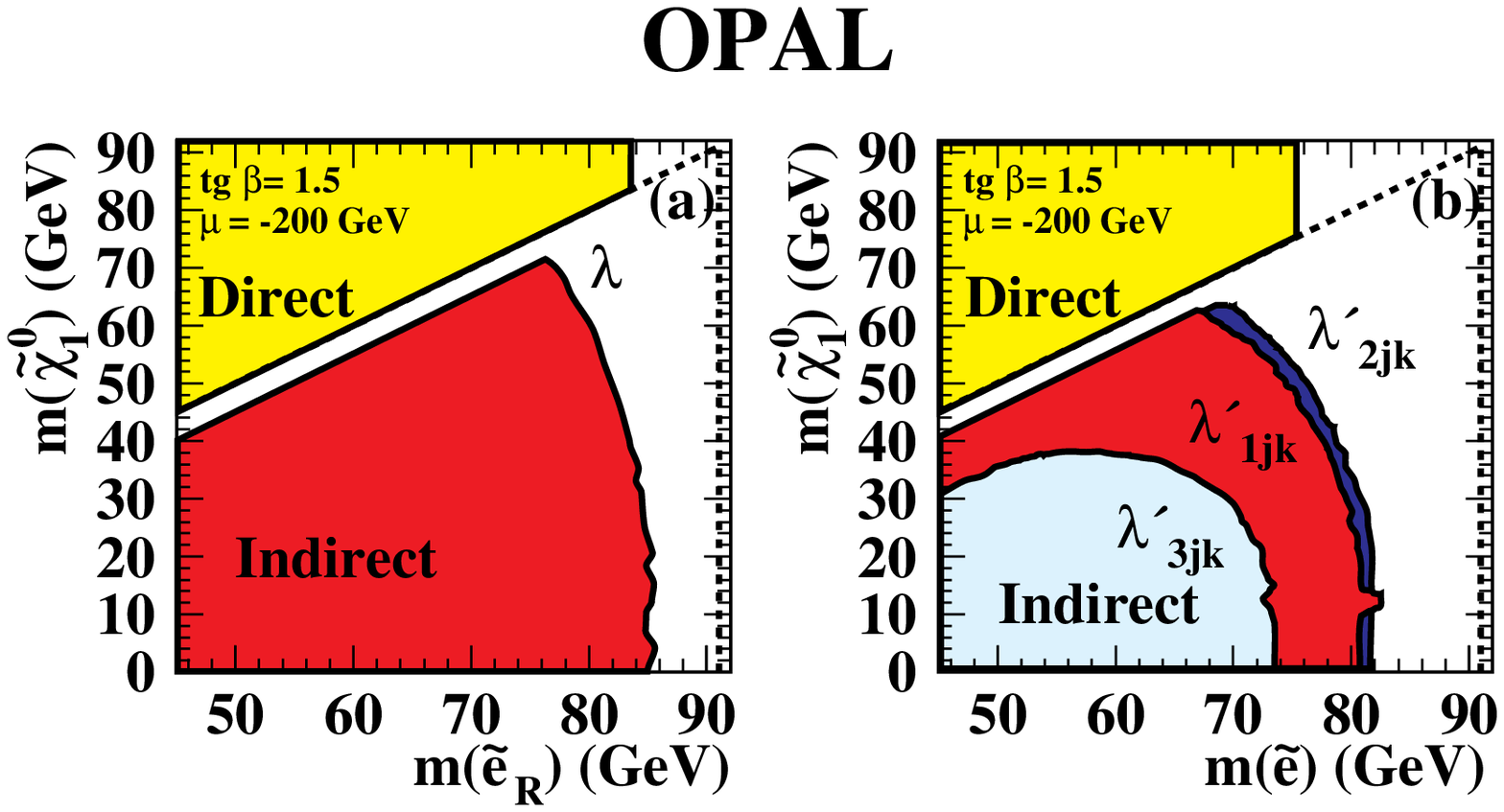, width=17.0cm} 
\caption[]{\sl
Selectron: 
MSSM exclusion region for $\sele^+ \sele^-$ 
production in 
the $(m_{\sele}, m_{\nt_1})$ plane at 95\% C.L.  
for  (a) a \lb\ coupling and (b) a \lbp\ coupling.
For the direct and indirect decays via \lb\ and the indirect decays 
via \lbp\, the exclusion region for $\sele_R \sele_R$ is shown. For 
the direct decays via \lbp\, the exclusion is shown for the only 
possible case of $\sele_L \sele_L$. 
The kinematic limit is shown as the dashed line. 
The gap between the excluded regions for direct and indirect decays 
corresponds to $\Delta m = m_{\sell} - m_{\nt_1} < 5$~GeV.}
\label{fig:mssm_selectron}
\end{figure}

In the MSSM, the $\sele$ pair-production cross-section
is enhanced by the presence of the $t$-channel diagram.
Figure~\ref{fig:mssm_selectron}(a)
shows the 95\% C.L. exclusion limits for right-handed selectrons
decaying directly or indirectly via a \lb\ coupling. 
In the region where the $\nt_1$ is heavier than
the $\sele$, only direct decays are possible. 
When the $\nt_1$ is lighter than the $\sele$, the
indirect decays are expected to be dominant. 
For indirect decays via a \lb\ coupling, a right-handed 
selectron with a mass smaller than 84~GeV 
is excluded at the 95\% C.L. in the case of a low-mass 
$\nt_1$. For direct decays via a \lb\ coupling, 
a right-handed selectron with a 
mass smaller than 84~GeV 
is excluded at the 95\% C.L.
Figure~\ref{fig:mssm_selectron}~(b)
shows the 95\% C.L. exclusion limits for selectrons
decaying via a \lbp\ coupling. The exclusion refers to right-handed 
selectrons for the indirect decays and to left-handed selectrons for 
direct decays. 
In the case of indirect decay,
a right-handed selectron with a mass smaller than 72~GeV 
is excluded at the 95\% C.L. in the case of a low-mass 
$\nt_1$ and a left-handed selectron with a mass smaller than 76~GeV 
is excluded in the case of direct decays.

\subsection{Smuon Limits}

\begin{figure}[htbp]
\centering
\begin{tabular}{c}
\epsfig{file=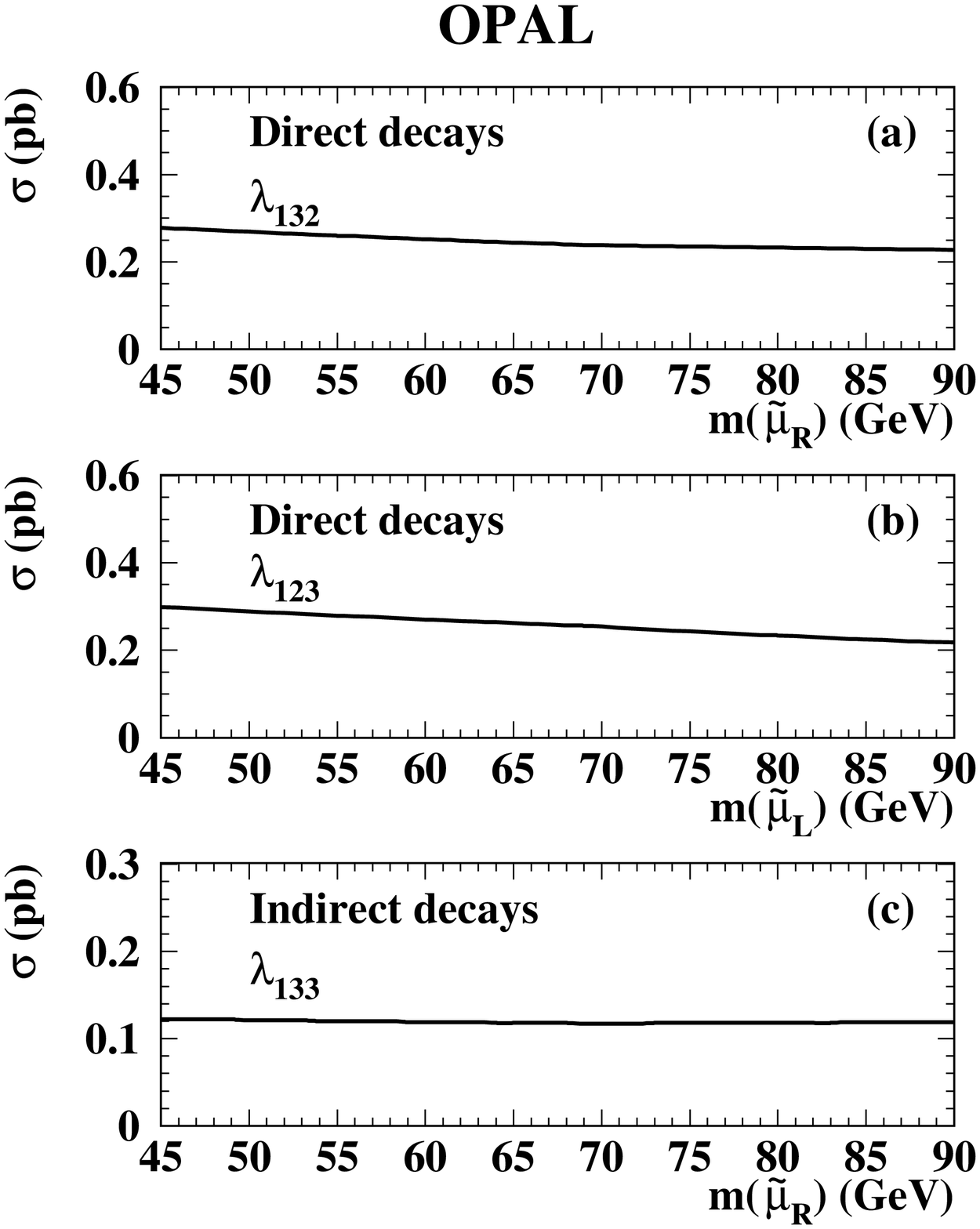     ,width=15.0cm} \\
\end{tabular}
\caption[]{\sl
Smuon decays via a \lb\ coupling: Upper limits at the 95\% C.L. 
on the pair production 
cross-sections of $\smu$
for (a) the direct decay of a right-handed $\smu_R$, 
(b) the direct decay of a left-handed $\smu_L$ and
(c) the indirect decay of a $\smu_R$.
Only the worst limit curve is shown and the \lb\ corresponding to it is 
indicated.} 
\label{fig:cross_smuon_lb}
\end{figure}

\begin{figure}[htbp]
\centering
\begin{tabular}{c}
\epsfig{file=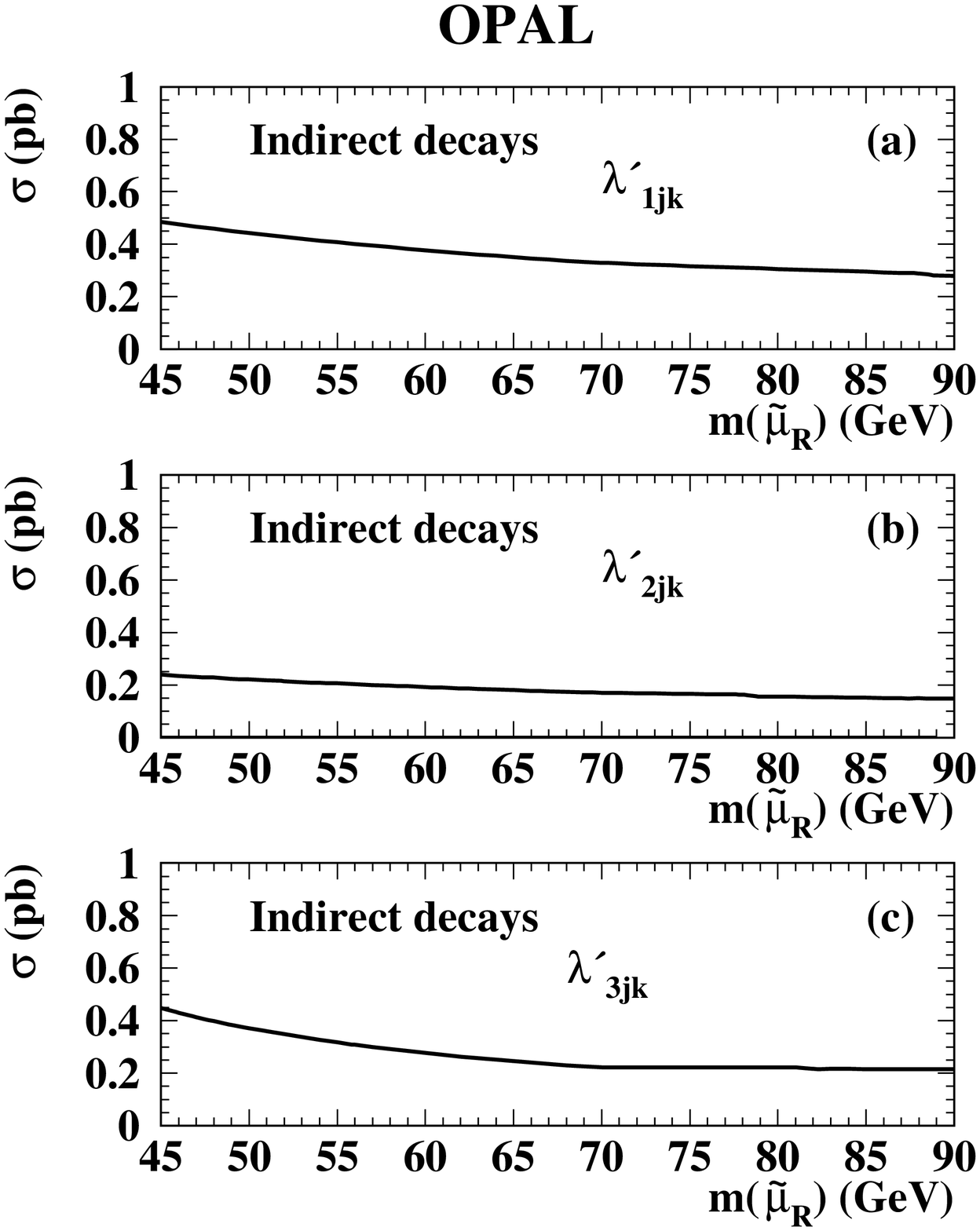     ,width=15.0cm} \\
\end{tabular}
\caption[]{\sl
Smuon decays via a \lbp\ coupling:
Upper limits at the 95\% C.L. on the pair-production cross-sections of $\smu$
for (a) the indirect decay of a $\smu_R$ in the electron
channel, 
(b) the indirect decay of a $\smu_R$ in the muon channel and
(c) the indirect decay of a $\smu_R$ in the tau channel. 
} 
\label{fig:cross_smuon_lbp}
\end{figure}

Figures~\ref{fig:cross_smuon_lb} and
\ref{fig:cross_smuon_lbp} 
show upper limits on the cross-sections for pair-produced $\smu$.
The weakest upper limit on the cross-section is 0.30~pb for the
\lb\ couplings and 0.48~pb for the \lbp\ couplings.

\begin{figure}[htbp]
\centering
\begin{tabular}{c}
\epsfig{file=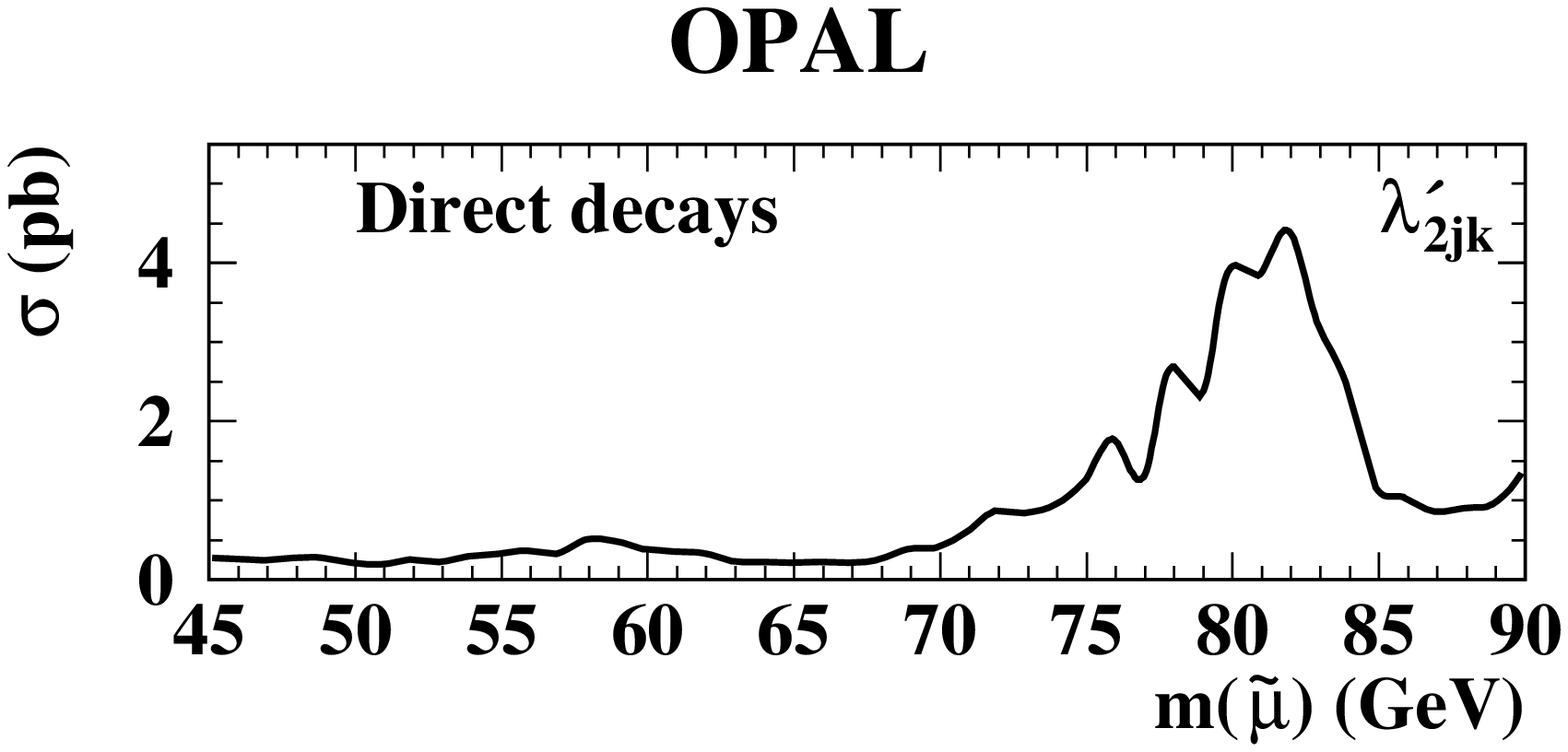     ,width=15.0cm} \\
\end{tabular}
\caption[]{\sl
Smuon direct decays via a \lbp\ coupling:
Upper limits at the 95\% C.L. on the pair-production cross-sections of $\smu$. 
} 
\label{fig:cross_smu_4jets}
\end{figure}

Figure~\ref{fig:cross_smu_4jets} shows 
upper limits on the cross-sections of pair produced $\smu$ directly
decaying via a \lbp\ coupling to a four-jet final state.

\begin{figure}[htbp]
\centering
\epsfig{file=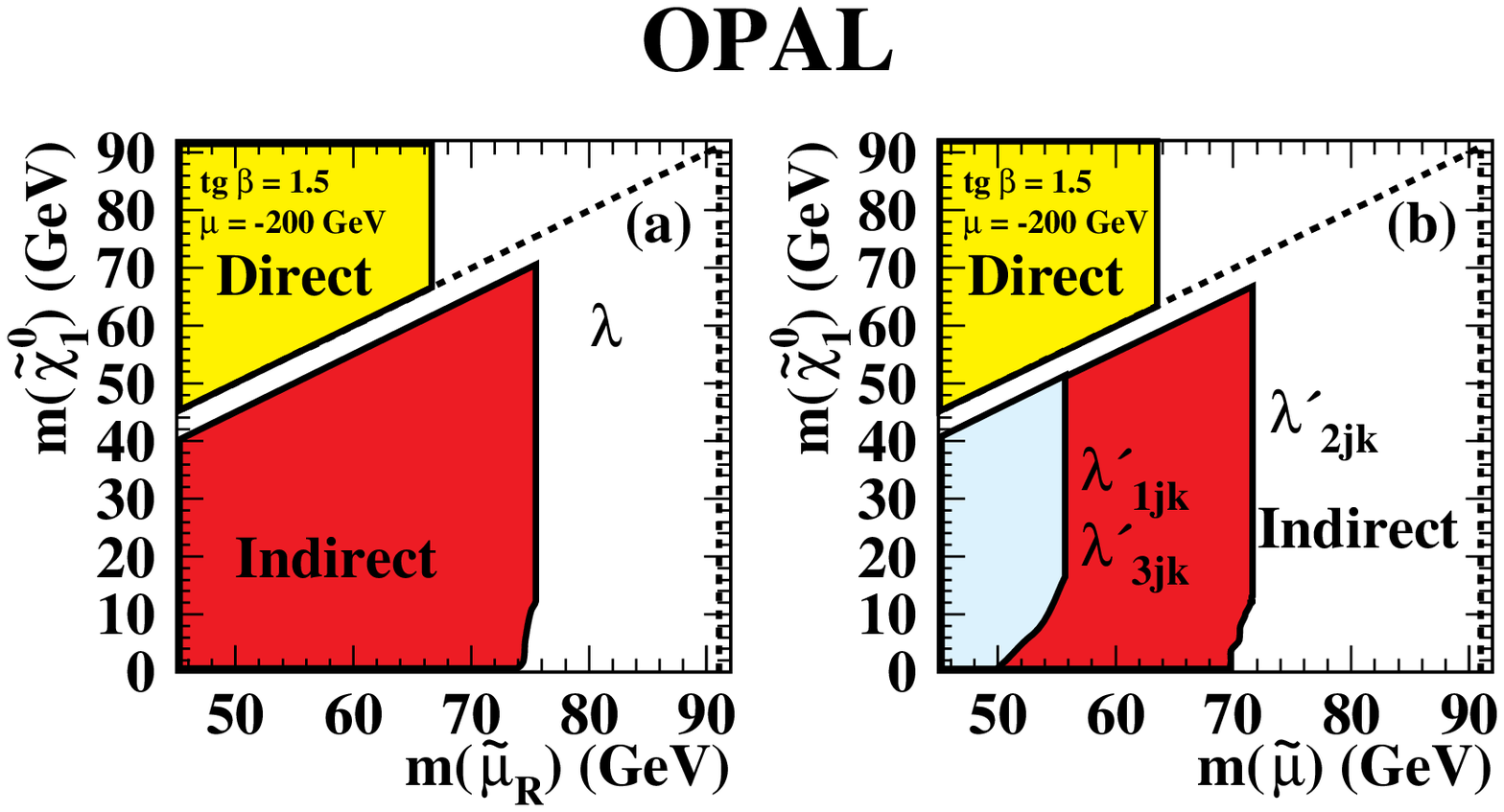, width=17.0cm} 
\caption[]{\sl
Smuon: 
MSSM exclusion region for $\smu^+ \smu^-$ 
production in 
the $(m_{\smu}, m_{\nt_1})$ plane at 95\% C.L.  
for  (a) a \lb\ coupling and (b) a \lbp\ coupling.
For the direct decays via \lbp\, the exclusion region is shown for 
the case $\smu_L \smu_L$. In the other cases, the exclusion regions 
for $\smu_R \smu_R$ are shown.
The kinematic limit is shown as the dashed line.  }
\label{fig:mssm_smuon}
\end{figure}

In the MSSM,
for indirect decays via a \lb\ coupling, a right-handed 
smuon with a mass smaller than 74~GeV 
is excluded at the 95\% C.L. in the case of a low-mass 
$\nt_1$, see Figure~\ref{fig:mssm_smuon}. 
For direct decays via a \lb\ coupling, 
a right-handed smuon with a 
mass smaller than 66~GeV 
is excluded at the 95\% C.L.
For indirect decays via a \lbp\ coupling, a right-handed smuon with a 
mass smaller than 50~GeV 
is excluded at the 95\% C.L. in the case of a low-mass 
$\nt_1$: for direct decays a left-handed smuon with a mass smaller 
than 64~GeV is excluded.

\subsection{Stau Limits}

\begin{figure}[htbp]
\centering
\begin{tabular}{c}
\epsfig{file=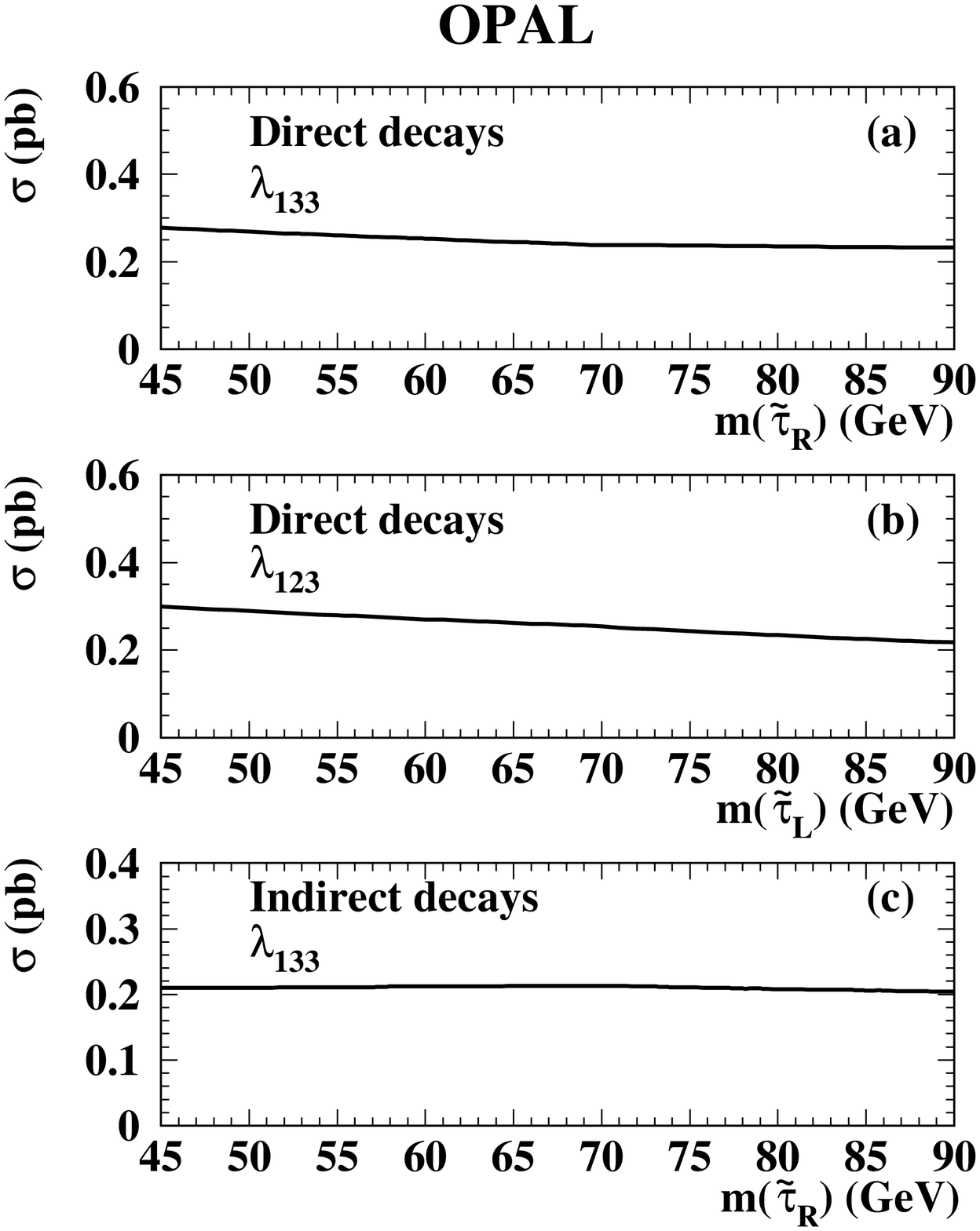     ,width=15.0cm} \\
\end{tabular}
\caption[]{\sl
Stau decays via a \lb\ coupling: upper limits on the pair-production 
cross-sections 
for (a) the direct decay of a right-handed $\stau_R$, 
(b) the direct decay of a left-handed $\stau_L$ and
(c) the indirect decay of a $\stau_R$.
Only the worst limit curve is shown and the \lb\ corresponding to it is 
indicated.}  
\label{fig:cross_stau_lb}
\end{figure}

\begin{figure}[htbp]
\centering
\begin{tabular}{c}
\epsfig{file=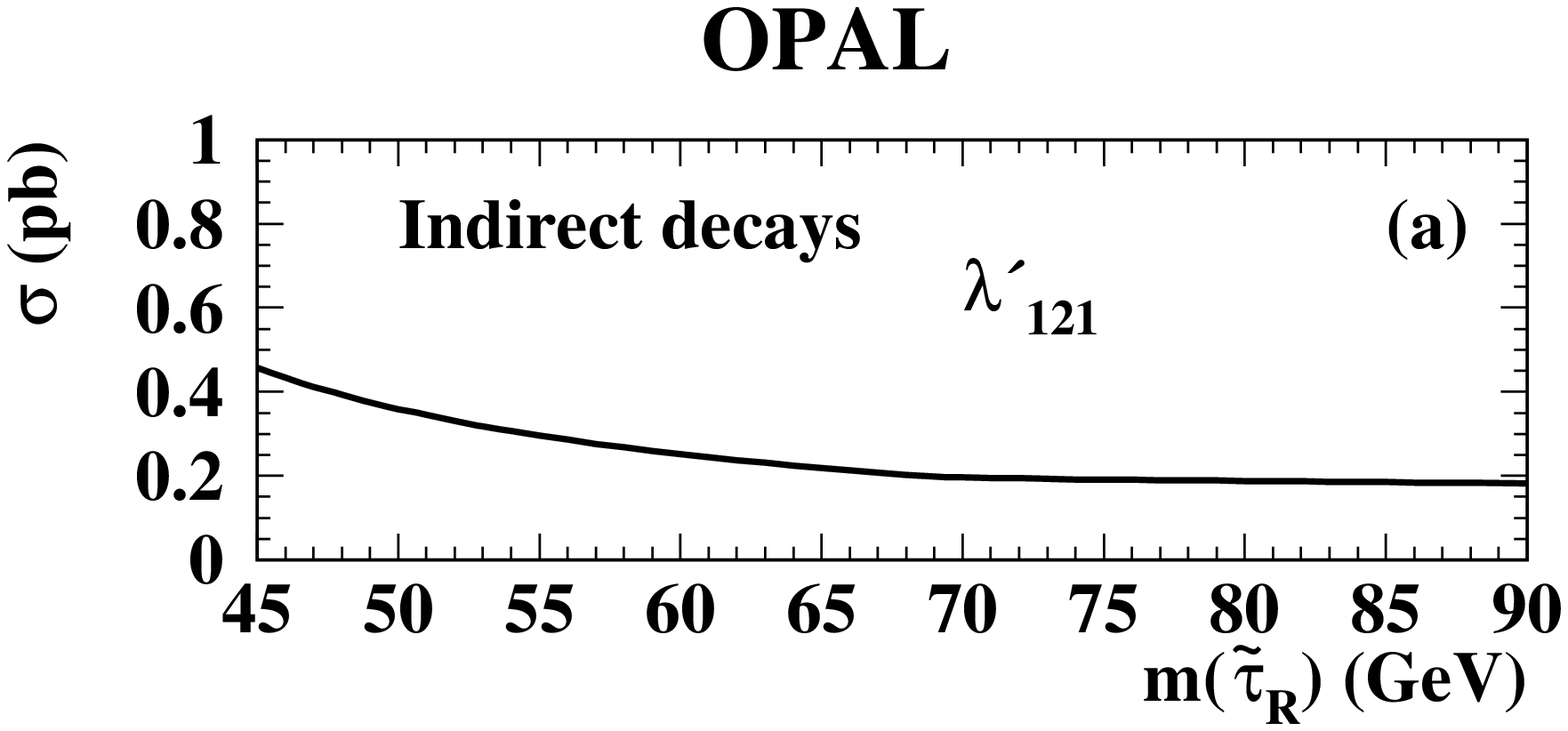     ,width=15.0cm} \\
\end{tabular}
\caption[]{\sl
Stau decays via a \lbp\ coupling:
Upper limits on the pair-production cross-sections
for the indirect decay of a $\stau_R$ in the electron
channel. 
The indirect decay of a $\stau_R$ in the muon channel and
the indirect decay of a $\stau_R$ in the tau channel yield identical 
results.
Only the worst limit curve is shown and the \lbp\ corresponding to it is 
indicated.} 
\label{fig:cross_stau_lbp}
\end{figure}

\begin{figure}[htbp]
\centering
\epsfig{file=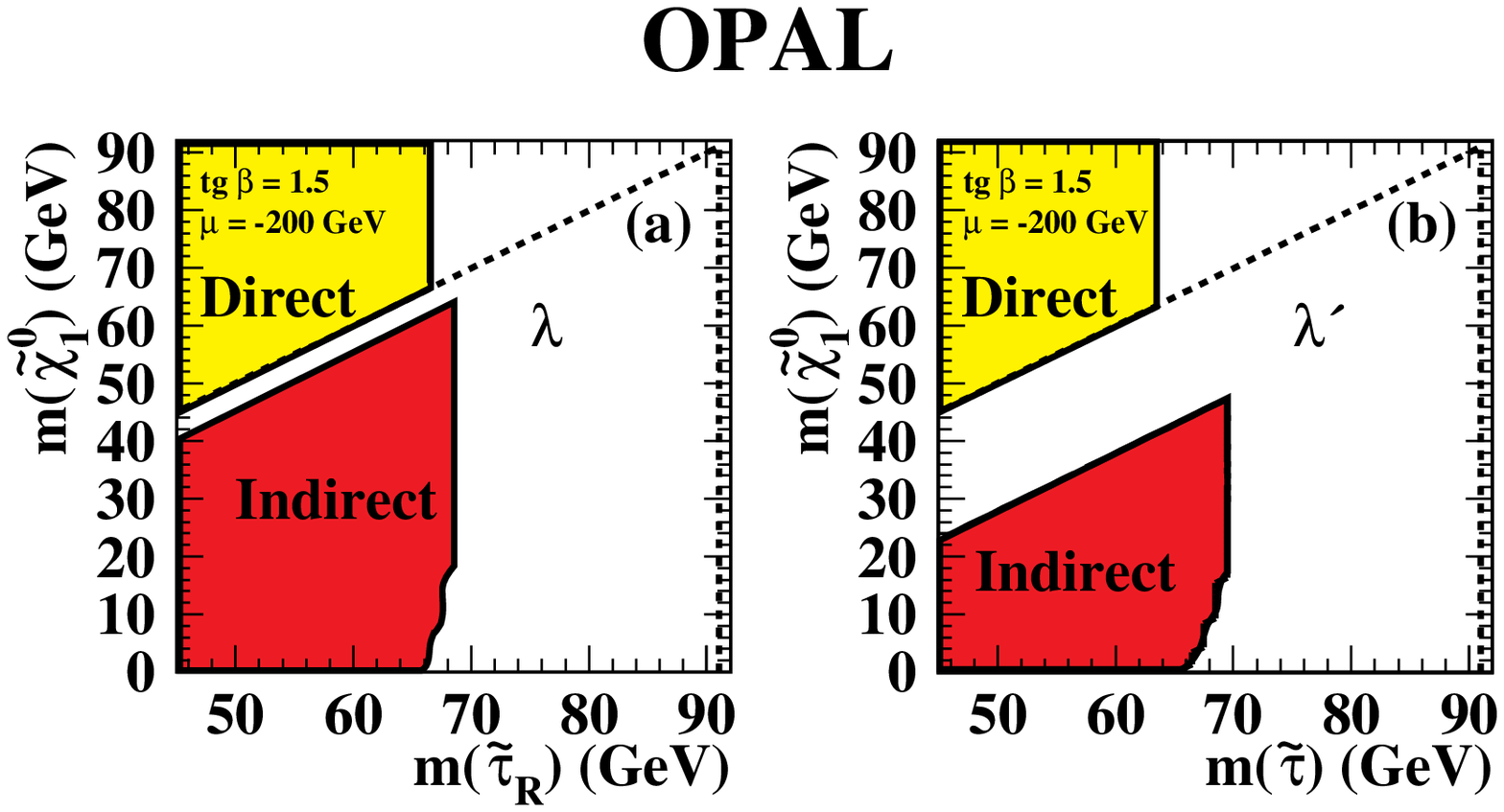, width=17.0cm} 
\caption[]{\sl
Stau: 
MSSM exclusion region for $\stau^+ \stau^-$ 
production in 
the $(m_{\stau}, m_{\nt_1})$ plane at 95\% C.L.  
for  (a) a \lb\ coupling
and (b) a \lbp\ coupling.
For direct decays via \lbp\, the exclusion region for $\stau_L 
\stau_L$ is shown. In the other cases, exclusion regions for $\stau_R 
\stau_R$ are shown.
The kinematic limit is shown as the dashed line.  }
\label{fig:mssm_stau}
\end{figure}

Figures~\ref{fig:cross_stau_lb}
to \ref{fig:mssm_stau}
show the exclusion plots for pair-produced $\stau$. 
The weakest upper limit on the cross-section is 0.30~pb for the
\lb\ couplings
and 0.45~pb for the \lbp\ couplings.

Pair-produced $\stau$ directly
decaying via a \lbp\ coupling to a four-jet final state yield 
identical results as shown for the $\smu$ case, 
see Figure~\ref{fig:cross_smu_4jets}.

In the MSSM, for indirect decays via a \lb\ coupling, a right-handed 
stau with a mass smaller than 66~GeV 
is excluded at the 95\% C.L. in the case of a low-mass 
$\nt_1$. For direct decays via a \lb\ coupling, a right-handed stau with a 
mass smaller than 66~GeV 
is excluded at the 95\% C.L.
For indirect decays via a \lbp\ coupling, a right-handed stau with a 
mass smaller than 66~GeV 
is excluded at the 95\% C.L. in the case of a low-mass 
$\nt_1$.
For direct decays, a left-handed stau with a mass smaller than 63~GeV 
is excluded.

\subsection{Sneutrino Limits}

\begin{figure}[htbp]
\centering
\begin{tabular}{c}
\epsfig{file=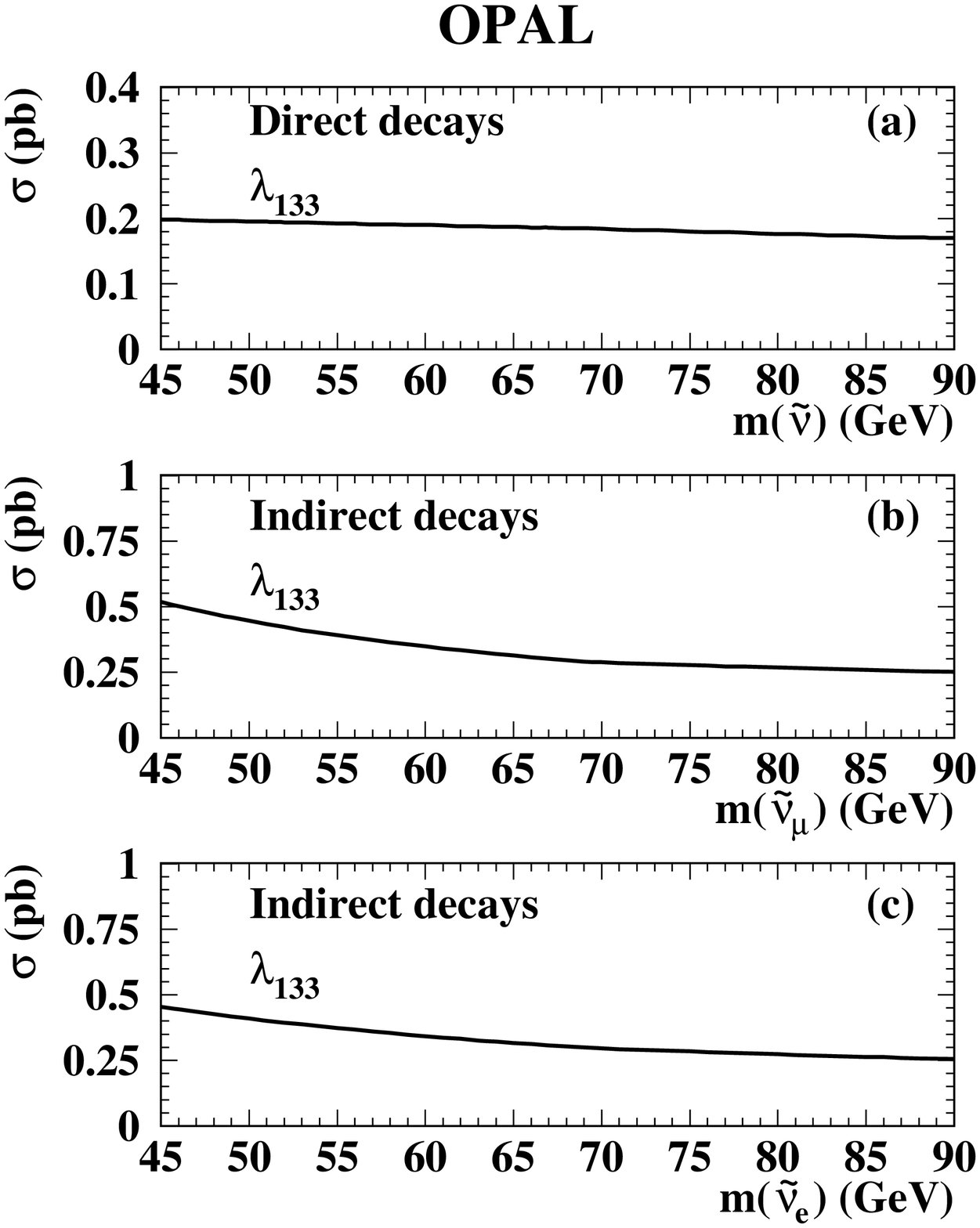       ,width=15.0cm} \\
\end{tabular}
\caption[]{\sl
Sneutrino decays via a \lb\ coupling: Upper limits at the 95\% C.L. 
on the pair-production 
cross-sections
for (a) the direct decay, 
(b) the indirect decay of $\snu_{\mu}$ (or $\snu_{\tau}$) and 
(c) the indirect decay of $\snu_e$.
Only the worst limit curve is shown and the \lb\ corresponding to it is 
indicated.} 
\label{fig:cross_snu_lb}
\end{figure}

\begin{figure}[htbp]
\centering
\begin{tabular}{c}
\epsfig{file=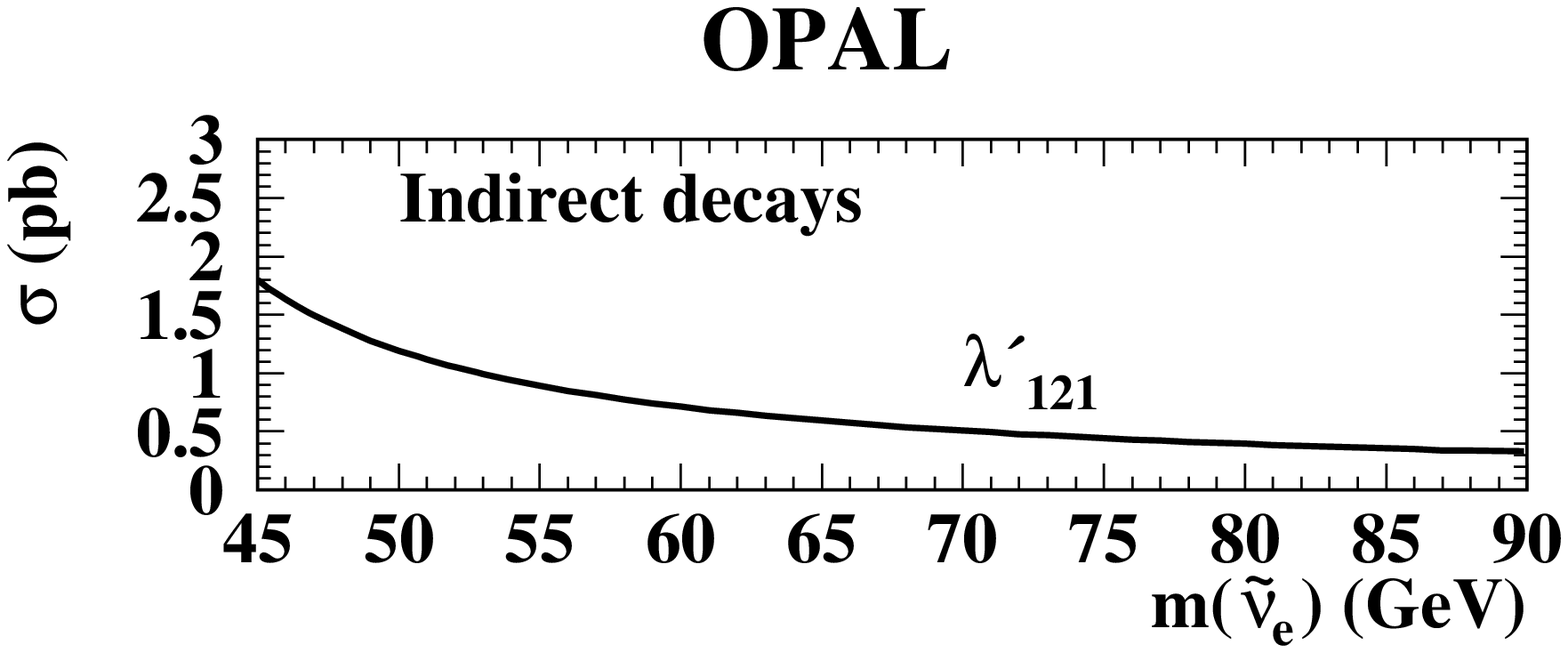    ,width=15.0cm} \\
\end{tabular}
\caption[]{\sl
Sneutrino decays via a \lbp\ coupling:
Upper limits at the 95\% C.L. on the pair-production cross-sections.
Only the worst limit curve is shown and the \lbp\ corresponding to it is 
indicated.} 
\label{fig:cross_snu_lbp}
\end{figure}

Figures~\ref{fig:cross_snu_lb} and
\ref{fig:cross_snu_lbp} 
show the exclusion plots for pair produced $\snu$. 
The weakest upper limit on the cross-section is 0.52~pb for the
\lb\ couplings and 1.8~pb for the \lbp\ couplings.

\begin{figure}[htbp]
\centering
\begin{tabular}{c}
\epsfig{file=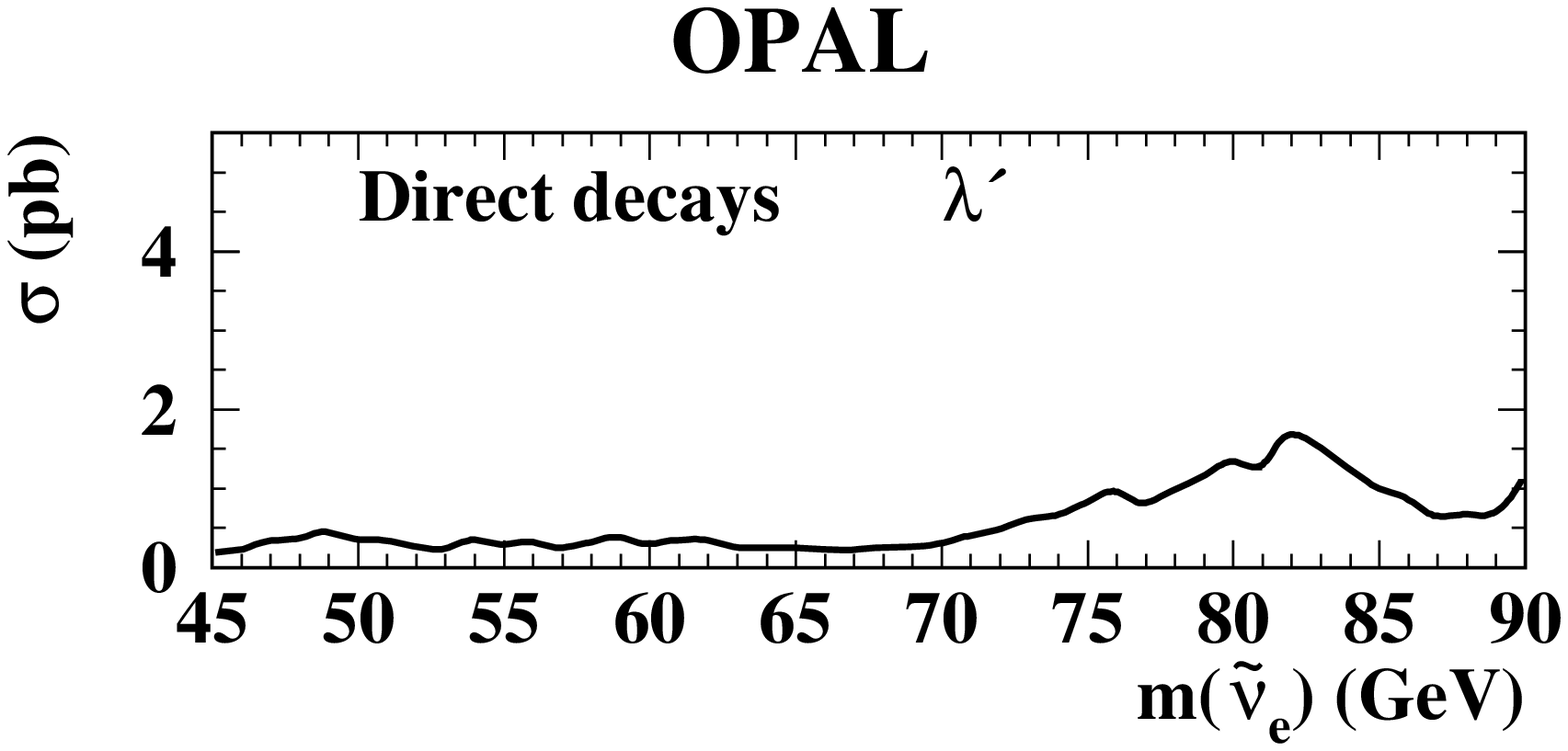     ,width=15.0cm} \\
\end{tabular}
\caption[]{\sl
Sneutrino direct decays via a \lbp\ coupling:
Upper limits at the 95\% C.L. on the 
pair-production cross-sections of $\snu_{\mathrm e}$. 
The search for $\snu_{\mu}$ and $\snu_{\tau}$ yield identical limits
to the ones shown in Figure 12.
} 
\label{fig:cross_sneu_4jets}
\end{figure}

Figure~\ref{fig:cross_sneu_4jets} shows 
upper limits on the cross-sections of pair-produced  
$\snu$ decaying directly via a \lbp\ coupling to a four-jet final state.
The searches for $\snu_{\mu}$ and $\snu_{\tau}$ yield identical limits.

\begin{figure}[htbp]
\centering
\epsfig{file=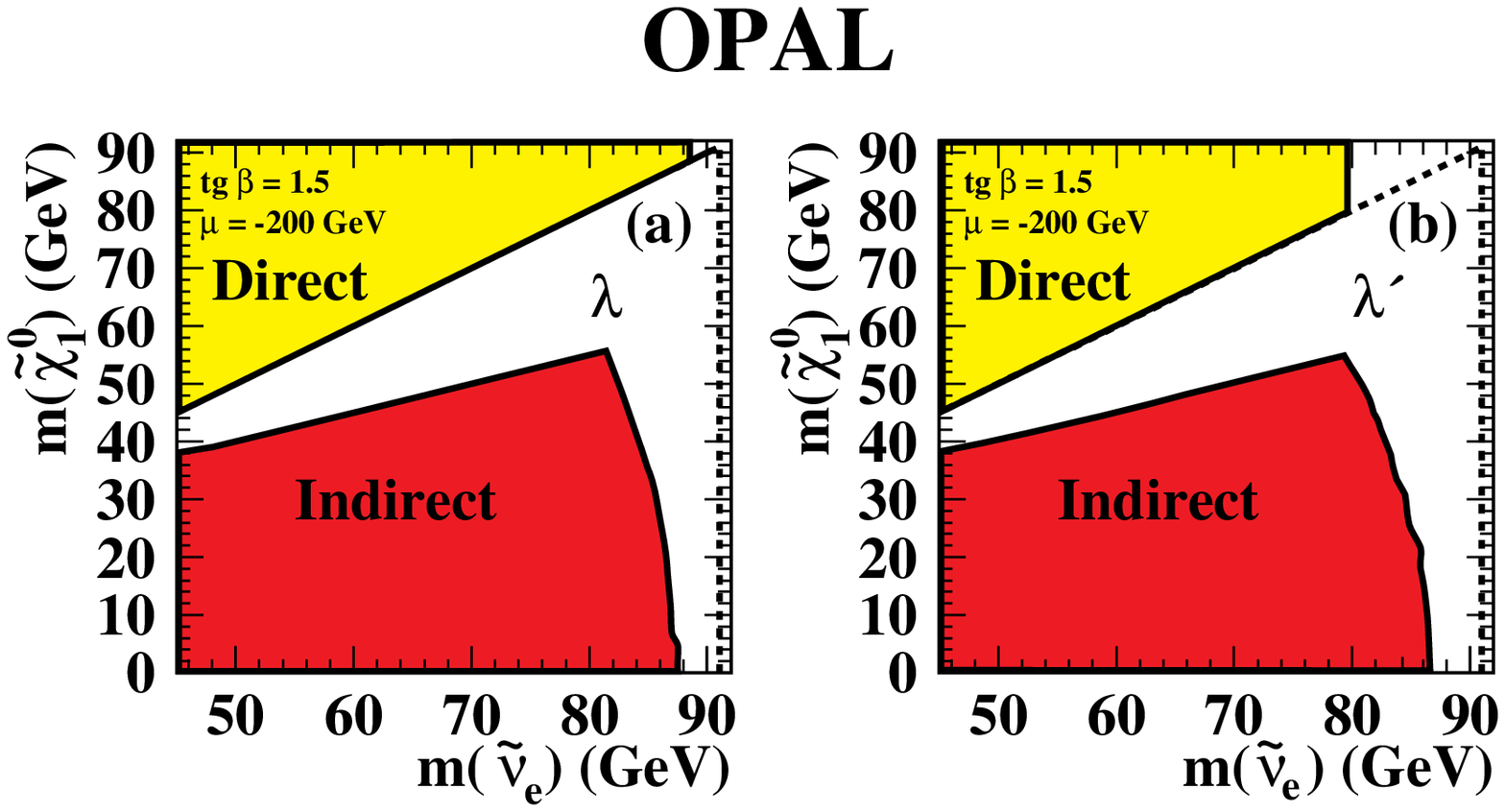, width=17.0cm} 
\caption[]{\sl
Sneutrino: 
MSSM exclusion region for $\snu_e \snu_e$ 
production in 
the $(m_{\snu_e}, m_{\nt_1})$ plane at 95\% C.L.  
for  (a) a \lb\ coupling and (b) a \lbp\ coupling.
The kinematic limit is shown as the dashed line.  }
\label{fig:mssm_sneutrino}
\end{figure}

In the MSSM, the $\snu_e$ pair-production cross-section  
is enhanced by the presence of the $t$-channel diagram.
Figure~\ref{fig:mssm_sneutrino}(a)
shows the 95\% C.L. exclusion limits for $\snu_e$
decaying directly or indirectly via a \lb\ coupling.  
For indirect decays via a \lb\ coupling, an electron sneutrino
with a mass smaller than 87~GeV 
is excluded at the 95\% C.L. in the case of a low-mass 
$\nt_1$. For direct decays via a \lb\ coupling, 
an electron sneutrino with a 
mass smaller than 88~GeV 
is excluded at the 95\% C.L.
Figure~\ref{fig:mssm_sneutrino}~(b)
shows the 95\% C.L. exclusion limits for electron sneutrinos
decaying indirectly via a \lbp\ coupling. 
In this case, an electron sneutrino with a mass smaller than 86~GeV 
is excluded at the 95\% C.L. in the case of a low-mass 
$\nt_1$.
For direct decays, a sneutrino with a mass smaller than 80~GeV is 
excluded.
MSSM exclusion plots for $\snu_{\mu}$ and $\snu_{\tau}$ are not shown 
because of their very small cross-section. 
For direct $\snu_{\mu}$ decay via a \lb\ coupling
a lower mass limit of 66 GeV is derived. For direct $\snu_{\mu}$ decay 
via a \lbp\ coupling a lower mass limit of 58 GeV is obtained.

\subsection{Stop Limits}

\begin{figure}
\centering
\begin{tabular}{c}
\epsfig{file=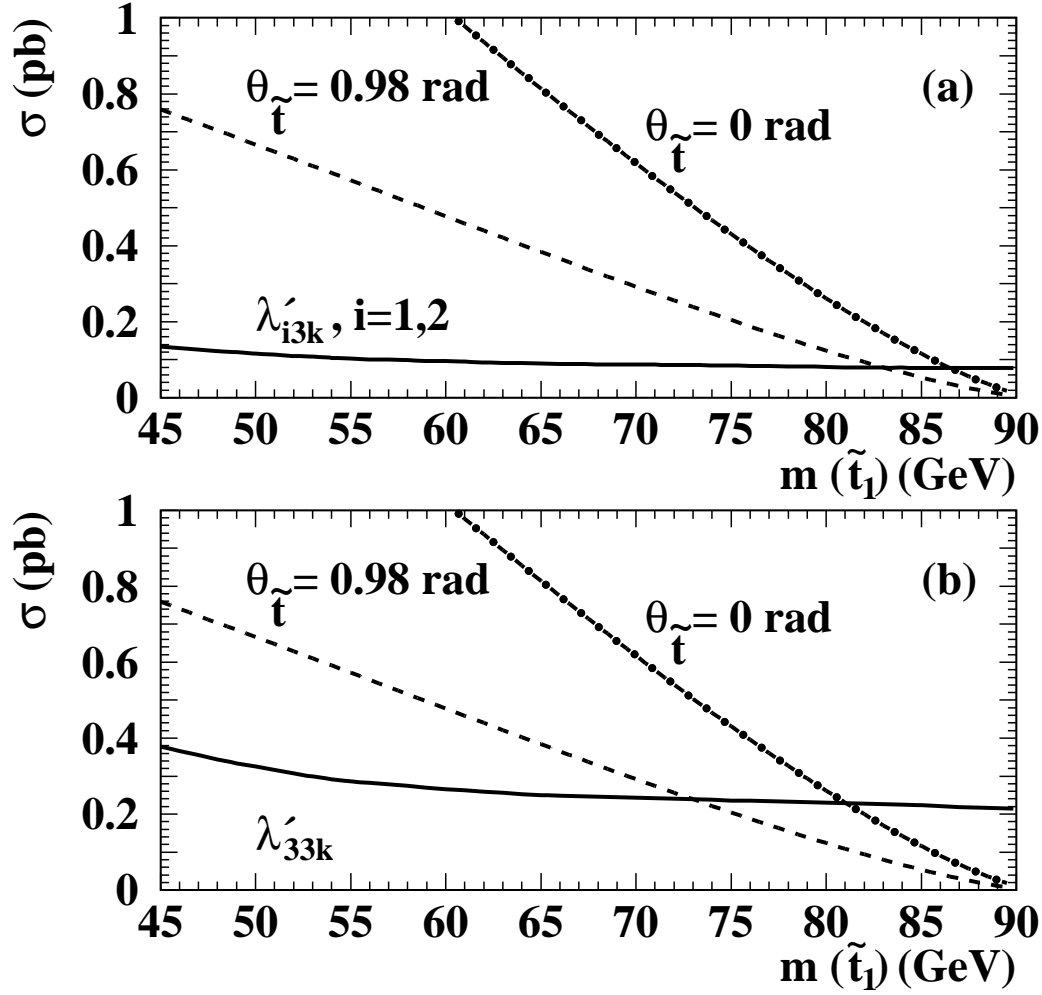,width=15cm} \\
\end{tabular}
\caption[]{\sl
Stop direct decays via a \lbp\ coupling: 
Cross-section limits at the 95\% C.L. 
in the electron and muon channels (a) and 
in the tau channel (b). Also shown are the maximum (dashed-dotted line) 
and minimum (dashed line)
cross-sections predicted by the MSSM,
corresponding to a mixing angle of 0 rad and 0.98 rad (decoupling 
limit). 
}
\label{fig:limit_stop}
\end{figure}

For the stop search in the electron and muon channel, 
no events satisfy the final selection cuts. 
A cross-section limit of 0.15~pb was derived for the pair-production
of stops decaying directly via \lbp$_{13k}$ or \lbp$_{23k}$,
in the mass region 45~GeV $< m_{\stopx} <$90~GeV. The
excluded cross-section as a function of the stop mass is shown in 
Fig.~\ref{fig:limit_stop}~(a).
If one assumes a stop production cross-section as predicted by the MSSM, 
masses lower than 82~GeV can be excluded for any mixing 
angle $\theta_{\stopx}$ under
the assumptions made above. 
For the stop search in the tau channel, two events have satisfied the
final selection cuts. 
A cross-section limit of 0.24~pb was derived for the pair-production
of the stops decaying directly via \lbp$_{i3k}$,
in the mass region 45~GeV $< m_{\stopx} <$ 90~GeV. The
excluded cross-section as a function of the stop mass is shown in 
Fig.~\ref{fig:limit_stop}~(b).
In the tau channel, 
masses lower than 73~GeV can be excluded for any mixing 
angle $\theta_{\stopx}$. 
More detailed exclusion limits 
are given in Table~\ref{tab:limit_stop}. 

\begin{figure}
\centering
\begin{tabular}{c}
\epsfig{file=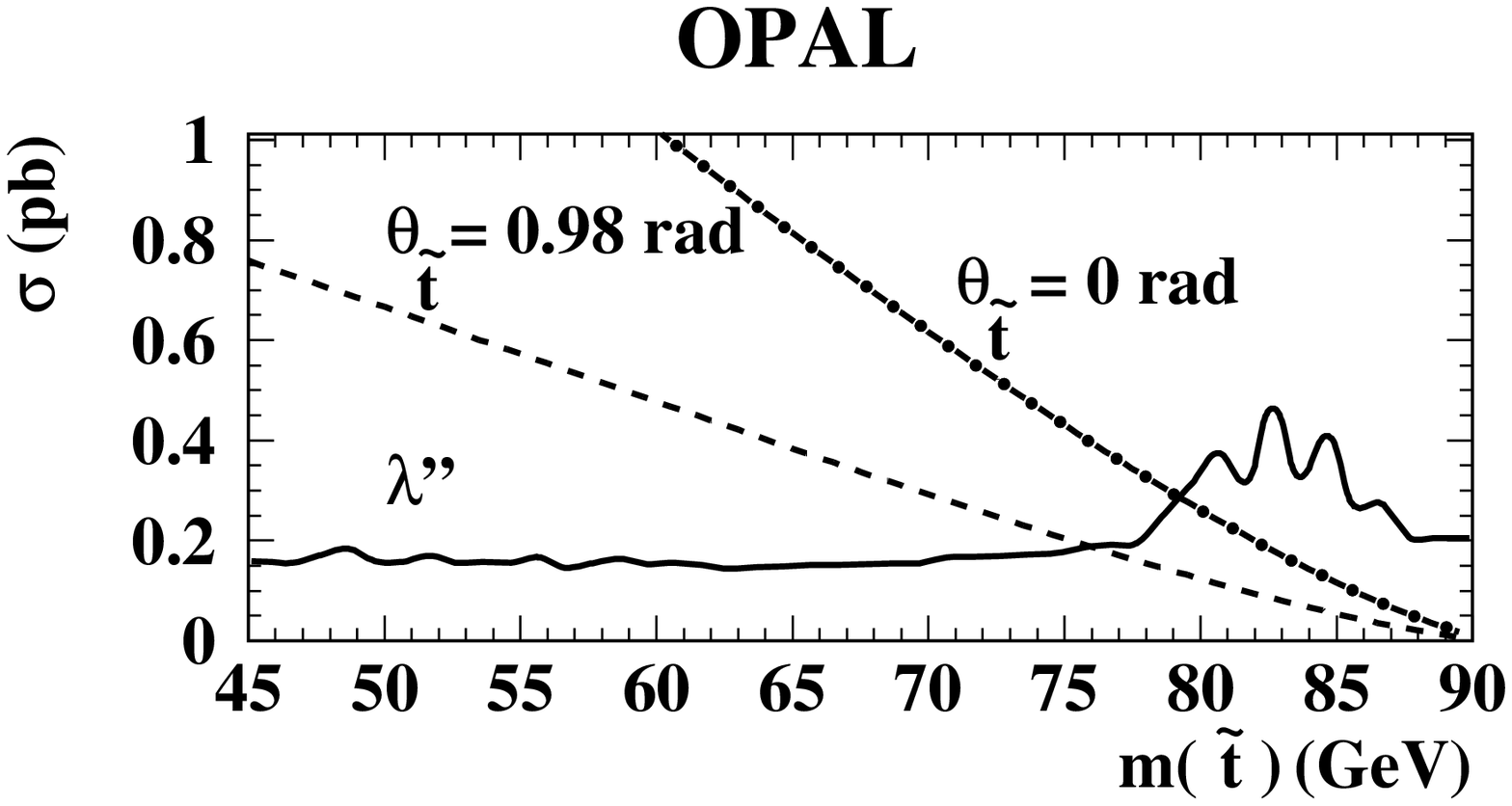,width=15cm} \\
\end{tabular}
\caption[]{\sl
Stop direct decays via a \lbpp\ coupling: 
Upper limits at the 95\% C.L. on the production cross-section.
Also shown are the maximum (dashed-dotted line) and minimum (dashed line)
cross-sections predicted by the MSSM,
corresponding to a mixing angle of 0 rad and 0.98 rad. }
\label{fig:limit_stop_4jets}
\end{figure}

For the stop decays via \lbpp couplings, 7 events satisfied the 
selection cuts. A cross-section limit of approximately 0.3~pb was 
derived for a stop mass up to $\approx$~75 GeV degrading slightly
in the range of the W mass as shown in Fig.~\ref{fig:limit_stop_4jets}.

\begin{table}[htb]
\begin{center}
\begin{tabular}{|c||c|c|}
\hline
Limits & $\theta_{\stopx} = 0$ rad & $\theta_{\stopx} = 0.98$ rad \\
\hline
\hline
$ \stopm \rightarrow {\rm e} +$ q & 86 GeV & 82 GeV \\
\hline
$ \stopm \rightarrow \mu +$ q & 86 GeV & 82 GeV \\
\hline
$ \stopm \rightarrow \tau +$ q & 81 GeV & 73 GeV \\
\hline
$ \stopm \rightarrow $ qq & 79 GeV & 76 GeV \\
\hline
\end{tabular}
\caption[]{\sl
Mass limits for stop for the two extreme values of the mixing
angle in the electron, muon and tau channels as well as in the 4-jet 
channel.}
\label{tab:limit_stop}
\end{center}
\end{table}

\section{Conclusions}
\label{sec:conclusions}

We have performed a search for pair produced 
sfermions with \Rparity\ violating 
decays using the data collected by the OPAL detector at 
$\sqrt{s} \simeq  183$~GeV corresponding to a luminosity of 
approximately 56
pb$^{-1}$.
Direct and indirect \Rparity\ violating decay modes of $\sell$, $\snu$ 
via the Yukawa-like
\lb\ and \lbp\ couplings as well as 
direct \Rparity\ violating decay modes of
$\stopx$ via \lbp\ and 
\lbpp\ were considered.

No significant excess of events has been observed in the data.
Upper limits on the pair production cross-sections for 
sfermions  
have been computed assuming that only \Rparity\ violating decays occur. 
These cross-section limits, within the MSSM frame used,
depend only on the mass of the sfermion and 
not on other SUSY parameters.
Mass limits were derived in the framework of the 
constrained Minimal Supersymmetric Standard Model 
whenever the predicted cross-sections were sufficiently large. 
\bigskip\bigskip\bigskip
\appendix
\par
{\Large\bf Acknowledgements}
\par
We particularly wish to thank the SL Division for the efficient operation
of the LEP accelerator at all energies
 and for their continuing close cooperation with
our experimental group.  We thank our colleagues from CEA, DAPNIA/SPP,
CE-Saclay for their efforts over the years on the time-of-flight and trigger
systems which we continue to use.  In addition to the support staff at our own
institutions we are pleased to acknowledge the  \\
Department of Energy, USA, \\
National Science Foundation, USA, \\
Particle Physics and Astronomy Research Council, UK, \\
Natural Sciences and Engineering Research Council, Canada, \\
Israel Science Foundation, administered by the Israel
Academy of Science and Humanities, \\
Minerva Gesellschaft, \\
Benoziyo Center for High Energy Physics,\\
Japanese Ministry of Education, Science and Culture (the
Monbusho) and a grant under the Monbusho International
Science Research Program,\\
Japanese Society for the Promotion of Science (JSPS),\\
German Israeli Bi-national Science Foundation (GIF), \\
Bundesministerium f\"ur Bildung, Wissenschaft,
Forschung und Technologie, Germany, \\
National Research Council of Canada, \\
Research Corporation, USA,\\
Hungarian Foundation for Scientific Research, OTKA T-016660, 
T023793 and OTKA F-023259.\\



\begin{thebibliography}{99}


\bibitem{ref:SUSY}
Y.~Gol'fand and E.~Likhtam, JETP Lett. {\bf 13} (1971) 323;\\
D.~Volkov and V.~Akulov, \PhysLett\ {\bf B46} (1973) 109;\\
J.~Wess and B.~Zumino, \NPhys\ {\bf B70} (1974) 39.

\bibitem{ref:MSSM}
H.P.~Nilles, \PhysRep\ {\bf 110} (1984) 1;\\
H.E.~Haber and G.L.~Kane, \PhysRep\ {\bf 117} (1985) 75.

\bibitem{ref:rparity}
P.~Fayet, in {\it ``Unification of the Fundamental Particle 
Interactions''},
eds. S.~Ferrara, J.~Ellis and P. Van Nieuewenhuizen, Plenum Press (1980) 
727.

\bibitem{ref:dreiner1}
H.~Dreiner, {\it ``An Introduction to Explicit \Rparity\ Violation''},
in {\it ``Perspectives on Supersymmetry''}, ed. G.L.~Kane (1997) 462,
hep-ph/9707435.

\bibitem{ref:opal_rpv_gauginos}
\OPALColl, G.~Abbiendi \etal {\it ``Searches for R-Parity Violating 
Decays of Gauginos at 183 GeV at LEP''},
CERN-EP/98-203, submitted to Eur. Phys. J. C.

\bibitem{ref:aleph_rpv_lle}
\ALEPHColl, R.~Barate \etal
\EuroPhys\ {\bf C4} (1998) 433.

\bibitem{ref:aleph_rpv_lqd}
\ALEPHColl, R.~Barate \etal 
\EuroPhys\ {\bf C7} (1999) 383.

\bibitem{ref:bhatta}
G.~Bhattacharyya, {\it ``\Rparity\ Violating Supersymmetric 
Yukawa Couplings: a mini-Review,''} \NPhys\ Proc. Suppl. {\bf A52} (1997) 83. 

\bibitem{ref:barger1}
V.~Barger, G.F.~Giudice and T.~Han,  \PhysRev\ {\bf D40} (1989) 2987.

\bibitem{ref:agashe}
K.~Agashe and M.~Graesser, \PhysRev\ {\bf D54} (1996) 4445.

\bibitem{ref:godbole}
R.M.~Godbole, P.~Roy and T.~Tata, \NPhys\ {\bf B401} (1993) 67.

\bibitem{ref:ellis}
G.~Bhattacharyya, J.~Ellis and K.~Sridhar, 
Mod.\ Phys.\ Lett.\ {\bf A10} (1995) 1583. 

\bibitem{ref:mohapatra}
R.N.~Mohapatra,  \PhysRev\ {\bf D34} (1986) 3457; \\
M.~Hirsh \etal \PRL\ {\bf 75} (1995) 17.

\bibitem{ref:goity}
J.L.~Goity and M.~Sher, \PhysLett\ {\bf B346} (1995) 69.

\bibitem{ref:smir}
A.Y.~Smirnow and F.~Vissani, \PhysLett\ {\bf B380} (1996) 317.

\bibitem{opal_2fermion} 
\OPALColl, K.~Ackerstaff \etal 
\EuroPhys\ {\bf C6} (1999) 1.

\bibitem{ref:stable-part}
\OPALColl, K.~Ackerstaff \etal 
\PhysLett\ {\bf B433} (1998) 195.


\bibitem{ref:barger2} 
V.~Barger, W.-Y.~Keung and R.J.N.~Phillips, \PhysLett\ {\bf B364} 
(1995) 27, Erratum-ibid {\bf B377} (1996) 486.  

\bibitem{ref:carena}
M.~Carena \etal  
\PhysLett\ {\bf B395} (1997) 225.

\bibitem{ref:dreiner2} 
H.~Dreiner, E.~Perez and Y.~Sirois, in {\it ``Future Physics at HERA''},
eds. G.~Ingelman, A.~De Roeck, R.~Klanner, DESY 96-235, vol.1, p.~295, 
hep-ph/9703444.



\bibitem{ref:GOPAL}
J.~Allison \etal 
\NIM\ {\bf A317} (1992) 47.

\bibitem{ref:SUSYGEN}
S.~Katsanevas and S.~Melachroinos, in {\it ``Physics at LEP2''},
eds. G.~Altarelli, T.~Sj\"{o}strand and
F.~Zwirner, CERN 96--01, vol. 2, p. 328; \\
S.~Katsanevas and P.~Morawitz, \CPC\ {\bf 112} (1998) 23.

\bibitem{ref:stoppaper}
\OPALColl, K.~Ackerstaff \etal 
\ZPhys\ {\bf C75} (1997) 409.

\bibitem{ref:JETSET1}
T.\ Sj\"{o}strand and M.\ Bengtsson, \CPC\ {\bf 43} (1987) 367;\\
{\it ``PYTHIA 5.6 and JETSET 7.3, Physics and Manual''}, CERN--TH.\ 6488/92; \\
T.\ Sj\"{o}strand, \CPC\ {\bf 82} (1994) 74.

\bibitem{ref:JETSET2} 
B.~Anderson \etal \PhysRep\ {\bf 97} (1993) 31.

\bibitem{ref:peterson} 
C.~Peterson, D.~Schlatter, I.~Schmitt and P.M.~Zerwas, 
\PhysRev\ {\bf D27} (1983) 105.


\bibitem{ref:PHOJET}
R.~Engel and J.~Ranft, \PhysRev\ {\bf D54} (1996) 4244;\\
R.~Engel, \ZPhys\ {\bf C66} (1995) 203. 

\bibitem{ref:herwig}
G.~Marchesini \etal \CPC\ {\bf 67} (1992) 465.

\bibitem{ref:VERMASEREN}
R.~Bhattacharya, J.~Smith and G.~Grammer, \PhysRev\ {\bf D15} (1977) 3267; \\
J.~Smith, J.A.M.~Vermaseren and G.~Grammer, \PhysRev\ {\bf D15} (1977) 3280.

\bibitem{ref:grace4f} 
J. Fujimoto \etal 
\CPC\ {\bf 100} (1997) 128.

\bibitem{ref:BHWIDE}
S.~Jadach \etal in  {\it ``Physics at LEP2''},
eds. G.~Altarelli, T.~Sj\"ostrand and 
F.~Zwirner, CERN 96--01, vol.2, p.~229;\\
S.~Jadach, W.~Placzek and B.F.L.~Ward, 
\PhysLett\ {\bf B390} (1997) 298.

\bibitem{ref:KORALZ}
S.~Jadach, B.F.L.~Ward and Z.~W\c{a}s, \CPC\ {\bf 79} (1994) 503.


\bibitem{ref:OPAL-detector}
\OPALColl, K.~Ahmet \etal \NIM\ {\bf A305} (1991) 275;\\
P.P.~Allport \etal \NIM\ {\bf A324} (1993) 34;\\
P.P.~Allport \etal \NIM\ {\bf A346} (1994) 476.

\bibitem{ref:SW}
B.E.~Anderson \etal IEEE Transactions on Nuclear Science {\bf 41} (1994) 845.


\bibitem{ref:leptpairs}
\OPALColl, R.~Akers \etal 
\ZPhys\ {\bf C61} (1994) 19.

\bibitem{ref:slept161} 
\OPALColl, K.~Ackerstaff \etal \PhysLett\ {\bf B396} (1997) 301.

\bibitem{ref:OPAL-Higgs}
\OPALColl, M.Z.~Akrawy \etal 
\PhysLett\ {\bf B253} (1991) 511.

\bibitem{ref:NN}
\OPALColl, R.~Akers  \etal 
\PhysLett\ {\bf B327} (1994) 411.

\bibitem{ref:elecbarrel}
\OPALColl, R.~Akers  \etal 
\ZPhys\ {\bf C60} (1993) 199.

\bibitem{ref:elecendcap}
\OPALColl, P.D.~Acton \etal 
\ZPhys\ {\bf C60} (1993) 19.

\bibitem{ref:conversion}
\OPALColl, G.~Alexander \etal 
\ZPhys\ {\bf C70} (1996) 357.


\bibitem{ref:wwpaper}
\OPALColl, K.~Ackerstaff \etal 
\PhysLett\ {\bf B389} (1996) 416.


\bibitem{ref:cousins}
R.D.~Cousins and V.L.~Highland, \NIM\ {\bf A320} (1992) 331.
 


\bibitem{ref:durham}
N. Brown and W.J. Stirling, \PhysLett\ {\bf B252} (1990) 657; \\
S. Bethke, Z. Kunszt, D. Soper and W.J. Stirling, Nucl. Phys. {\bf B370}
(1992) 310; \\
S. Catani \etal \PhysLett\  {\bf B269} (1991) 432; \\
N. Brown and W.J. Stirling, \ZPhys\ {\bf C53} (1992) 629.


\bibitem{ref:opalstop}
\OPALColl, K.~Ackerstaff \etal 
\EuroPhys\ {\bf C6} (1999) 225.

\bibitem{ref:bowler} 
M.G.~Bowler, \ZPhys\ {\bf C11} (1981) 169.


\bibitem{ref:smpaper}
\OPALColl, K. Ackerstaff \etal
\EuroPhys\ {\bf C1} (1998) 425.

\bibitem{ref:mssmpaper}
\OPALColl, K. Ackerstaff \etal 
\EuroPhys\ {\bf C5} (1998) 19.

\bibitem{ref:lumino}
OPAL Collaboration, K. Ackerstaff \etal Phys. Lett. {\bf B391} (1997) 221.






























\bibitem{tauID}
\OPALColl, G. Abbiendi \etal 
\EuroPhys\ {\bf C7} (1999) 407.


\bibitem{LEP2MH} 
\OPALColl, K.~Ackerstaff \etal 
\EuroPhys\ {\bf C2} (1998) 441.

\bibitem{dpar} 
R.K.~Ellis, D.A.~Ross and A.E.~Terrano, 
\NPhys\  {\bf B178} (1981) 421.


\bibitem{sprime}
\OPALColl, G.~Alexander \etal 
\PhysLett\ {\bf B376} (1996) 232.

\bibitem{ref:stopgen}
E. Accomando \etal, in {\it ``Physics at LEP2''},, 
eds. G.~Altarelli, T.~Sj\"{o}strand and F.~Zwirner,
CERN 96-01, vol. 2, 299.

\bibitem{ch172}
\OPALColl, K.~Ackerstaff \etal 
\PhysLett\ {\bf B426} (1998) 180.





\bibitem{likelihood}
A.G. Frodesen, O. Skeggestad, and H. Tofte, 
{\it ``Probability and Statistics in
Particle Physics''}, Universitetsforlaget, 1979, ISBN 82-00-01-01906-3; \\
S.L. Meyer, {\it ``Data Analysis for Scientists and Engineers''}, John Wiley and
Sons, 1975, ISBN 0-471-59995-6.



\end{thebibliography}
\end{document}